\pdfoutput=1
\documentclass[aps,prb,twocolumn,groupedaddress,showpacs,preprintnumbers,superscriptaddress]{revtex4-2}
\usepackage{graphicx}
\usepackage{color}
\usepackage{xcolor}
\usepackage{hyperref}
\usepackage{array}
\usepackage{mathtools}
\usepackage[bitstream-charter, greekfamily=default]{mathdesign}
\usepackage{physics}
\usepackage[capitalize]{cleveref}

\newcommand{\de}{\mathrm{d}}
\newcommand{\im}{\mathrm{Im}}
\newcommand{\avg}[1]{\left\langle #1 \right\rangle}

\newcommand{\ft}{f_{\rm typ}}
\newcommand{\fm}{f_{\rm min}}
\newcommand{\at}{a_{\rm typ}}
\newcommand{\am}{a_{\rm min}}
\newsavebox{\panelbox}
\newcommand{\panel}[2]{\sbox{\panelbox}{#2}\usebox{\panelbox}\llap{\makebox[\wd\panelbox][l]{\raisebox{\dimexpr\ht\panelbox+2pt\relax}{\small (#1)}}}}

\begin{document}

\title{Anomalous corrections to the inverse participation ratio in the localized phase on the Bethe lattice: \\ insights from a solvable ``toy'' non-linear $\sigma$-model}

\author{Tommaso Rizzo}
\affiliation{Dip. Fisica, Universit\`a ``Sapienza'', Piazzale A. Moro 2, I-00185, Rome, Italy}
\affiliation{ISC-CNR, UOS Rome, Universit\`a ``Sapienza'', Piazzale A. Moro 2, I-00185, Rome, Italy}
\author{Marco Tarzia}
\affiliation{LPTMC, CNRS-UMR 7600, Sorbonne Universit\'e, 4 Pl. Jussieu, F-75005 Paris, France}
\affiliation{Institut Universitaire de France, 1 rue Descartes, 75231 Paris Cedex 05, France}
\author{Angelica Vecchiarelli}
\affiliation{Dip. Fisica, Universit\`a ``Sapienza'', Piazzale A. Moro 2, I-00185, Rome, Italy}
\affiliation{LPTMC, CNRS-UMR 7600, Sorbonne Universit\'e, 4 Pl. Jussieu, F-75005 Paris, France}

\begin{abstract}
In the localized phase of the Anderson model on the Bethe lattice the inverse participation ratio (IPR) approaches its asymptotic value as the imaginary spectral regulator $\eta$ is sent to zero with an anomalous power law, $I_2(\eta) - I_2^\infty \propto \eta^{2\beta - 1}$, where $\beta \in (1/2,1)$ is the disorder-dependent tail exponent of the local density of states distribution. In this paper we derive this result analytically. We do so within a ``toy'' non-linear $\sigma$-model with non-compact graded symmetry defined on the hyperbolic superplane, which shares the exotic criticality of the Anderson transition on the Bethe lattice and whose cavity equation in the presence of a symmetry-breaking field $H$ --- the counterpart of $\eta$ --- can be analyzed in full detail. The small-$H$ solution organizes into a bulk regime, where the correction to the field distribution factorizes into an $H$-dependent amplitude times a universal function $\ft$, and a boundary layer at the scale set by $H$. Matching the two fixes the amplitude to scale as $H^{2\beta-1}$: the exponent is controlled by the subdominant root $1-\beta$ of the linearized cavity equation, which acquires here a direct physical meaning. At the critical point the two roots merge and the correction becomes logarithmic, $\propto 1/|\ln H|$. We then show that, under a few natural assumptions, the same mechanism operates in the cavity equations of the Anderson model: the anomalous correction is a zero mode of the $\eta=0$ recursion, with tail $\propto \hat m^{-(2-\beta)}$, whose amplitude $\propto \eta^{2\beta-1}$ is fixed by matching to the boundary layer at $\hat m \sim 1/\eta$, beyond which the distribution is the inverted image of the $\eta=0$ one and decays as $\hat m^{\beta-3}$. These predictions are tested against the numerical solution of the cavity equations for the full distribution of the local density of states, and shown to carry over, in the appropriate variables, to the finite-size corrections to the distribution of the eigenstates' amplitudes. We also show that, through a spectral identity and an exact sum rule, both the finite-regulator and the finite-size corrections are fixed by the correlations of pairs of eigenstates close in energy, so that the same exponent governs exactly the decay $\omega^{-\mu}$, with $\mu = 2(1-\beta)$, of the correlation function of two eigenstates at energy distance $\omega$ and of its generalizations to higher powers of the amplitudes. We confirm this relation, including its amplitude, by exact diagonalization and by population dynamics directly on the infinite lattice, and analytically in the toy model. A phenomenological counting of the Mott resonances between distant localized eigenstates, based on the exact statistics of the eigenstate amplitudes, whose small-amplitude tail is controlled by $\beta$, gives it a physical interpretation. On finite graphs, exact diagonalization shows that the finite-size corrections reflect a deficit of hybridized, doubly-peaked eigenstates: those whose two centers would lie farther apart than the graph allows. Finally, we derive the finite-size-scaling form of the IPR implied by this mechanism, in which the scaling variable is $(2\beta-1)\ln N$ and the critical point is approached as $1/\ln N$, and show that it describes both the finite-$\eta$ and the finite-$N$ data. This form yields analytically the renormalization-group flow of the running fractal dimension proposed phenomenologically in previous work, which turns out to be exactly of the Kosterlitz type, with $2\beta - 1$ as the conserved quantity that labels the trajectories and fixes the line of fixed points, and shows that finite fractal dimensions of the localized states at accessible sizes are a logarithmic finite-size effect.
\end{abstract}

\maketitle

\section{Introduction} \label{sec:intro}

Anderson localization (AL)~\cite{anderson1958absence} is one of the most spectacular phenomena in condensed matter physics~\cite{lee1985disordered,evers2008anderson,lagendijk2009fifty}, and its reach today extends far beyond the physics of disordered non-interacting electrons, from quantum chaos and thermalization~\cite{basko2006metal,gornyi2005interacting,altshuler1997quasiparticle} to random matrix theory and the physics of many-body localized systems~\cite{altman2015universal,nandkishore2015many,abanin2017recent,alet2018many,abanin2019colloquium,sierant2024many}. Within this broad landscape, the Anderson model on the Bethe lattice (BL) holds a privileged position. Owing to its tree-like structure, the model is exactly solvable~\cite{anderson1973selfconsistent,abou1974self,efetov1985anderson,efetov1987density,efetov1987anderson,zirnbauer1986localization,zirnbauer1986anderson,verbaarschot1988graded,mirlin1991localization,mirlin1991universality,fyodorov1991localization,fyodorov1992novel,mirlin1994statistical,mirlin1994distribution,tikhonov2019statistics,tikhonov2019critical,biroli2010anderson,parisi2019anderson,pascazio2023anderson,rizzo2024localized}, and it has served for five decades as the mathematical sandbox that has shaped our physical intuition about the localization transition. Moreover, since the upper critical dimension of AL is infinite~\cite{castellani1986upper,mirlin1994distribution,tarquini2017critical,baroni2024corrections}, the BL constitutes the natural mean-field starting point for localization problems both in real space and in the Hilbert space of interacting systems~\cite{altshuler1997quasiparticle}.

The order parameter of the localized phase is the inverse participation ratio (IPR) of the eigenstates, $I_2 = \sum_i |\psi_\alpha(i)|^4$, which measures the inverse of the volume over which an eigenstate is spread. On the BL, $I_2$ vanishes as $1/N$ in the delocalized phase, jumps discontinuously to a finite value at the critical disorder $W_c$~\cite{efetov1985anderson,efetov1987density,efetov1987anderson,zirnbauer1986localization,zirnbauer1986anderson,verbaarschot1988graded,mirlin1991localization,mirlin1991universality,fyodorov1991localization,fyodorov1992novel,mirlin1994statistical,tikhonov2019statistics}, and then grows with a square-root singularity in the localized phase~\cite{rizzo2024localized}. On the infinite BL the asymptotic value $I_2^\infty$ is accessible directly, from the $\eta \to 0^+$ limit of the cavity equations~\cite{mirlin1994distribution,rizzo2024localized}. On any finite graph, instead, it is approached with corrections, which are due to the loops that the BL limit discards~\cite{montanari2005compute,parisi2006loop,chertkov2006loop,baroni2024corrections,ferrari2013finite,lucibello2014finite,metz2014finite}. Controlling these corrections matters well beyond the extrapolation of numerical data: since, as recalled above, the BL is the mean-field starting point for localization both in finite dimension and on the Hilbert-space graphs of interacting systems, they are the first ingredient of a systematic $1/d$ expansion around the BL solution~\cite{castellani1986upper,tarquini2017critical,baroni2024corrections}, and of a quantitative description of many-body localization in terms of localization on these graphs~\cite{altshuler1997quasiparticle,de2013ergodicity,logan2019many,tikhonov2021anderson,biroli2017delocalized,biroli2020anomalous,herre2023ergodicity,biroli2023large,sierant2024many}. A second, related issue is that on a finite graph there is, strictly speaking, no eigenvalue at the target energy, so that local observables are usually computed with an imaginary regulator $\eta$ added to the energy, the limit $\eta \to 0^+$ being taken after the thermodynamic limit. How $I_2$ approaches $I_2^\infty$, as $N \to \infty$ and as $\eta \to 0^+$, turns out to be far from trivial.

In a companion Letter~\cite{letter} we have shown that in the localized phase of the Anderson model on the BL both corrections are anomalous: they vanish as power laws with a {\it continuously varying}, disorder-dependent exponent,
\begin{equation} \label{eq:kappa_intro}
    I_2(\eta) - I_2^\infty \simeq c_\eta \, \eta^{\kappa(W)} \, , \qquad
    I_2(N) - I_2^\infty \simeq c_N \, N^{-\kappa(W)} \, ,
\end{equation}
with the same exponent $\kappa = 2 \beta - 1 \in (0,1)$ in the two cases, where $\beta(W)$ is the tail exponent of the probability distribution of the local density of states (LDoS) in the localized phase~\cite{anderson1973selfconsistent,abou1974self,mirlin1991localization,mirlin1991universality,mirlin1994distribution,rizzo2024localized}, which behaves as $\beta \simeq 1/2 + c_\beta \sqrt{W - W_c}$ close to the transition and tends to one at strong disorder. A non-trivial exponent that varies continuously throughout an entire phase is reminiscent of the low-temperature phase of the Berezinskii--Kosterlitz--Thouless (BKT) transition~\cite{berezinskii1971destruction,kosterlitz1973ordering}, where however it stems from quasi-long-range order, which is absent here (the analogy becomes exact, however, for the finite-size flow of the running fractal dimension, Sec.~\ref{sec:previous}), and more closely of Griffiths phases~\cite{griffiths1969nonanalytic,vojta2006rare}, where non-universal power laws arise from rare regions in the tails of the disorder distribution. The latter is in fact the closest analogy, since, as we shall see, the corrections to the IPR originate from the power-law tail of the LDoS distribution, i.e.\ from rare resonant configurations, and $\kappa$ is inherited from the tail exponent $\beta$; to our knowledge, however, this is the first instance in which such a mechanism governs the finite-size (or finite-regulator) corrections of an order parameter far from the critical point.

Beyond its conceptual interest, this result has a direct practical consequence, since the finite-size scaling analysis of the IPR on the BL has been the subject of considerable debate~\cite{tikhonov2019statistics,garcia2022critical,10.21468/SciPostPhys.15.2.045,garcia2017scaling,chen2023quantum,chen2024describing,vanoni2023renormalization,garcia2026connecting,de2014anderson,altshuler2016nonergodic,tikhonov2016anderson}: in the localized phase the corrections are not controlled by the critical point alone, but by an exponent that varies continuously with $W$, $\kappa = 2\beta - 1 \propto \sqrt{W - W_c}$ close to $W_c$. Scaling ansatzes with a disorder-dependent scaling length can accommodate this behavior --- the linear-scaling form of Refs.~\cite{garcia2017scaling,garcia2022critical}, for instance, reproduces $N^{-\kappa}$ if its scaling length is identified with $1/\kappa$ --- but the length that controls the corrections is then $1/\kappa \propto (W - W_c)^{-1/2}$, and our data favor it over a length diverging with $\nu = 1$. In Sec.~\ref{sec:FSS} we derive the corresponding scaling form, $I_2 - I_2^\infty \simeq \Phi\big((2\beta-1)\ln N\big)/\ln N$, which crosses over to a $1/\ln N$ correction at $W_c$, and show that it describes both the finite-$\eta$ and the finite-$N$ data. We then use it to reexamine previous finite-size analyses (Sec.~\ref{sec:previous}). It shows that the two-parameter renormalization-group flow of the running fractal dimension proposed phenomenologically in Ref.~\cite{vanoni2023renormalization}, by analogy with the BKT transition, is exact in the localized phase and at the critical point: the exponent $\kappa = 2\beta - 1$ is conserved along each trajectory, as the invariant of the Kosterlitz equations, so that the flow can be integrated exactly, with constants fixed by $\kappa$, by $I_2^\infty$ and by the amplitude of the scaling function. It also shows that finite fractal dimensions of the localized states at accessible sizes, and the absence of a divergence of their derivative with respect to disorder at $W_c$~\cite{pino2020scaling}, are logarithmic finite-size effects.

The exponent also has a direct meaning in terms of eigenstate correlations. Through a spectral identity and an exact sum rule, we show that both corrections are entirely due to the spatial overlap of pairs of eigenstates close in energy --- pairs that a finite $\eta$ does not resolve, or that are missing on a finite graph. This relation is exact and requires no assumption on the nature of these pairs: it implies $\kappa = 1 - \mu$, where $\mu$ is the exponent of the power-law decay $\omega^{-\mu}$ of the correlation function of two eigenstates at energy distance $\omega$~\cite{tikhonov2019statistics,tikhonov2021eigenstate}, and of its generalizations to higher powers of the wavefunction amplitudes, which control the corrections to the higher moments $I_q$, and it fixes the amplitude of the correlation function in terms of that of the correction to the IPR. Since we derive $\kappa = 2\beta - 1$ analytically, the exponent $\mu = 2(1-\beta)$ and the amplitude of the eigenstate correlations follow with no further input. A direct computation of the eigenstate correlations can only confirm this prediction, and it does: numerically for the Anderson model, by exact diagonalization and by population dynamics directly on the infinite BL, and analytically for the toy model introduced below, by continuing its cavity solution to a complex field.

Neither the computation of $I_2(\eta)$ nor that of the eigenstate correlations, however, tells which pairs of eigenstates are involved. This physical picture is provided by a phenomenological, Mott-type counting of resonances in real space: the tail of the LDoS distribution is produced by eigenstates localized far from a given site, whose small weight on it is distributed with the exponent $\beta$, and counting their resonances with the states localized at that site gives the exponent $\mu = 2(1-\beta)$, though not the amplitude; the same counting shows that $\beta > 1/2$ is the condition for these resonances not to proliferate, i.e.\ for localization. Exact diagonalization confirms this picture directly. Two eigenstates that hybridize resonantly are both spread over the same two localization centers, as the symmetric and antisymmetric combinations of the states localized on each of them, and the distance between the two centers grows as $\ln(1/\omega)$ as their splitting $\omega$ decreases. On a finite graph, the hybridized pairs whose two centers would lie farther apart than the graph allows cannot form: their eigenstates remain localized on a single center, with a larger IPR, and this deficit of doubly-peaked eigenstates accounts for most of the finite-size correction.

The purpose of the present paper is twofold. First, we provide an analytical derivation of the first of Eqs.~\eqref{eq:kappa_intro}, i.e.\ of the finite-regulator corrections, and identify the mechanism behind the anomalous exponent. Second, we explain why the finite-regulator and finite-size corrections share the same exponent: through a spectral identity and an exact sum rule, both are expressed in terms of the same correlation function of eigenstates close in energy, a finite $\eta$ and a finite $N$ acting as two infrared cutoffs on it, at splittings of order $\eta$ and of order of the mean level spacing, respectively --- the latter property being established numerically (Sec.~\ref{sec:finiteN}). The exact solution of the Anderson model on the BL is formulated in terms of a self-consistent integral equation for the joint probability distribution of the real and imaginary parts of the local Green's functions~\cite{anderson1973selfconsistent,abou1974self}, which is not amenable to a complete analytic treatment at finite $\eta$. Even its field-theoretical reformulation as a non-linear $\sigma$-model (NL$\sigma$M) with non-compact supersymmetry~\cite{schafer1980disordered,efetov1985anderson,efetov1987density,efetov1987anderson,zirnbauer1986localization,zirnbauer1986anderson,verbaarschot1988graded,mirlin1991localization,mirlin1991universality,fyodorov1991localization,fyodorov1992novel,mirlin1994statistical}, which has been solved on the BL in the $\eta \to 0^+$ limit, has never been analyzed, to the best of our knowledge, in the presence of a finite regulator. We therefore follow a different route, and study a ``toy'' NL$\sigma$M defined on the hyperbolic superplane, introduced in Refs.~\cite{zirnbauer1990fourier,zirnbauer1991fourier} precisely as a simplified model of Anderson localization and further studied in Refs.~\cite{Gruzberg_1996,dupre1996localization}. The model retains the non-compact graded symmetry structure of the full problem, and Gruzberg and Mirlin~\cite{Gruzberg_1996} showed that on the BL it displays exactly the same exotic criticality as the true Anderson transition, including the structure of the tail exponent $\beta$ and its square-root singularity. Its cavity equation involves a single real function, and the symmetry-breaking field $H$ that regularizes the non-compact zero mode plays the role of $\eta$. This makes it possible to carry out the small-$H$ analysis in complete detail.

The outcome is a transparent mechanism. The small-$H$ solution of the cavity equation organizes into two regimes: a bulk regime, where the correction to the $H=0$ fixed point factorizes into an $H$-dependent amplitude times a universal scaling function, and a boundary layer at the scale set by $H$. The $H=0$ cavity equation, linearized around its trivial fixed point, admits two power-law solutions, with exponents $\beta$ and $1-\beta$ symmetric about $1/2$; the $H=0$ solution is controlled by the larger root $\beta$, while the smaller one normally plays no role. The matching between the bulk and the boundary layer shows that the bulk correction is carried precisely by this subdominant root, and that its amplitude scales as $H^{2\beta - 1}$. At the critical point the two roots merge and the matching produces a logarithmic correction, $\propto 1/|\ln H|$; logarithmically slow corrections had been observed numerically in both models in Ref.~\cite{tikhonov2019statistics}, where they were fitted with effective exponents that we argue to be preasymptotic (Sec.~\ref{sec:critical}). We then argue that the same structure is present in the cavity equations of the Anderson model: there, the recursion for the joint distribution of $g$ and $\hat g$ depends on $\eta$ only through $\eta^2$, so that a non-analytic correction can only come from a zero mode of the $\eta = 0$ recursion linearized around its fixed point, whose tail is governed by the second root $1-\beta$ of the Abou-Chacra--Anderson--Thouless condition~\cite{anderson1973selfconsistent,abou1974self}; matching to the boundary layer at $\hat m \sim 1/\eta$ fixes again the amplitude $\propto \eta^{2\beta-1}$, and the structure of the recursion beyond the boundary layer predicts a new power law $\hat m^{\beta-3}$ for the far tail of the LDoS distribution at the resonant sites. These predictions go well beyond the scaling of the average IPR, since they concern the full probability distributions, and we test them against the numerical solution of the cavity equations. Finally, since the bulk correction is a zero mode of the same linear operator whatever the perturbation that injects it, the same scaling function must describe the finite-size corrections to the distribution of the eigenstates' amplitudes, once expressed in the appropriate variables; we show that this is indeed the case.

The paper is organized as follows. In Sec.~\ref{sec:problem} we set up the problem, recalling the relation between the IPR and the local Green's functions of the Anderson model on the BL, the exact identities that relate the finite-$\eta$ and finite-$N$ corrections to the correlations of eigenstates close in energy, the structure of the distribution of the local Green's functions in the localized phase, and the numerical evidence for the anomalous corrections. In Sec.~\ref{sec:realspace} we show that the spectral identity fixes the exponent and the amplitude of the correlations of eigenstates close in energy, check this by exact diagonalization and population dynamics, interpret it through a resonance counting based on the statistics of the eigenstate amplitudes, whose picture we test numerically, and show that the finite-size corrections are due to the hybridized pairs missing on a finite graph. In Sec.~\ref{sec:toy} we introduce the toy NL$\sigma$M, its cavity equation and its IPR proxy. Section~\ref{sec:scaling} contains the small-$H$ scaling analysis away from criticality, and Sec.~\ref{sec:critical} the analysis at and near the critical point. In Sec.~\ref{sec:anderson} we show how the same mechanism operates in the Anderson model under a few explicit assumptions, and derive the far-tail exponent $\beta - 3$. Section~\ref{sec:numerics} presents the numerical tests on the full distributions, including the extension to the eigenstates' amplitudes, and Sec.~\ref{sec:FSS} the finite-size-scaling form of the IPR, its test on the finite-$\eta$ and finite-$N$ data, and its consequences for previous finite-size analyses. We conclude in Sec.~\ref{sec:conclusions}.

\section{The problem: finite-regulator and finite-size corrections to the IPR on the Bethe lattice} \label{sec:problem}

\subsection{Model, observables and an exact sum rule}

We consider the Anderson tight-binding Hamiltonian
\begin{equation} \label{eq:H}
    \mathcal{H} = - t \sum_{\langle i, j \rangle} \left( \vert i \rangle \langle j \vert + \vert j \rangle \langle i \vert  \right) + \sum_{i=1}^N \epsilon_i \vert i \rangle \langle i \vert \, ,
\end{equation}
with unit hopping $t=1$ between nearest neighbors and on-site energies $\epsilon_i$ independently drawn from the uniform distribution $p(\epsilon)$ of width $W$ on $[-W/2,W/2]$. The lattice is the BL of branching ratio $k$ (connectivity $k+1$), realized at finite $N$ by random regular graphs; we present results for $k=2$, for which the localization transition sits at $W_c \simeq 18.17$~\cite{tikhonov2019critical,parisi2019anderson}. We focus on the band center, $E = 0$.

Averaging over the eigenstates at a given energy requires, at finite $N$, a regularization of the discrete spectrum. On the infinite BL, the natural regulated quantity is expressed in terms of the resolvent $G(z) = [\mathcal{H} - z]^{-1}$ at $z = {\rm i}\eta$ through the local Green's functions $G_{ii}$~\cite{mirlin2000statistics,evers2008anderson,rizzo2024localized},
\begin{equation} \label{eq:I2G}
    I_2(\eta) = \frac{\eta \avg{ |G_{ii}|^2}}{\avg{\im G_{ii}}} \, ,
\end{equation}
where the average is over the disorder (and, equivalently, over the sites), $\avg{\rho} = \avg{\im G_{ii}}/\pi$ is the average density of states per site, and $I_2^\infty = \lim_{\eta \to 0^+} I_2(\eta)$. Note that Eq.~\eqref{eq:I2G} is a ratio of averages. Its numerator is, up to the factor $\eta$, the correlation function of local Green's functions $K_1(r,\eta)$ of Ref.~\cite{tikhonov2019statistics} at coinciding points, $\avg{|G_{ii}|^2} = K_1(r=0,\eta)$, so that $I_2(\eta) = \eta K_1(0,\eta)/(\pi\avg{\rho})$.

The content of Eq.~\eqref{eq:I2G} is made explicit by inserting the spectral representation $G_{ii} = \sum_\alpha |\psi_\alpha(i)|^2/(E_\alpha - {\rm i}\eta)$ (at $E=0$). The denominator becomes $N^{-1}\sum_\alpha \eta/(E_\alpha^2+\eta^2)$, while the numerator separates into a diagonal piece, involving a single eigenstate $\alpha$, and an off-diagonal piece, involving pairs $\alpha \neq \alpha'$:
\begin{equation} \label{eq:split}
\begin{aligned}
    \eta \frac{1}{N}\sum_i |G_{ii}|^2 =
    \frac{1}{N}\sum_\alpha \Big(\sum_i |\psi_\alpha(i)|^4\Big)
    \frac{\eta}{E_\alpha^2+\eta^2} \\
    {}+ \frac{1}{N}\sum_{\alpha\neq\alpha'} \Big(\sum_i
    |\psi_\alpha(i)|^2|\psi_{\alpha'}(i)|^2\Big) \eta\,
    \frac{E_\alpha E_{\alpha'} + \eta^2}{(E_\alpha^2+\eta^2)(E_{\alpha'}^2+\eta^2)} \, .
\end{aligned}
\end{equation}
The diagonal piece contributes to $I_2(\eta)$
\begin{equation} \label{eq:I2eta}
    \frac{\avg{\sum_{\alpha} \sum_{i} \left|\psi_\alpha (i) \right|^{4} \eta/(E_\alpha^2 + \eta^2)}}{\avg{\sum_{\alpha} \eta/(E_\alpha^2 + \eta^2)}} \, ,
\end{equation}
i.e.\ the IPR of the individual eigenstates averaged with a Lorentzian weight of width $\eta$ around $E=0$ (each eigenstate, not each sample, carrying its own weight), which is finite and regular as $\eta \to 0$ and differs from $I_2^\infty$ only by analytic terms $O(\eta)$. The off-diagonal piece involves pairs of near-resonant eigenstates, each weighted by a factor $O(1/\eta)$, and would naively diverge as $\eta \to 0$ unless the spatial overlap $\sum_i |\psi_\alpha(i)|^2 |\psi_{\alpha'}(i)|^2$ of two such states itself vanishes. Averaged over a window of energies much wider than $\eta$, it takes the form
\begin{equation} \label{eq:spectral}
    I_2(\eta) - I_2^\infty \simeq \avg{\rho} \int_{-\infty}^{\infty} \de\omega \, N^2\mathcal{C}(\omega)\, \frac{4\eta^2}{\omega^2 + 4\eta^2} \, ,
\end{equation}
up to analytic $O(\eta)$ terms, where
\begin{equation} \label{eq:Cdef}
    N^2\mathcal{C}(\omega) = N \Big\langle \sum_i |\psi_\alpha(i)|^2 |\psi_{\alpha'}(i)|^2 \Big\rangle_{E_\alpha = E,\; E_{\alpha'} = E+\omega} \, ,
\end{equation}
the average running over the pairs of eigenstates with energies $E_\alpha = E$ (the reference energy, $E = 0$ here) and $E_{\alpha'} = E + \omega$, is the correlation function of two eigenstates at energy distance $\omega$~\cite{chalker1990scaling,fyodorov1997strong,cuevas2007two}, studied on the BL in Refs.~\cite{tikhonov2019statistics,tikhonov2021eigenstate} [denoted there by $\beta(\omega)$], which is obtained from $K_1$ by analytic continuation to real frequency~\cite{tikhonov2019statistics}. The {\it finite-regulator correction} $I_2(\eta) - I_2^\infty$, the object of this paper, is thus entirely due to the spatial overlap of pairs of eigenstates within $\sim \eta$ of each other in energy.

Equation~\eqref{eq:spectral} has an exact counterpart at finite $N$ and $\eta = 0$, which follows from a simple sum rule. Since $\sum_{\alpha'} |\psi_{\alpha'}(i)|^2 = 1$ on every site, the overlaps $O_{\alpha\alpha'} = \sum_i |\psi_\alpha(i)|^2 |\psi_{\alpha'}(i)|^2$ of a given eigenstate with all the eigenstates sum to one, and its IPR is exactly $I_\alpha = O_{\alpha\alpha} = 1 - \sum_{\alpha' \neq \alpha} O_{\alpha\alpha'}$. Averaging over the eigenstates at energy $E$, on a graph of $N$ sites and at $\eta = 0$,
\begin{equation} \label{eq:sumrule}
    I_2(N) = 1 - \int_{-\infty}^{\infty} \de\omega \, \avg{\rho(E+\omega)} \, N^2\mathcal{C}_N(\omega) \, ,
\end{equation}
where $\mathcal{C}_N$ is the eigenstate correlation function of Eq.~\eqref{eq:Cdef} on the finite graph; $N^2\mathcal{C}_N(\omega)$ has a finite limit $N^2\mathcal{C}_\infty(\omega)$ as $N \to \infty$, which is the function $N^2\mathcal{C}(\omega)$ of the infinite BL entering Eq.~\eqref{eq:spectral}, and $\omega = E_{\alpha'} - E_\alpha$ takes both signs, so that all the integrals over $\omega$ run over the whole real axis. The IPR is thus fixed by the total overlap of an eigenstate with all the others, and its finite-size correction is exactly the deficit of this total overlap with respect to the infinite system; since the deficit is concentrated at small $\omega$ (Sec.~\ref{sec:finiteN}), where $\avg{\rho(E+\omega)} \simeq \avg{\rho}$,
\begin{equation} \label{eq:sumruleN}
    I_2(N) - I_2^\infty \simeq \avg{\rho} \int_{-\infty}^{\infty} \de\omega \left[ N^2\mathcal{C}_\infty(\omega) - N^2\mathcal{C}_N(\omega) \right] .
\end{equation}
Equation~\eqref{eq:spectral} can be written in the same form, $I_2(\eta) \simeq 1 - \int_{-\infty}^{\infty} \de\omega \, \avg{\rho(E+\omega)}\, N^2\mathcal{C}(\omega)\, \omega^2/(\omega^2 + 4\eta^2)$, up to analytic terms: a finite $\eta$ removes from the sum rule the pairs with splittings $\omega \lesssim \eta$, which it does not resolve, while a finite $N$ depletes them. Both corrections thus measure the overlap of pairs of eigenstates close in energy that are either not resolved or missing, and both have the same sign: they make the IPR larger.

On the infinite BL the local Green's functions can be computed exactly through a recursive construction~\cite{anderson1973selfconsistent,abou1974self}. Introducing the ``cavity'' Green's functions $G_{i \to j}$, defined as the diagonal elements of the resolvent of the graph from which the neighbor $j$ of $i$ has been removed, one has
\begin{subequations} \label{eq:cavity}
\begin{align}
    G_{i \to j}^{-1} &= \epsilon_i - {\rm i}\eta - \sum_{m \in \partial i \setminus j} G_{m \to i} \, , \label{eq:cavity_a} \\
    G_{ii}^{-1} &= \epsilon_i - {\rm i}\eta - \sum_{m \in \partial i} G_{m \to i} \, . \label{eq:cavity_b}
\end{align}
\end{subequations}
After averaging over the disorder, Eq.~\eqref{eq:cavity_a} becomes a closed, self-consistent integral equation for the joint probability distribution $P_\eta(g, \hat g)$ of the real and imaginary parts of the cavity Green's functions, from which the distribution of the local Green's functions, and hence $I_2(\eta)$ through Eq.~\eqref{eq:I2G}, follows. These equations can be solved numerically with arbitrary precision by population dynamics~\cite{anderson1973selfconsistent,abou1974self,mezard2001bethe,biroli2010anderson,parisi2019anderson,rizzo2024localized}, and we use them throughout as the ``exact'' reference for the Anderson model.

\subsection{Structure of the solution in the localized phase} \label{sec:linearized}

In the localized phase the imaginary part of the Green's functions vanishes linearly with $\eta$ on the vast majority of sites, while it is anomalously large on the rare sites close to the localization center of an eigenstate at energy $E \approx 0$. It is therefore convenient to write $G_{i \to j} = g_{i \to j} + {\rm i} \eta \hat g_{i \to j}$, and similarly for the local functions. Following Ref.~\cite{rizzo2024localized}, we introduce the inverse local Green's function
\begin{equation} \label{eq:M}
\begin{aligned}
    M_{ii} \equiv G_{ii}^{-1} &= m_{ii} - {\rm i} \eta \hat m_{ii} \, , \\
    m_{ii} = \epsilon_i - \!\! \sum_{m \in \partial i} \!\! g_{m \to i} \, , & \qquad
    \hat m_{ii} = 1 + \!\! \sum_{m \in \partial i} \!\! \hat g_{m \to i} \, ,
\end{aligned}
\end{equation}
and its joint distribution $Q_\eta(m,\hat m)$. The moments of the local Green's functions entering Eq.~\eqref{eq:I2G} are dominated by the resonant region $m \approx 0$; changing variable to $m = \eta \hat m x$ one finds
\begin{equation} \label{eq:G2andImG}
\begin{aligned}
    \avg{|G_{ii}|^2} &= \int \frac{Q_\eta (\eta \hat m x, \hat m)}{\eta \hat m \, (1 + x^2)} \, \de x \, \de \hat m \, , \\
    \avg{\im G_{ii}} &= \int \frac{Q_\eta (\eta \hat m x, \hat m)}{1 + x^2} \, \de x \, \de \hat m \, .
\end{aligned}
\end{equation}
In the $\eta \to 0^+$ limit the cavity equations~\eqref{eq:cavity_a} simplify,
\begin{equation} \label{eq:linearized}
\begin{aligned}
    g_{i \to j} &= \frac{1}{\epsilon_i - \sum_{m \in \partial i \setminus j} g_{m \to i}} \, , \\
    \hat g_{i \to j} &= g_{i \to j}^2 \Big( 1 + \sum_{m \in \partial i \setminus j} \hat g_{m \to i} \Big) \, ,
\end{aligned}
\end{equation}
the equation for the real parts decouples, the one for the imaginary parts becomes linear and independent of $\eta$, and $Q_\eta(\eta \hat m x, \hat m)$ can be replaced by $Q_{0^+}(0, \hat m)$ in Eq.~\eqref{eq:G2andImG}. One then obtains the asymptotic value of the IPR~\cite{mirlin1994distribution,rizzo2024localized}
\begin{equation} \label{eq:I2infQ}
    I_2^{\infty} = \frac{ \int_0^\infty  Q_{0^+}(0 ,\hat m)  \, \hat m^{-1} \, \de \hat m}{\int_0^\infty  Q_{0^+}(0 ,\hat m)  \, \de \hat m}
    = \int_0^\infty \overline{Q}_{0^+}(\hat m) \, \hat m^{-1} \, \de \hat m \, ,
\end{equation}
where $\overline{Q}_\eta(\hat m) \equiv Q_\eta(0,\hat m) / \int_0^\infty Q_{0^+}(0,\hat m) \de \hat m$ is the distribution of $\hat m$ at $m=0$, normalized to one in the $\eta \to 0^+$ limit. The denominator of Eq.~\eqref{eq:I2infQ} is the average density of states, $\int_0^\infty Q_{0^+}(0,\hat m)\,\de\hat m = \avg{\rho} = \avg{\im G_{ii}}/\pi$.

At finite $\eta$ the same decomposition gives the correction. Splitting $Q_\eta(m,\hat m) = Q_\eta(0,\hat m) + [Q_\eta(m,\hat m) - Q_\eta(0,\hat m)]$ in Eq.~\eqref{eq:G2andImG}, the second piece is regular in $m$ and contributes to $\eta \avg{|G_{ii}|^2}$ only at order $\eta$ [its contribution is $\eta \int \de m \, [\tilde{Q}_{0^+}(m) - \tilde{Q}_{0^+}(0)]/m^2 + o(\eta)$, with $\tilde Q$ the marginal distribution of $m$], while the first piece gives $\pi \int Q_\eta(0,\hat m) \hat m^{-1} \de \hat m$ exactly. Assuming that no singular corrections arise in the average density of states in the denominator of Eq.~\eqref{eq:I2G}, one concludes that, to leading order,
\begin{equation} \label{eq:I2deltaQ}
    I_2(\eta) - I_2^\infty \approx \int_0^\infty \left[ \overline{Q}_\eta (\hat m) - \overline{Q}_{0^+} (\hat m) \right] \hat m^{-1} \, \de \hat m \, .
\end{equation}
Anomalous corrections to the IPR must therefore reflect an anomalous dependence on $\eta$ of the distribution of $\hat m$ at $m=0$ in the region $\hat m = O(1)$ that dominates the integral --- that is, of the {\it typical} values of the LDoS at the resonant sites, not of its far tail.

The distribution $Q_{0^+}(0,\hat m)$ has a concrete physical meaning: at a resonant site, $\im G_{ii} = 1/(\eta \hat m)$, so that $\hat m^{-1}$ is the LDoS in units of $1/\eta$, i.e.\ the squared amplitude $u = |\psi_\alpha(i)|^2$ of the resonant eigenstate on site $i$. In fact, the whole distribution $T_\infty(u)$ of the amplitudes of the $E=0$ eigenstates in the thermodynamic limit is given by~\cite{mirlin1994distribution,rizzo2024localized}
\begin{equation} \label{eq:Tuinf}
    T_\infty(u) = \overline{Q}_{0^+} \! \left( \frac{1}{u} \right) \frac{1}{u^3} \, ,
\end{equation}
normalized such that $\int_0^1 T_N(u) \, \de u = N$ at finite $N$, and $I_2^\infty = \int_0^1 T_\infty(u) \, u^2 \, \de u$ in agreement with Eq.~\eqref{eq:I2infQ}. We shall come back to this relation in Sec.~\ref{sec:Tu}.

The crucial feature of the localized phase is that $Q_{0^+}(0,\hat m)$ has a power-law tail,
\begin{equation} \label{eq:Qmhattails}
    Q_{0^+} (0,\hat m) \simeq \frac{C_Q}{\hat m^{1 + \beta}} \, , \qquad \hat m \to \infty \, ,
\end{equation}
with a disorder-dependent exponent $\beta \in (1/2, 1)$ which behaves as~\cite{anderson1973selfconsistent,abou1974self,efetov1985anderson,efetov1987density,efetov1987anderson,zirnbauer1986localization,zirnbauer1986anderson,verbaarschot1988graded,mirlin1991localization,mirlin1991universality,fyodorov1991localization,fyodorov1992novel,mirlin1994statistical,mirlin1994distribution,tikhonov2019critical,tikhonov2019statistics,rizzo2024localized}
\begin{equation} \label{eq:beta}
    \beta \simeq \frac{1}{2} + c_\beta \sqrt{W - W_c}
\end{equation}
close to the transition and tends to one for $W \to \infty$. The exponent $\beta$ is determined by the linear recursion~\eqref{eq:linearized} for the imaginary parts: writing the tail of the cavity distribution as $P_0(g,\hat g) \simeq f_\beta(g) \hat g^{-1-\beta}$ and noting that, for $\hat g$ large, $\hat g \simeq g^2 \hat g_1$ is dominated by the largest of the $\hat g_m$'s, one finds that $f_\beta$ must satisfy
\begin{equation} \label{eq:fbeta}
\begin{aligned}
    f_\beta(g) = k \int \de \epsilon \, p(\epsilon) & \left[ \prod_{i=2}^k \de g_i \, P_0(g_i) \right] \de g_1 \, |g|^{2\beta} f_\beta(g_1) \\
    & \times \delta \Big( g - \frac{1}{\epsilon - \sum_{i=1}^k g_i} \Big) \, ,
\end{aligned}
\end{equation}
where $P_0(g)$ is the (known) distribution of the real parts at $\eta=0$. Equation~\eqref{eq:fbeta} is the celebrated Abou-Chacra--Anderson--Thouless condition~\cite{anderson1973selfconsistent,abou1974self,mirlin1991localization}: denoting by $\lambda(\beta)$ the largest eigenvalue of the linear operator on its right-hand side, it requires $\lambda(\beta) = 1$. The function $\lambda(\beta)$ is symmetric under $\beta \to 1 - \beta$~\cite{anderson1973selfconsistent,abou1974self,mirlin1991localization,mirlin1991universality,kravtsov2018non,parisi2019anderson}, so that in the localized phase Eq.~\eqref{eq:fbeta} has {\it two} solutions, $\beta$ and $1-\beta$, which merge at $\beta = 1/2$ at the transition; the tail of $P_0$ is governed by the larger of the two. Equation~\eqref{eq:Qmhattails} corresponds, through Eq.~\eqref{eq:Tuinf}, to $T_\infty(u) \propto u^{-(2-\beta)}$ at small $u$, and the asymptotic IPR displays the square-root singularity $I_2^\infty \simeq I_2^{\infty,{\rm crit}} + c_{I_2} \sqrt{W - W_c}$ above the jump at $W_c$~\cite{rizzo2024localized}. In real space, a large $\hat m$ at a resonant site signals an eigenstate localized far from the site, whose exponential tail reaches it with the small weight $u = 1/\hat m$: the tail~\eqref{eq:Qmhattails} counts these distant states, as discussed in Sec.~\ref{sec:resonances}.

\begin{figure}[t]
\centering
\includegraphics[width=0.46\textwidth]{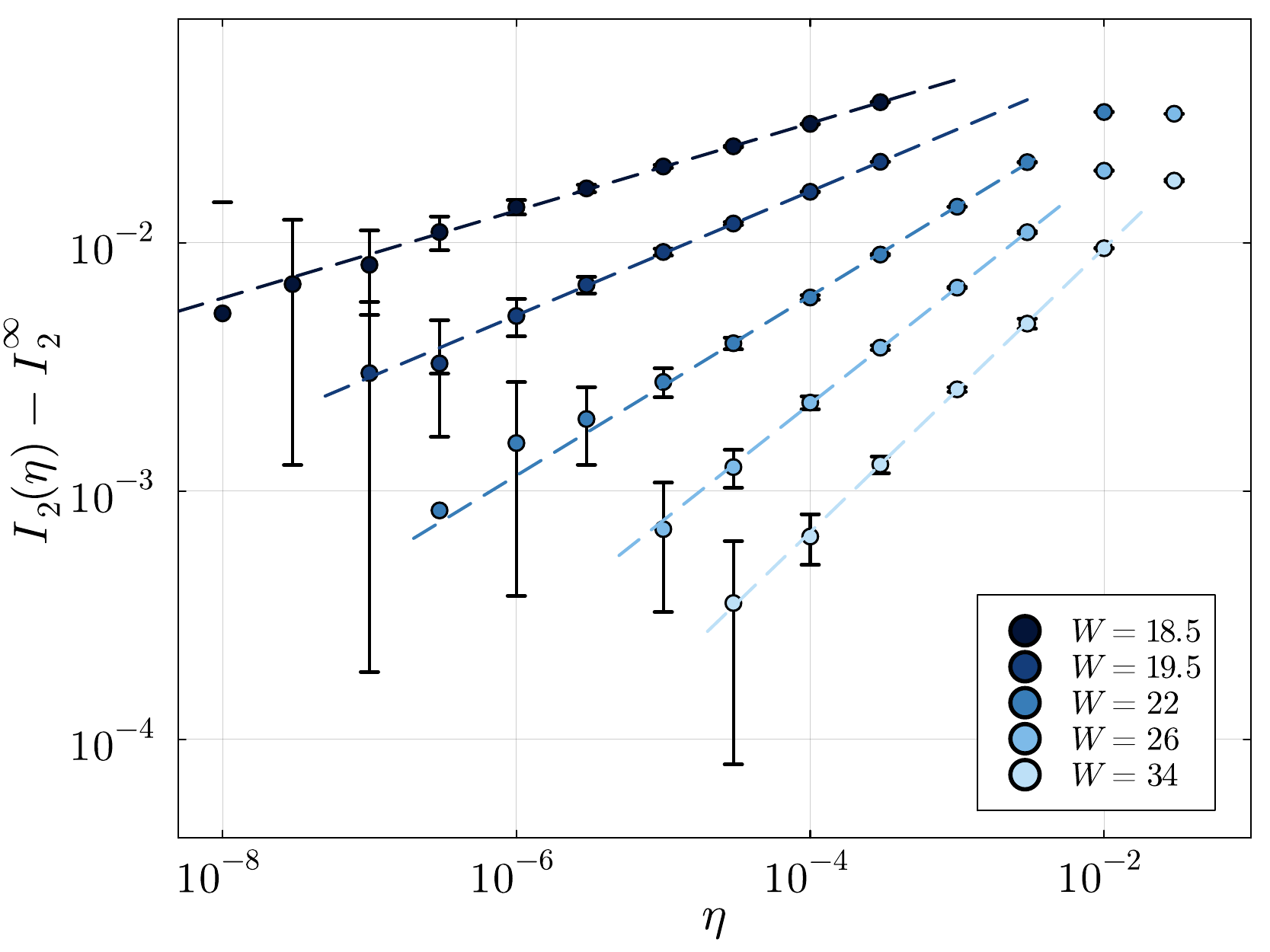}
\caption{Finite-$\eta$ corrections to the asymptotic value of the IPR: $I_2(\eta) - I_2^\infty$ as a function of $\eta$, from the population-dynamics solution of the cavity equations~\eqref{eq:cavity} on the infinite Bethe lattice ($k=2$), for several values of the disorder strength in the localized phase. The dashed lines are power-law fits to Eq.~\eqref{eq:DeltaI2eta}.}
\label{fig:I2eta}
\end{figure}

\begin{figure}[t]
\centering
\includegraphics[width=0.46\textwidth]{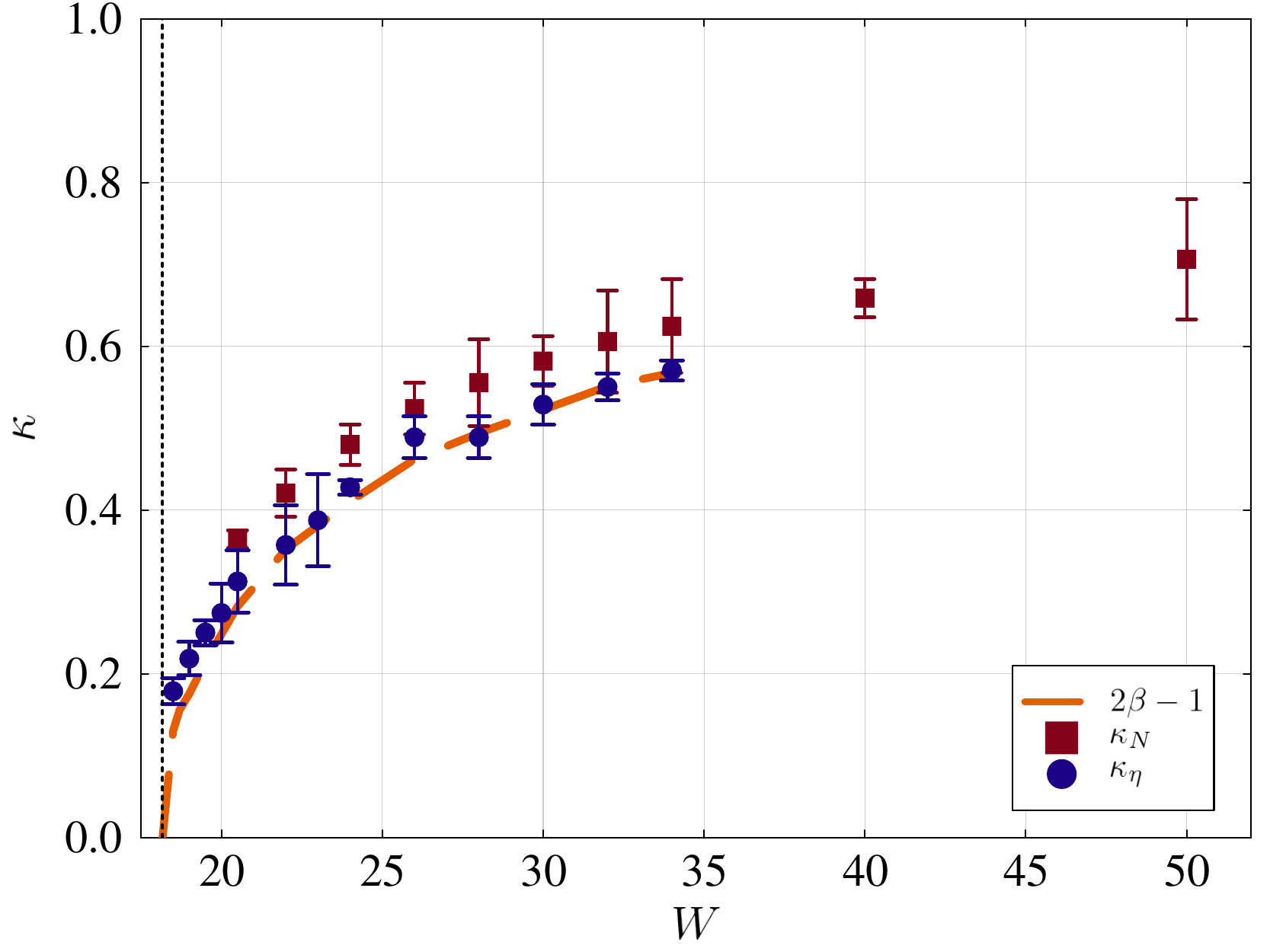}
\caption{Exponent $\kappa_\eta$ of the finite-$\eta$ corrections to the IPR (circles), extracted from the power-law fits of Fig.~\ref{fig:I2eta}, as a function of the disorder strength $W$. The dashed line is the prediction $\kappa = 2 \beta - 1$ derived in this paper, with $\beta(W)$ obtained independently from the tail~\eqref{eq:Qmhattails} of the LDoS distribution. The squares are the exponent $\kappa_N$ of the finite-size corrections at $\eta=0$ obtained in Ref.~\cite{letter} from exact diagonalization, shown for comparison. Close to $W_c$ the pure power-law fits overestimate the exponent, as the corrections cross over to the logarithmic critical behavior (Sec.~\ref{sec:FSS}).}
\label{fig:kappa}
\end{figure}

\subsection{Numerical evidence for anomalous finite-$\eta$ corrections} \label{sec:evidence}

The most direct way to access the finite-regulator corrections is to solve the full cavity equations~\eqref{eq:cavity_a} at finite $\eta$ by population dynamics, and to evaluate Eq.~\eqref{eq:I2G} by direct sampling. The results are shown in Fig.~\ref{fig:I2eta} for several values of $W$ in the localized phase and over several orders of magnitude in $\eta$, using the very precise determination of $I_2^\infty$ provided by the method of Ref.~\cite{rizzo2024localized}. Naively, since $\eta$ enters the cavity equations only through $\eta^2$ [see Eq.~\eqref{eq:gghateta} below], one would expect analytic corrections of order $\eta^2$, or at most of order $\eta$. Instead, the corrections follow a power law
\begin{equation} \label{eq:DeltaI2eta}
    I_2(\eta) - I_2^\infty \simeq c_\eta \, \eta^{\kappa_\eta} \, ,
\end{equation}
with an exponent $\kappa_\eta \in (0,1)$ that increases continuously with $W$, and a prefactor $c_\eta$ that depends only weakly on $W$ deep in the localized phase ($c_\eta \approx 0.13$--$0.15$ for $22 \le W \le 34$). This power law is the asymptotic regime $\kappa_\eta \ln(1/\eta) \gg 1$ of the corrections: as $W \to W_c$ it moves to ever smaller $\eta$, its prefactor vanishes linearly with the exponent (Fig.~\ref{fig:cprefactor} in Sec.~\ref{sec:FSS}), as predicted in Sec.~\ref{sec:nearcritical}, and the corrections cross over to the logarithmic critical behavior (Sec.~\ref{sec:FSS}). As shown in Fig.~\ref{fig:kappa}, the exponent is remarkably well described by
\begin{equation} \label{eq:kappa}
    \kappa_\eta = 2 \beta - 1 \, ,
\end{equation}
with $\beta(W)$ determined independently from the tail~\eqref{eq:Qmhattails}. The exponent vanishes at the critical point, where the corrections are expected to become logarithmic~\cite{tikhonov2019statistics}, and tends to one for $W \to \infty$. The finite-size corrections at $\eta = 0$ obtained by exact diagonalization in Ref.~\cite{letter} have the same exponent, $\kappa_N \simeq \kappa_\eta$ (squares in Fig.~\ref{fig:kappa}).\footnote{As in Ref.~\cite{letter}, the IPR at finite $N$ is computed from the two eigenstates closest to $E = 0$ in each sample. Averaging instead over all the eigenstates in a fixed energy window around $E = 0$ weights the samples differently, since the number of eigenstates in the window fluctuates from sample to sample in a way that is correlated with how strongly the sample is localized. This changes the amplitude of the correction, by up to a factor of two, and, over the accessible sizes, gives a somewhat larger effective exponent.} Close to $W_c$ the pure power-law fits overestimate the exponent: there the corrections are entering the crossover to the logarithmic critical behavior, which the scaling form of Sec.~\ref{sec:FSS} accounts for with $\kappa = 2\beta - 1$; $\kappa_N$ also lies somewhat above $\kappa_\eta$, by an amount that decreases slowly with $N$ (Sec.~\ref{sec:FSS}).

\section{Eigenstate correlations and the real-space picture} \label{sec:realspace}

The spectral identity~\eqref{eq:spectral} and the sum rule~\eqref{eq:sumrule} express both corrections through the correlations of pairs of eigenstates close in energy. In this section we first show that they fix exactly the exponent $\mu$ of the eigenstate correlations, $\mu = 1 - \kappa$, together with their amplitude (Sec.~\ref{sec:kappamu}), and confirm this by computing the correlations directly (Sec.~\ref{sec:corrtests}). We then give the corrections a real-space interpretation: a resonance counting based on the statistics of the eigenstate amplitudes reproduces $\mu = 2(1-\beta)$ (Sec.~\ref{sec:resonances}), and exact diagonalization confirms its picture of the resonant pairs (Sec.~\ref{sec:pairtests}); the finite-size corrections are due to the hybridized pairs missing on a finite graph (Sec.~\ref{sec:finiteN}), and $\beta = 1/2$ marks the proliferation of the resonances (Sec.~\ref{sec:proliferation}).

\subsection{Finite-regulator corrections and eigenstate correlations: $\kappa = 1 - \mu$} \label{sec:kappamu}

Equation~\eqref{eq:spectral} relates the exponent $\kappa$ to the eigenstate correlations exactly, with no assumption on the nature of the pairs of eigenstates that contribute. Writing $N^2\mathcal{C}(\omega) \simeq B\,|\omega|^{-\mu(W)}$, with $0 < \mu < 1$, and inserting this form into Eq.~\eqref{eq:spectral} gives Eq.~\eqref{eq:DeltaI2eta} with
\begin{equation} \label{eq:kappamu}
    \kappa = 1 - \mu \, , \qquad c_\eta = \frac{\pi \avg{\rho} B\, 2^{\kappa}}{\sin(\pi\kappa/2)} \, .
\end{equation}
Conversely, since $N^2\mathcal{C}(\omega)$ and the kernel of Eq.~\eqref{eq:spectral} are both positive, the power law~\eqref{eq:DeltaI2eta} fixes the weight of the eigenstate correlations at small splittings, $\avg{\rho}\int_{-\omega}^{\omega} N^2\mathcal{C}(\omega')\, \de\omega' \simeq 2\avg{\rho} B\, \omega^{\kappa}/\kappa$ with $B$ given by Eq.~\eqref{eq:kappamu}, and hence $N^2\mathcal{C}(\omega) \simeq B\,|\omega|^{-\mu}$ with $\mu = 1 - \kappa$, provided only that $N^2\mathcal{C}(\omega)$ varies monotonically at small $\omega$.\footnote{In the variable $t = \omega^2$ the right-hand side of Eq.~\eqref{eq:spectral} is proportional to $4\eta^2 \int \de F(t)/(t + 4\eta^2)$, where $F(\omega^2) = \int_{-\omega}^{\omega} N^2\mathcal{C}(\omega')\, \de\omega'$ is non-decreasing, i.e.\ to $4\eta^2$ times the Stieltjes transform of a positive measure. By Karamata's Tauberian theorem for Stieltjes transforms~\cite{bingham1987regular}, for $0 < \kappa < 1$ the transform behaves as $A\,\Gamma(1+\kappa/2)\,\Gamma(1-\kappa/2)\,(4\eta^2)^{\kappa/2-1}$ for $\eta \to 0$ if and only if $F(t) \simeq A\,t^{\kappa/2}$ for $t \to 0$; the monotone density theorem~\cite{bingham1987regular} then gives the power law of $N^2\mathcal{C}(\omega)$ itself.}\nocite{bingham1987regular} Equation~\eqref{eq:kappa} therefore fixes both the exponent of the eigenstate correlations, $\mu = 2(1-\beta)$, and their amplitude, $B = c_\eta \sin(\pi\kappa/2)/(\pi\avg{\rho}\, 2^\kappa)$, with no further input: since $\kappa = 2\beta - 1$ is derived analytically in this paper, for the toy model in Sec.~\ref{sec:scaling} and, under a few assumptions, for the Anderson model in Sec.~\ref{sec:anderson}, so is $\mu$.

The same reasoning identifies the eigenstate correlations that control the finite-$\eta$ corrections of the higher moments of the eigenfunctions. Defining $I_q(\eta) = \pi \eta^{q-1} \avg{(\im G_{ii})^q}/(n_q \avg{\im G_{ii}})$, with $n_q = \sqrt{\pi}\,\Gamma(q-1/2)/\Gamma(q)$, which tends to $I_q^\infty = \avg{\sum_i |\psi_\alpha(i)|^{2q}}$ as $\eta \to 0^+$ (for $q = 2$ it has the same leading correction as Eq.~\eqref{eq:I2G}), and expanding $(\im G_{ii})^q$ on the eigenstates, the terms involving two distinct eigenstates give the generalization of Eq.~\eqref{eq:spectral} in terms of the generalized pair correlations
\begin{equation} \label{eq:Opsdef}
    \mathcal{O}_{p,s}(\omega) = \avg{\sum_i |\psi_\alpha(i)|^{2p} |\psi_{\alpha'}(i)|^{2s}}_{E_{\alpha'} - E_\alpha = \omega} \, ,
\end{equation}
with $N\mathcal{O}_{1,1} = N^2\mathcal{C}$:
\begin{equation} \label{eq:spectralq}
\begin{aligned}
    I_q(\eta) - I_q^\infty \simeq \frac{\avg{\rho}}{2n_q} \sum_{p=1}^{q-1} \binom{q}{p} \int_{-\infty}^{\infty} \de\omega \, & N\mathcal{O}_{p,q-p}(\omega) \\
    & \times K_{p,q-p}(\omega/\eta) \, ,
\end{aligned}
\end{equation}
with $K_{p,s}(x) = \int \de y \, (1+y^2)^{-p} [1+(y+x)^2]^{-s}$ [for $q=2$, $K_{1,1}(x) = 2\pi/(4+x^2)$ and Eq.~\eqref{eq:spectral} is recovered], while the terms involving three or more distinct eigenstates require two simultaneous resonances and are of higher order, $O(\eta^{2\kappa})$. All the moments $I_q$ therefore share the exponent $\kappa = 1 - \mu$ if all the $\mathcal{O}_{p,s}$ decay with the same exponent $\mu$, which is what the resonance picture predicts (Sec.~\ref{sec:resonances}) and the population dynamics confirms (Sec.~\ref{sec:corrtests}).

Since Eq.~\eqref{eq:kappamu} is exact, a direct computation of $N^2\mathcal{C}(\omega)$ can only confirm it. We carry it out numerically for the Anderson model, by exact diagonalization and by population dynamics directly on the infinite BL (Sec.~\ref{sec:corrtests}), and analytically for the toy model, by continuing its cavity solution to a complex field (Sec.~\ref{sec:complexH}). What Eq.~\eqref{eq:spectral} does not provide is a physical picture of the pairs of eigenstates that carry the correlations. In the localized phase two states close in energy are typically localized far apart and overlap exponentially little, and the average overlap and its higher moments are dominated by rare Mott-like resonances~\cite{mott1968conduction,cuevas2007two,ivanov2012hybridization}, as argued for the BL in Ref.~\cite{tikhonov2021eigenstate}; Sec.~\ref{sec:resonances} makes this picture quantitative.

\begin{figure*}[t]
\panel{a}{\includegraphics[width=0.48\textwidth]{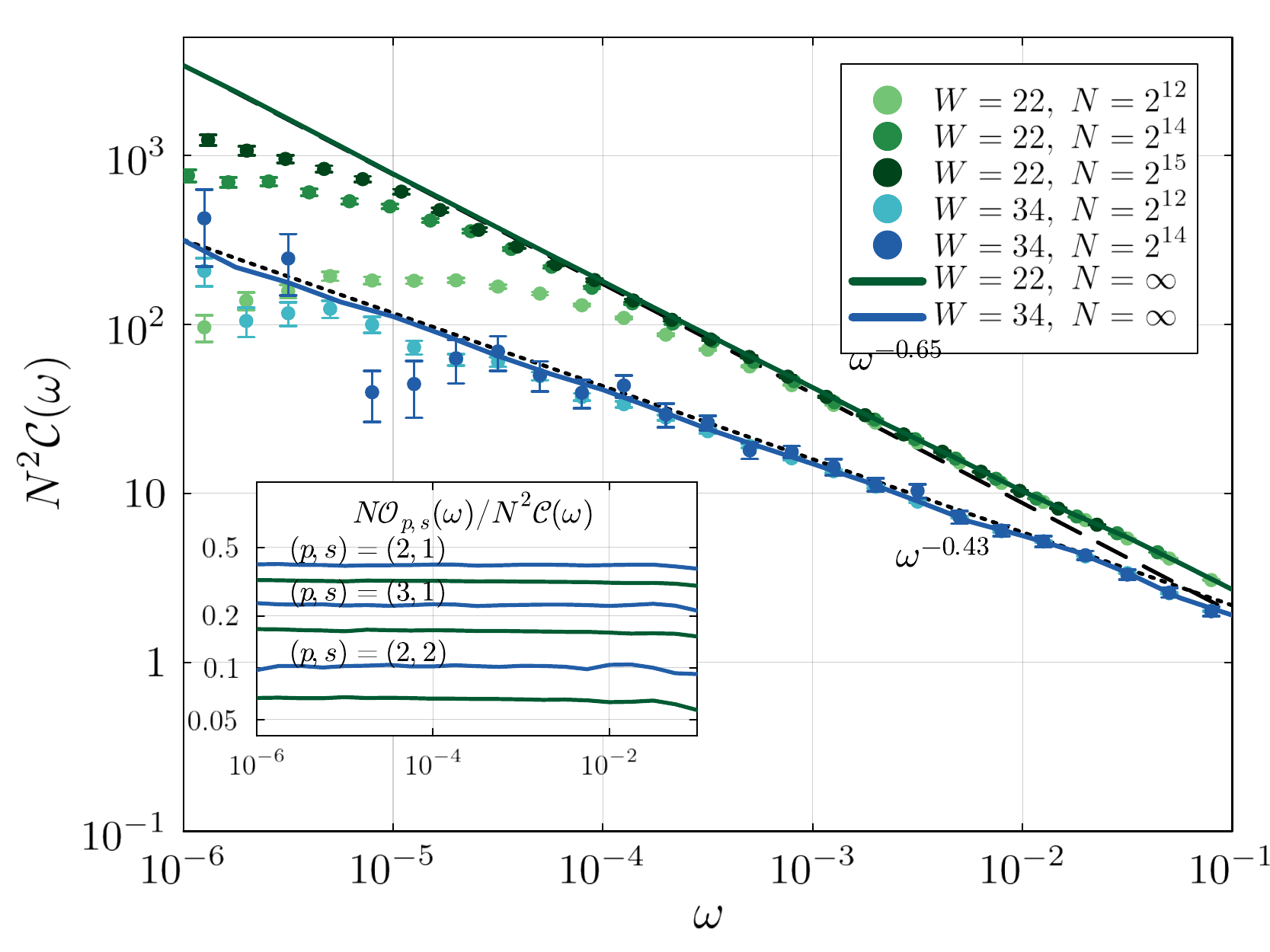}}\hfill
\panel{b}{\includegraphics[width=0.48\textwidth]{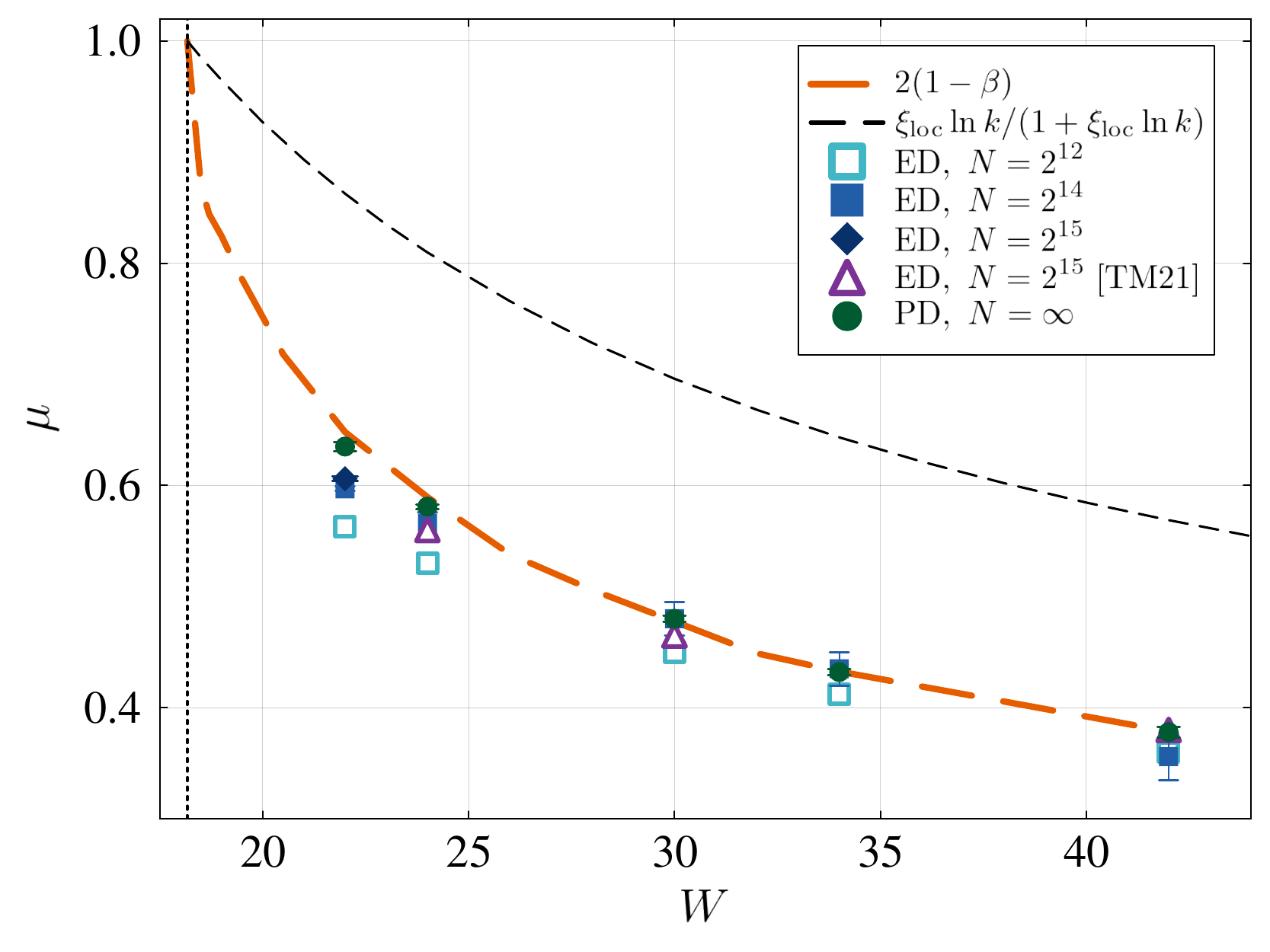}}
\caption{(a) Correlation function of two eigenstates at energy distance $\omega$, $N^2\mathcal{C}(\omega)$, from exact diagonalization of random regular graphs with $N = 2^{12}$ (light symbols) and $2^{14}$ (dark symbols) sites, at $W = 22$ (greens) and $W = 34$ (blues), and with $N = 2^{15}$ sites at $W = 22$ (darkest green); all the eigenstates with $|E| < 0.2$ are kept in each sample ($|E| < 0.11$ and $0.056$ for $N = 2^{14}$ and $2^{15}$ at $W = 22$), and the error bars are statistical. Lines: parameter-free prediction $B\,\omega^{-\mu}$ of Eq.~\eqref{eq:kappamu}, with $\mu = 2(1-\beta)$ from the tail exponent $\beta$ and $B = c_\eta \sin(\pi\kappa/2)/(\pi\avg{\rho}\,2^\kappa)$ from the finite-$\eta$ corrections to the IPR [dashed: $W=22$, $\mu = 0.65$; dotted: $W = 34$, $\mu = 0.43$]. The saturation at small $\omega$ is a finite-size effect, which moves to smaller $\omega$ as $N$ grows. Solid lines: population dynamics on the infinite BL directly at $\eta = 0^+$, Eq.~\eqref{eq:OpsPD} ($N = \infty$; statistical errors smaller than the line width). Inset: generalized correlations $N\mathcal{O}_{p,s}(\omega)$, Eq.~\eqref{eq:Opsdef}, divided by $N^2\mathcal{C}(\omega)$, from the same population dynamics, for $(p,s) = (2,1)$, $(3,1)$ and $(2,2)$ (top to bottom; dark green: $W=22$, blue: $W = 34$): all the pair correlations decay with the same exponent $\mu$. (b) Exponent $\mu$ of $N^2\mathcal{C}(\omega) \propto \omega^{-\mu}$ against the disorder strength, from power-law fits of the exact diagonalization data over $10^{-4} < \omega < 2\times 10^{-2}$ ($N = 2^{12}$, open squares; $N = 2^{14}$, full squares; $N = 2^{15}$ at $W = 22$, diamond), of the exact diagonalization data of Ref.~\cite{tikhonov2021eigenstate} ($N = 2^{15}$, triangles), and of the population dynamics at $\eta = 0^+$ over $10^{-6} < \omega < 10^{-3}$ ($N = \infty$, circles). Dashed line: prediction $\mu = 2(1-\beta)$, with $\beta(W)$ from the tail of the LDoS distribution, which reaches $\mu = 1$ at $W_c$ (dotted line). Thin dashed line: estimate $\mu = \xi_{\rm loc}\ln k/(1 + \xi_{\rm loc}\ln k)$ based on the average decay of the eigenstates~\cite{tikhonov2021eigenstate} (Sec.~\ref{sec:resonances}), with $\xi_{\rm loc}$ from $\lambda(1/2) = e^{-1/\xi_{\rm loc}}$ computed numerically, which close to $W_c$ reproduces $\xi_{\rm loc}^{-1} \simeq 0.031\,(W - W_c)$, consistent with the analysis of Ref.~\cite{tikhonov2019critical}.}
\label{fig:N2C}
\end{figure*}

\subsection{Direct computation of the eigenstate correlations} \label{sec:corrtests}

The prediction $\mu = 2(1-\beta)$, with $\beta(W)$ from the tail of $\overline Q_{0^+}$, agrees well with the exponents obtained by exact diagonalization for $N = 2^{15}$ at $W = 24$, $30$ and $42$ in Ref.~\cite{tikhonov2021eigenstate} [Fig.~\ref{fig:N2C}(b)]. We have also checked Eq.~\eqref{eq:kappamu} directly, by computing $N^2\mathcal{C}(\omega)$ by exact diagonalization of random regular graphs with $N = 2^{12}$ and $2^{14}$ sites, and with $N = 2^{15}$ sites at $W = 22$ (all the eigenstates with $|E| < 0.2$ in each sample, pairs with $|\omega| < 0.1$ around reference states with $|E| < 0.1$; at $W = 22$ these windows are reduced to $|E| < 0.11$ and $0.056$, with $|\omega| < 0.055$ and $0.028$, for $N = 2^{14}$ and $2^{15}$). Power-law fits over $10^{-4} < \omega < 2 \times 10^{-2}$ at $W = 22$, $24$, $30$, $34$ and $42$ give exponents close to $2(1-\beta)$ [Fig.~\ref{fig:N2C}(b)]; at $W < 30$ the effective exponents grow towards the predicted values with $N$ (at $W = 22$, $\mu = 0.563$, $0.597$ and $0.606$ for $N = 2^{12}$, $2^{14}$ and $2^{15}$). The residual deficit at $W = 22$ is, however, mostly not a finite-size effect: the population dynamics on the infinite BL described below, fitted over the same window, gives $\mu = 0.611$, since over this window $N^2\mathcal{C}(\omega)$ has not yet reached its asymptotic power law. The amplitude agrees as well [Fig.~\ref{fig:N2C}(a)]: at $W = 34$ the measured correlation function follows the parameter-free prediction $B\,\omega^{-2(1-\beta)}$, with $B = c_\eta \sin(\pi\kappa/2)/(\pi\avg{\rho}\, 2^\kappa)$ fixed by the finite-$\eta$ data, closely over more than three decades in $\omega$; at $W = 22$ it does so for $10^{-4} \lesssim \omega \lesssim 3 \times 10^{-3}$ at $N = 2^{15}$, with a finite-size saturation below, which moves to smaller $\omega$ as $N$ grows, and non-asymptotic corrections above, which the population dynamics at $N = \infty$ shows as well. Consistently, inserting the measured $N^2\mathcal{C}(\omega)$ into Eq.~\eqref{eq:spectral} reproduces $I_2(\eta) - I_2^\infty$ from population dynamics reasonably well for $\eta \gtrsim 10^{-3}$ at $N = 2^{14}$, smaller values of $\eta$ being affected by the finite-size saturation.

The finite-size saturation can be avoided altogether by computing the eigenstate correlations directly on the infinite BL at $\eta = 0^+$. In the localized phase the LDoS at a site is a sum of $\delta$ peaks: at $\eta = 0^+$ and energy $E$ the cavity Green's functions $g_{m \to i}(E)$ are real, and $G_{ii}(E)$ has a pole whenever $m_{ii}(E) = \epsilon_i - E - \sum_{m \in \partial i} g_{m \to i}(E)$ vanishes, with residue $|\psi_\alpha(i)|^2 = 1/\hat m_{ii}(E)$, where $\hat m_{ii}(E) = 1 + \sum_{m\in\partial i} \hat g_{m \to i}(E)$ and $\hat g_{m\to i}(E) = \partial_E g_{m\to i}(E)$ is the $\eta \to 0^+$ limit of $\im G_{m\to i}/\eta$ at energy $E$ [since $G(E+{\rm i}\eta) = g(E) + {\rm i}\eta\,\partial_E g(E) + O(\eta^2)$], which obeys the linear recursion in Eq.~\eqref{eq:linearized}. [At a single energy this is the statement that $I_2^\infty$ is the average of $1/\hat m$ at $m = 0$~\cite{rizzo2024localized}.] Following the cavity variables at two energies $E_{1,2} = \mp\omega/2$ on the same tree, the generalized pair correlations~\eqref{eq:Opsdef} are given by
\begin{equation} \label{eq:OpsPD}
    \avg{\rho}^2 N \mathcal{O}_{p,s}(\omega) = \avg{\hat m_1^{1-p}\, \hat m_2^{1-s}\, \delta(m_1)\, \delta(m_2)} \, ,
\end{equation}
where $m_j = m_{ii}(E_j)$, $\hat m_j = \hat m_{ii}(E_j)$ (each $\delta(m_j)$ contributes the weight $1/\hat m_j$ of the eigenstate at $E_j$). Since $\epsilon_i$ enters $m_1$ and $m_2$ in the same way, the average over $\epsilon_i$ can be performed analytically, leaving $W^{-1}\delta(m_1 - m_2)$, restricted to the support of $p(\epsilon)$; the joint distribution of the real cavity variables at the two energies is obtained by population dynamics of the $\eta = 0^+$ recursion. We have solved it with populations of $2^{21}$ elements for $10^{-6} \le \omega \le 10^{-1}$, representing $\delta(m_1 - m_2)$ by a box of width $0.04\,\omega$ (a box five times wider gives the same results within statistical errors). The resulting $N^2\mathcal{C}(\omega)$ [solid lines in Fig.~\ref{fig:N2C}(a)] shows no saturation, and is described by a power law down to the smallest $\omega$: a fit over $10^{-6} < \omega < 10^{-3}$ gives exponents in close agreement with $2(1-\beta)$ at all five disorder strengths, with statistical errors $\le 0.004$ [Fig.~\ref{fig:N2C}(b)], the small deficit at $W = 22$ being due to the slow approach to the asymptotic regime visible in the local slope, which reaches $0.645$ for $\omega < 10^{-5}$, against $2(1-\beta) = 0.648$. The amplitude agrees with the parameter-free prediction $B\,\omega^{-2(1-\beta)}$ within statistical errors at $W = 22$ for $\omega \lesssim 10^{-5}$, above which non-asymptotic corrections become visible, and at $W = 34$ within the accuracy with which $c_\eta$ is known, over the whole range of $\omega$.

The population dynamics also shows that all the generalized correlations decay with the same exponent: the ratios $\mathcal{O}_{p,s}/\mathcal{O}_{1,1}$ for $(p,s) = (2,1)$, $(3,1)$ and $(2,2)$ are independent of $\omega$, within the statistical accuracy, over five decades [inset of Fig.~\ref{fig:N2C}(a)]. Finally, inserting the population-dynamics $N\mathcal{O}_{p,s}(\omega)$ into Eq.~\eqref{eq:spectralq} reproduces, without any adjustable parameter, the corrections $I_q(\eta) - I_q^\infty$ obtained from the finite-$\eta$ population dynamics [Eq.~\eqref{eq:I2deltaQ} and its analogue with $\hat m^{1-q}$] for $q = 2$, $3$ and $4$, over the whole range $10^{-6} \le \eta \le 10^{-4}$ at both $W = 22$ and $W = 34$ (and up to $\eta = 10^{-3}$ at $W = 22$), with small residual deviations that grow only at the smallest $\eta$, where the finite-$\eta$ data are least precise. The analysis is thus closed: the anomalous finite-$\eta$ corrections of all the moments of the eigenfunctions are quantitatively accounted for by the correlations of pairs of eigenstates close in energy.

\subsection{Resonance counting from the statistics of the amplitudes} \label{sec:resonances}

The physical origin of the power law of $N^2\mathcal{C}(\omega)$ is revealed by counting the Mott resonances between localized states. The argument is phenomenological, of the type used for localized systems since Mott~\cite{mott1968conduction,cuevas2007two,ivanov2012hybridization,tikhonov2021eigenstate}: one starts from unhybridized localized states, estimates the coupling of two of them from the region where the peak of one overlaps the tail of the other, and assumes that they hybridize when this coupling exceeds their energy separation. It is not a controlled derivation, and it fixes the exponent of $N^2\mathcal{C}(\omega)$ but not its amplitude (see below); its picture of the resonant pairs is, however, confirmed in detail by exact diagonalization (Secs.~\ref{sec:pairtests} and~\ref{sec:finiteN}). If the amplitude of an eigenstate decays on average as $|\psi(r)| \sim k^{-r} e^{-r/\xi_{\rm loc}}$ with the distance $r$ from its localization center, two states whose centers are at distance $R$ are coupled by a matrix element $M \sim k^{-R} e^{-R/\xi_{\rm loc}}$, and they hybridize into a resonant pair, with overlap $\sum_i |\psi_\alpha(i)|^2|\psi_{\alpha'}(i)|^2 = O(1)$, when $M$ is of the order of their energy separation $\omega$, i.e.\ at $R(\omega) \simeq \ln(1/\omega)/(\ln k + \xi_{\rm loc}^{-1})$. The number of states at this distance is $k^{R(\omega)} \sim \omega^{-\mu}$, with $\mu = \xi_{\rm loc} \ln k/(1 + \xi_{\rm loc}\ln k)$~\cite{tikhonov2021eigenstate}. Here the localization length is defined by $\lambda(1/2) = e^{-1/\xi_{\rm loc}}$, where $\lambda(s)$ is the largest eigenvalue of the linear operator of Eq.~\eqref{eq:fbeta} with $\beta$ replaced by $s$, so that $\beta$ itself is fixed by $\lambda(\beta) = 1$~\cite{tikhonov2019critical}. The probability that one of them lies within $\omega$ of a given state, $\omega\, k^{R(\omega)} \sim \omega^{1-\mu}$, gives $N^2\mathcal{C}(\omega) \sim \omega^{-\mu}$. This simple estimate yields $\mu \to 1$ at $W_c$ and $\mu \sim 1/\ln W$ at strong disorder, but it lies well above the measured exponents at all the disorder strengths accessible numerically [thin dashed line in Fig.~\ref{fig:N2C}(b)]; a more refined resonance counting, which accounts for the strong fluctuations of the wavefunctions, confirms the power law, and exact diagonalization gives $\mu(W)$ decreasing from $\approx 0.7$ near $W_c$ to $0.38$ at $W = 42$~\cite{tikhonov2021eigenstate}.

This estimate uses the average decay of the wavefunctions. We now show that the same counting gives $\mu = 2(1-\beta)$ once the average decay is replaced by the exact statistics of the amplitudes. We first identify the real-space configurations responsible for the tail~\eqref{eq:Qmhattails}. At $\eta = 0^+$ the cavity variables have the spectral representations $g = \sum_\alpha u_\alpha/(E_\alpha - E)$ and $\hat g = \sum_\alpha u_\alpha/(E_\alpha - E)^2$, where $u_\alpha$ is the weight on the root of the eigenstate $\alpha$ of the cavity graph. A single eigenstate with weight $u$ and energy separation $\delta = E_\alpha - E$ contributes $u/\delta$ to $g$ and $u/\delta^2$ to $\hat g$. If $u = O(1)$, a large $\hat g$ requires a small $\delta$, and hence a large $g$; if instead $u$ is small and $\delta \sim u$, the contribution to $g$ remains of order one, while that to $\hat g$, $\sim 1/u$, is large. The large values of $\hat g$ at finite $g$ that build the tail $P_0(g,\hat g) \simeq f_\beta(g)\hat g^{-1-\beta}$ can thus be produced by a single eigenstate localized far from the root, whose exponential tail reaches it with a small weight, and which is quasi-resonant with $E$ --- the real-space counterpart of the propagation $\hat g \simeq g^2 \hat g_1$ along a single branch. In the language of Eq.~\eqref{eq:Tuinf}, $u = 1/\hat m$ is the weight of the resonant eigenstate on the site, and the tail~\eqref{eq:Qmhattails} is the small-amplitude tail $T_\infty(u) \propto u^{-(2-\beta)}$ of the distribution of the amplitudes~\cite{mirlin1994distribution}: $\beta$ counts the distant localized states whose tail reaches a given site with a small weight $u$.

We have checked this picture by exact diagonalization of random regular graphs with $N = 2^{11}$, $2^{12}$ and $2^{13}$ sites at $W = 22$ and $34$, decomposing $G_{ii}(E) = \sum_\alpha |\psi_\alpha(i)|^2/(E_\alpha - E)$ and $\partial_E G_{ii}(E) = \sum_\alpha |\psi_\alpha(i)|^2/(E_\alpha - E)^2$ at $E = 0$ over the eigenstates, at the sites where $|G_{ii}| < 1/2$. This condition, satisfied by the large majority of the sites, removes the trivial resonances of the site itself, which dominate the unconditioned distribution (see Sec.~\ref{sec:fartail}). Along the tail, the share of $\partial_E G_{ii}$ carried by the dominant eigenstate grows from about one half in the bulk to $0.98$ at $\partial_E G_{ii} \approx 30$ and $0.999$ at $\partial_E G_{ii} \approx 500$, while that of the second one becomes negligible. The weight of the dominant eigenstate on $i$ decreases roughly as $1/\partial_E G_{ii}$, and its maximum lies at a graph distance from $i$ that grows logarithmically with $\partial_E G_{ii}$, up to the diameter of the graph. A maximum-likelihood fit of the tail of the distribution of $\partial_E G_{ii}$ at $|G_{ii}| < 1/2$, over $3 < \partial_E G_{ii} < 300$, gives the exponents $-1.69$ and $-1.70$ at $W = 22$ ($N = 2^{11}$ and $2^{12}$), against $-(1+\beta) = -1.675$, and $-1.80$ and $-1.79$ at $W = 34$, against $-1.78$, with statistical uncertainties of $0.02$--$0.03$; the fewer samples available at $N = 2^{13}$ give $-1.60 \pm 0.04$ and $-1.74 \pm 0.05$.

Counting the Mott resonances with this distribution gives $\mu$ directly. Consider an eigenstate $\alpha$ localized at site $i$, with weight $O(1)$ there, and a distant eigenstate $\alpha'$ with a small weight $u$ on $i$; by Eq.~\eqref{eq:Tuinf} and the normalization below it, the mean number of such states per unit energy is $\avg{\rho}\, T_\infty(u)\, \de u$. The two states are coupled by a matrix element $M \sim \sqrt{u}$, the amplitude of $\alpha'$ in the region where $\alpha$ is localized. If their energy separation is smaller than $M$ they hybridize into two eigenstates $\simeq (\psi_\alpha \pm \psi_{\alpha'})/\sqrt{2}$, split by $\omega \sim \sqrt{u}$ and both with weight $O(1)$ on $i$, so that their overlap is of order one. The density of such resonant pairs at splitting $\omega$ is therefore
\begin{equation} \label{eq:pairdensity}
    N^2\mathcal{C}(\omega) \, \de\omega \propto T_\infty(u) \, \sqrt{u} \, \de u \, \Big|_{u \sim \omega^2} \propto \omega^{2\beta - 2} \, \de \omega \, ,
\end{equation}
i.e.\ $\mu = 2(1-\beta)$, the exponent fixed exactly by Eqs.~\eqref{eq:kappa} and~\eqref{eq:kappamu}. Pairs with $\omega \gg \sqrt{u}$ acquire only a perturbative overlap $\sim u/\omega^2$, whose contribution, $\omega^{-2} \int_0^{\omega^2} T_\infty(u)\, u \, \de u \propto \omega^{2\beta-2}$, has the same scaling. The same counting applies to the generalized overlaps~\eqref{eq:Opsdef}, which all decay with the same exponent: a pair of states at distance $\omega$ coupled by $M$ has $\sum_i |\psi_\alpha(i)|^{2p}|\psi_{\alpha'}(i)|^{2s} = O(1)$ if $M \gtrsim \omega$ and $\sim (M/\omega)^{2\min(p,s)}$ otherwise, and since $2\min(p,s) > \mu$ the count of partners is dominated by $M \sim \omega$ for all $p,s \ge 1$, so that $N\mathcal{O}_{p,s}(\omega) \simeq B_{p,s}\,\omega^{-\mu}$. Equation~\eqref{eq:spectralq} then implies that all the moments $I_q$ share the exponent $\kappa = 1-\mu$, consistently with the zero-mode analysis of the following sections.

The rare large overlaps that dominate the average $N^2\mathcal{C}(\omega)$ and its generalizations are thus controlled by the statistics of the small ones: a resonance occurs when the small weight $u$ of a distant state on the site exceeds the square of its detuning, so that the same tail $T_\infty(u) \propto u^{\beta-2}$ that makes the typical overlap of two states close in energy exponentially small also sets the probability of the rare hybridized pairs, and of the perturbative admixtures.

The counting fixes the exponent, but not the amplitude $B$, because it treats the state $\alpha$ as if all its weight were on the site $i$. Its overlap with $\alpha'$ would then reduce exactly to the weight of $\alpha'$ on $i$, $\sum_j |\psi_\alpha(j)|^2 |\psi_{\alpha'}(j)|^2 = |\psi_{\alpha'}(i)|^2 = u$, and their coupling would be fixed by $u$ alone. In reality $|\psi_\alpha(j)|^2$ is of order one, but smaller than one, over a finite region around $i$, over which the weight of $\alpha'$ varies: the overlap and the coupling are sums over this region, and differ from their estimates in terms of $u$ by factors of order one that fluctuate from pair to pair. These factors, whose statistics is set by the region around $i$ and not by the small weight $u$, change the amplitude of Eq.~\eqref{eq:pairdensity} but not its exponent, which is set by the tail of $T_\infty(u)$; the amplitude is instead fixed exactly by Eq.~\eqref{eq:kappamu}, and agrees with the numerics of Sec.~\ref{sec:corrtests}.

This is the argument of Ref.~\cite{tikhonov2021eigenstate} recalled above, with the average decay of the wavefunctions replaced by the full statistics of the amplitudes. Indeed, if all the $k^r$ sites at distance $r$ carried the weight $u = k^{-2r} e^{-2r/\xi_{\rm loc}}$ given by the average decay, one would have $T_\infty(u)\, u \propto k^r \propto u^{-\ln k/[2(\ln k + \xi_{\rm loc}^{-1})]}$, and Eq.~\eqref{eq:pairdensity} would reproduce $\mu = \xi_{\rm loc} \ln k/(1 + \xi_{\rm loc} \ln k)$. The two estimates agree only to leading logarithmic order, asymptotically at infinite disorder (see below); at finite $W$ the estimate based on the average decay lies well above $2(1-\beta)$ [Fig.~\ref{fig:N2C}(b)], and close to $W_c$ the two differ qualitatively: the estimate based on the average decay gives $1 - \mu \propto 1/\xi_{\rm loc}$, which vanishes linearly in $W - W_c$ if $\xi_{\rm loc}$ is the length that controls the decay along the rare branches and diverges with exponent $\nu = 1$~\cite{garcia2020two}, whereas the exact statistics gives $1 - \mu = 2\beta - 1 \propto \sqrt{W - W_c}$. At strong disorder the Abou-Chacra--Anderson--Thouless condition gives $1-\beta \simeq \ln k/(2\ln W)$ to leading logarithmic order, i.e.\ $\mu \simeq \ln k/\ln W$, the same result as the resonance estimate with $|\psi(r)| \sim W^{-r}$.

\subsection{Numerical tests of the resonance picture} \label{sec:pairtests}

\begin{figure*}[t]
\centering
\panel{a}{\includegraphics[width=0.46\textwidth]{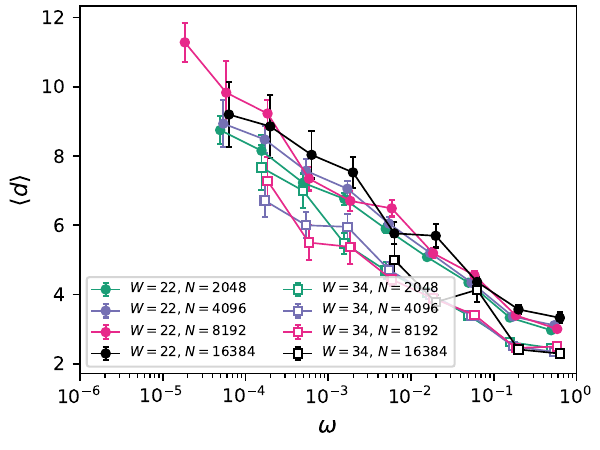}}\hfill
\panel{b}{\includegraphics[width=0.46\textwidth]{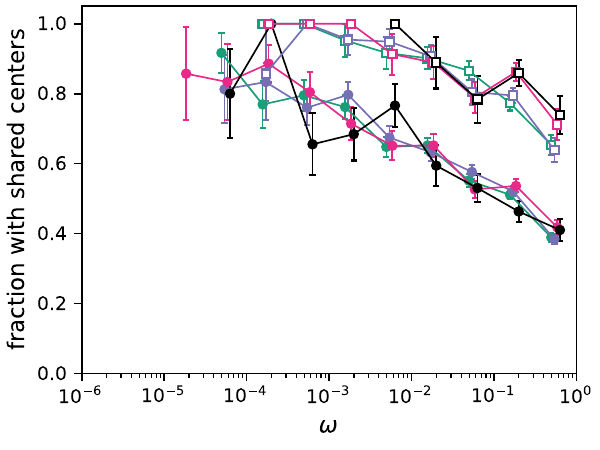}}\\[2mm]
\panel{c}{\includegraphics[width=0.46\textwidth]{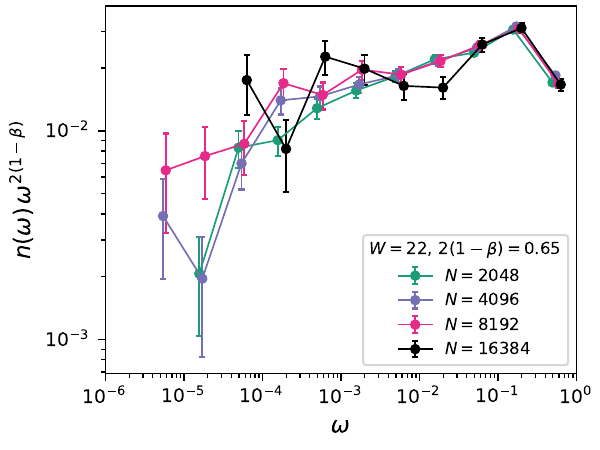}}\hfill
\panel{d}{\includegraphics[width=0.46\textwidth]{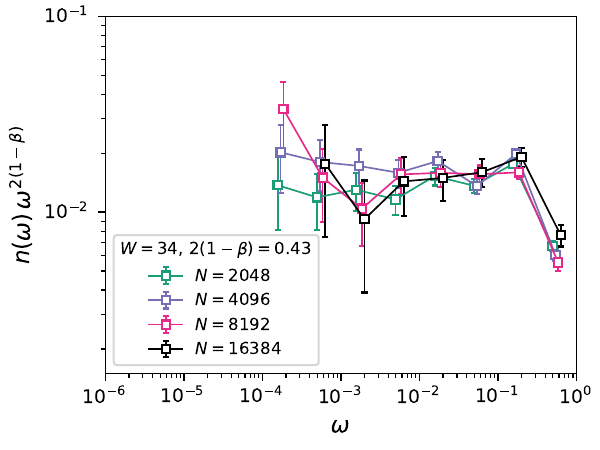}}
\caption{Geometry of the resonant pairs, from exact diagonalization of random regular graphs with $N = 2^{11}$ to $2^{14}$ sites, at $W = 22$ (full symbols) and $W = 34$ (open symbols); eigenstates with $|E| < 1$. Strongly hybridized pairs: overlap $\sum_i |\psi_\alpha(i)|^2 |\psi_{\alpha'}(i)|^2 > 0.05$ and, in both states, a second peak (largest weight on a ball of radius one around a site at distance at least two from the maximum) at least half the first. (a) Mean graph distance $\langle d \rangle$ between the two localization centers vs the splitting $\omega$. (b) Fraction of pairs whose two states share both centers. (c), (d) Number $n(\omega)$ of such partners per eigenstate and per unit $\omega$, multiplied by $\omega^{2(1-\beta)}$, at $W = 22$ and $34$: a plateau corresponds to Eq.~\eqref{eq:pairdensity}. Errors are statistical.}
\label{fig:pairs}
\end{figure*}

We have checked the geometry of the resonant pairs directly, by exact diagonalization of random regular graphs with $N = 2^{11}$ to $2^{14}$ sites at $W = 22$ and $34$ (Fig.~\ref{fig:pairs}). For each eigenstate with $|E| < 1$ we locate its maximum and a second peak, the largest weight on a ball of radius one around a site at distance at least two from the maximum, and we call a pair of eigenstates strongly hybridized if their overlap exceeds $0.05$ and, in both states, the second peak carries at least half the weight of the first. The distance between the two localization centers grows logarithmically with $1/\omega$, from about $3$ at $\omega \approx 0.3$ to about $9$ at $\omega \approx 10^{-4}$ for $W = 22$, and is shorter at $W = 34$, where the wavefunctions decay faster [Fig.~\ref{fig:pairs}(a)]. The fraction of these pairs whose two states share the same two centers grows towards one as $\omega$ decreases [Fig.~\ref{fig:pairs}(b)]: at small $\omega$ they are the hybridized pairs $\simeq (\psi_\alpha \pm \psi_{\alpha'})/\sqrt{2}$ assumed above. Their number per eigenstate and per unit $\omega$ follows $\omega^{-2(1-\beta)}$, as predicted by Eq.~\eqref{eq:pairdensity}: once multiplied by $\omega^{2(1-\beta)}$ it is flat at $W = 34$ and nearly so at $W = 22$, where the effective exponent is slightly smaller than $2(1-\beta) = 0.65$ [Fig.~\ref{fig:pairs}(c), (d)]. It is independent of $N$ for $\omega \gtrsim 10^{-3}$, while at smaller $\omega$ it is lower on the smallest graphs: this is the deficit of the rarest resonances on finite graphs, to which we now turn.

\subsection{Finite size: the missing hybridized pairs} \label{sec:finiteN}

By the sum rule~\eqref{eq:sumruleN}, the finite-size correction to the IPR is the deficit, on a finite graph, of the total overlap of an eigenstate with the others. To test Eq.~\eqref{eq:sumruleN} we have computed, from the same exact-diagonalization data, the total overlap of an eigenstate with the eigenstates within $\omega$ of it, $S_N(\omega) = \avg{\sum_{\alpha' \neq \alpha,\, |E_{\alpha'} - E_\alpha| < \omega} O_{\alpha\alpha'}}$. At $W = 22$, $S_N(10^{-2})$ grows from $N = 2^{11}$ to $N = 2^{12}$ and $2^{13}$ by $0.0023$ and $0.0038$, with statistical uncertainties of a few $10^{-4}$, against $I_2(2^{11}) - I_2(N) = c_N (2^{-11\kappa} - N^{-\kappa}) = 0.0022$ and $0.0040$ with $c_N = 0.15$~\cite{letter}; the difference builds up at $\omega$ below a few mean level spacings $1/(N\avg{\rho})$ of the smaller graph and does not grow further at larger $\omega$.

\begin{figure*}[t]
\centering
\panel{a}{\includegraphics[width=0.325\textwidth]{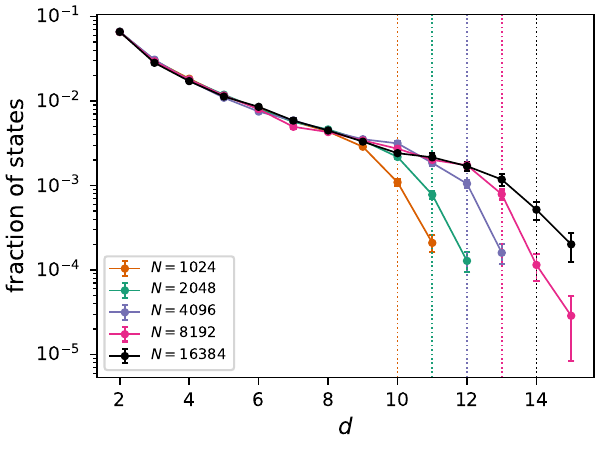}}\hfill
\panel{b}{\includegraphics[width=0.325\textwidth]{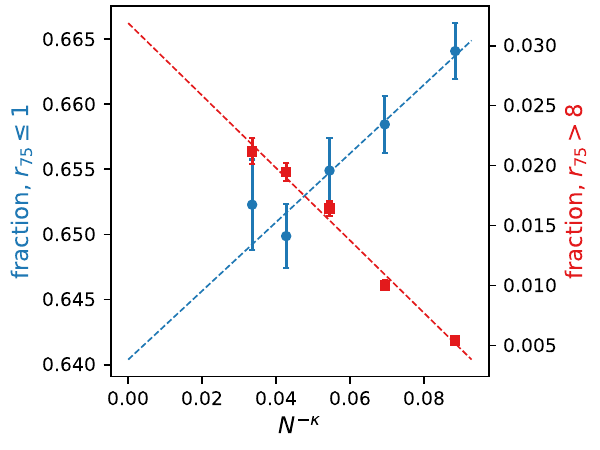}}\hfill
\panel{c}{\includegraphics[width=0.325\textwidth]{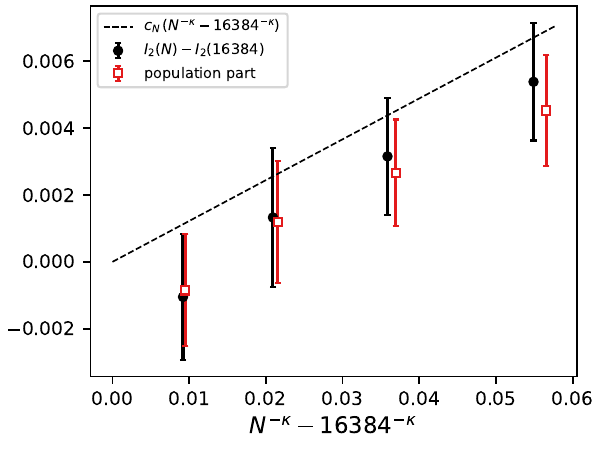}}
\caption{Finite-size corrections to the IPR and the doubly-peaked eigenstates, from exact diagonalization of random regular graphs with $N = 2^{10}$ to $2^{14}$ sites at $W = 22$; eigenstates with $|E| < 1$, errors from a bootstrap over samples. (a) Fraction of the eigenstates that are doubly peaked (second peak at least half the first, peaks defined as in Fig.~\ref{fig:pairs}) with the two peaks at graph distance $d$; dotted lines: $d = \ln N/\ln k = \log_2 N$. (b) Fractions of compact eigenstates, with $75\%$ of the weight within distance $r_{75} \leq 1$ of the maximum (left axis), and of eigenstates whose weight is split over two centers more than $8$ apart, $r_{75} > 8$ (right axis), vs $N^{-\kappa}$, $\kappa = 2\beta - 1 = 0.35$; dashed lines: linear fits. (c) $I_2(N) - I_2(2^{14})$ (full symbols) and its population part $\sum_c [p_c(N) - p_c(2^{14})]\, \bar I_c(2^{14})$, with the classes $c$ defined by $r_{75}$ (open symbols); dashed line: $c_N (N^{-\kappa} - 2^{-14\kappa})$, with $c_N = 0.12$ from a fit of $I_2(N)$ on the same data.}
\label{fig:bicentric}
\end{figure*}

The deficit can be traced to individual eigenstates. We have stored, for each eigenstate with $|E| < 1$ of random regular graphs with $N = 2^{10}$ to $2^{14}$ sites at $W = 22$, its IPR and the positions and weights of its two peaks, defined as above, and we call it doubly peaked if the second peak carries at least half the weight of the first (Fig.~\ref{fig:bicentric}). On these data $I_2(N)$ follows the second of Eqs.~\eqref{eq:kappa_intro} with $\kappa = 2\beta - 1$, with $I_2^\infty = 0.4895 \pm 0.0019$ and $c_N = 0.12 \pm 0.03$, compatible with the values found at $E = 0$. The distribution of the distance $d$ between the two peaks of the doubly-peaked states is the same at all $N$ up to a cutoff at $d \approx \ln N/\ln k$ ($= \log_2 N$ for $k = 2$), which moves by one unit each time $N$ doubles [Fig.~\ref{fig:bicentric}(a)]; this is of the order of the typical distance between two sites of the graph, whose mean grows from $8.1$ at $N = 2^{10}$ to $12.1$ at $N = 2^{14}$. The states that are missing on a finite graph are therefore the hybridized pairs whose two centers would lie farther apart than the graph allows; for the doubly-peaked states with $d \geq 5$, the eigenstate with which they have the largest overlap shares both centers in most cases. Classifying the eigenstates by the radius $r_{75}$ of the ball around their maximum that carries $75\%$ of their weight, the fraction of states whose weight is split over two centers more than $8$ apart ($r_{75} > 8$) grows from $0.005$ at $N = 2^{10}$ to $0.021$ at $N = 2^{14}$, while that of the compact states ($r_{75} \leq 1$) decreases by a similar amount, both linearly in $N^{-\kappa}$ [Fig.~\ref{fig:bicentric}(b)]. The mean IPR within each class is independent of $N$, about $0.75$ for the most compact states and $0.21$ for those with two distant centers, so that writing $I_2 = \sum_c p_c \bar I_c$, with $p_c$ the fraction of states in class $c$ and $\bar I_c$ their mean IPR, the change of the populations, $\sum_c [p_c(N) - p_c(2^{14})] \bar I_c(2^{14})$, accounts for most of $I_2(N) - I_2(2^{14})$ at all $N$ [Fig.~\ref{fig:bicentric}(c)]. The finite-size correction to the IPR is thus a population effect: on a finite graph the doubly-peaked states with distant centers are replaced by compact ones, which have a larger IPR. At $W = 34$ the finite-size corrections between $N = 2^{10}$ and $2^{14}$ are smaller than the statistical errors of these data, and the analysis is inconclusive. The same finite-size corrections are analyzed in Sec.~\ref{sec:Tu} from the point of view of the distribution of the amplitudes, where they are related to the zero mode of the cavity equations.

\subsection{$\beta = 1/2$ and the proliferation of resonances} \label{sec:proliferation}

The same counting makes the role of $\beta = 1/2$ transparent. The mean number of resonant partners of a given state with splitting smaller than $\eta$, $\avg{\rho}\int_{-\eta}^{\eta} N^2\mathcal{C}(\omega)\, \de\omega \propto \eta^{2\beta-1}$, is finite and vanishes as $\eta \to 0$ only for $\beta > 1/2$: localized states are then stable against hybridization with distant ones, and the finite-$\eta$ correction~\eqref{eq:spectral} simply counts the rare resonant pairs whose splitting is not resolved by the regulator. At $W_c$, where the two roots $\beta$ and $1 - \beta$ merge at $1/2$, the count becomes marginal: $\mu \to 1$, and the $1/|\ln H|$ correction derived below for the toy model (Sec.~\ref{sec:critical}), i.e.\ $1/|\ln\eta|$ here, corresponds to $N^2\mathcal{C}(\omega) \propto 1/[\omega \ln^2(1/\omega)]$, whose integral up to $\eta$ decays only logarithmically; in the delocalized phase, where the two roots become complex, the resonances proliferate. This is the resonance-proliferation criterion of Anderson's original argument~\cite{anderson1958absence} and of the self-consistent theory of Abou-Chacra, Anderson and Thouless~\cite{anderson1973selfconsistent,abou1974self}, expressed here through the exponent of the amplitude statistics; its rigorous counterpart on trees is the fractional-moment criterion for resonant delocalization~\cite{aizenman1993localization,aizenman2013resonant}.

\subsection{From the real-space picture to the cavity equations} \label{sec:bridge}

The real-space picture explains why the corrections are controlled by $\beta$, but it takes the tail of $T_\infty(u)$ as an input. The rest of this paper derives Eq.~\eqref{eq:kappa} from the cavity equations themselves; through Eq.~\eqref{eq:kappamu}, this derivation can be read equivalently as a derivation of the exponent of the eigenstate correlations in the localized phase. By Eq.~\eqref{eq:I2deltaQ}, the corrections to be derived are those of the typical values of the LDoS at the resonant sites, in the bulk region $\hat m = O(1)$, and not those of its far tail.

\section{A solvable toy model: the non-linear $\sigma$-model on the hyperbolic superplane} \label{sec:toy}

\subsection{Why a toy model, and why this one}

An effective field-theoretical description of the Anderson localization transition was introduced in Refs.~\cite{wegner1979mobility,schafer1980disordered,efetov1983supersymmetry}, where the problem was mapped onto a NL$\sigma$M with non-compact symmetry. The NL$\sigma$M representation arises in the $n \to \infty$ limit of an $n$-orbital generalization of the model~\cite{wegner1979norbitals}, describing an electron hopping between metallic granules with $n$ orbitals each located at the nodes of the lattice; for $n=1$ one recovers the Anderson tight-binding model with random Gaussian hopping. The $n$-orbital generalization shares the qualitative features and the critical behavior of the original model, with the advantage that analytical calculations are more tractable. The NL$\sigma$M on the infinite BL was solved with the supersymmetry method in Refs.~\cite{efetov1985anderson,efetov1987density,efetov1987anderson,zirnbauer1986localization,zirnbauer1986anderson,verbaarschot1988graded} (see also Refs.~\cite{mirlin1991localization,mirlin1991universality,fyodorov1991localization,fyodorov1992novel,mirlin1994statistical}).

Although this formulation simplifies certain aspects of the problem, it remains highly intricate: the order-parameter field is an $8 \times 8$ supermatrix with a non-compact symmetry structure, and the self-consistent equation for its distribution has, to our knowledge, only been analyzed in the $\eta \to 0^+$ limit. Evaluating the finite-$\eta$ corrections analytically within this framework is a formidable task. We therefore simplify the problem further and consider a ``toy'' model defined on a non-compact symmetric space known as the hyperbolic superplane, introduced in Refs.~\cite{zirnbauer1990fourier,zirnbauer1991fourier} as a simplified model of Anderson localization and further studied in Refs.~\cite{Gruzberg_1996,dupre1996localization}. The physical justification rests on the fact that the critical properties of NL$\sigma$Ms depend crucially on their global symmetry structure~\cite{zirnbauer1986anderson,zirnbauer1986localization,efetov1987anderson,efetov1987density}: the toy model mimics the symmetry structure of the full NL$\sigma$M in the orthogonal class, in particular the non-compact zero mode that must be regularized by a symmetry-breaking field, and it is therefore expected to capture the correct critical behavior. This expectation was confirmed on the BL by Gruzberg and Mirlin~\cite{Gruzberg_1996}, who showed that the toy model shares the exact same, exotic criticality of the Anderson transition on the BL, including the structure of the exponent $\beta$ (two roots $\beta$, $1-\beta$ merging at $1/2$) and its square-root behavior~\eqref{eq:beta}. [An attempt to test whether the correspondence persists in three dimensions was made in Ref.~\cite{dupre1996localization} by Monte Carlo simulations; the results were inconclusive, though not incompatible with it. The model, also known as the $H^{2|2}$ model, has since been studied rigorously: localization at strong disorder in any dimension~\cite{disertori2010anderson}, a delocalized phase in three dimensions~\cite{disertori2010quasi}, and an exact mapping onto the vertex-reinforced jump process~\cite{sabot2015edge} have been established.] For our purposes the essential point is that the toy model has a cavity equation for a {\it single} real function, in which the symmetry-breaking field plays exactly the role of $\eta$, and which can be analyzed in full detail at small field.

\subsection{Definition of the model and cavity equation}

The fields take values on the hyperbolic superplane, parameterized by five-component supervectors
\begin{equation}
    \overline{\psi} = (\sigma, \pi_1, \pi_2, \overline{\xi}, - \xi ) \, ,
\end{equation}
where the first three components are commuting (bosonic) variables and the last two are anticommuting Grassmann variables. The scalar product is $\overline{\psi}_a \cdot \psi_b = \overline{\psi}_a v \, \psi_b$ with metric $v = \text{diag}(1, -1, -1, -1, -1)$, so that the squared norm reads $\vert \vert \psi \vert \vert^2 = \sigma^2 - \pi_1^2 - \pi_2^2 - 2 \overline{\xi} \xi$, and the non-linear constraint $\vert \vert \psi \vert \vert^2 = 1$ is imposed on every site. The bosonic components can be parameterized by ``polar'' coordinates on the hyperbolic superplane, $\pi_1 = \sinh \theta \cos \phi$, $\pi_2 = \sinh \theta \sin \phi$, $\sigma = \cosh \theta + \overline{\xi} \xi / \cosh \theta$, with $0 \le \theta < \infty$ and $0 \le \phi < 2 \pi$, which solve the constraint identically. The Hamiltonian is
\begin{equation} \label{eq:Hhs}
    H_{\rm hs} =  J \sum_{\langle i,j\rangle} \overline{\psi}_i \cdot \psi_j + H \sum_i \sigma_i \, ,
\end{equation}
where $\langle i,j \rangle$ runs over the nearest-neighbor pairs of a BL of branching ratio $k$. The coupling $J$ plays the role of the ratio $1/W$ of hopping to disorder in the Anderson Hamiltonian: the ordered phase at large $J$, characterized by the spontaneous breaking of the non-compact symmetry (in which the IPR proxy defined below vanishes as $H \to 0$), is the counterpart of the delocalized phase, while the high-temperature paramagnetic phase at $J < J_c$ is the counterpart of the localized phase, on which we focus. The symmetry-breaking field $H$ regularizes the non-compact zero mode exactly as the imaginary regulator $\eta$ does in the Anderson model, and the Boltzmann weight of a configuration is $\propto \prod_i \de \psi_i \, e^{-H_{\rm hs}}$, with the invariant integration measure
\begin{equation}
    \de \psi = a \left( 1 - \frac{\overline{\xi} \xi}{\cosh^2 \theta} \right ) \de \theta \, \sinh \theta \, \de \phi \, \de \overline{\xi} \, \de \xi \, ,
\end{equation}
where the constant $a$ is fixed so that $\tilde P \equiv 1$ solves the cavity equation below at $H = 0$ (it absorbs, in particular, the factor $e^{J}$ that appears in Eq.~\eqref{eq:Ptheta_cavity}).

As shown in Ref.~\cite{Gruzberg_1996}, the model is solved exactly on the BL by a cavity recursion for the local distribution $\tilde{P}(\psi)$ of the field at a node, in terms of the distributions of its neighbors on the cavity graph where one neighbor has been removed. Since the model has no quenched disorder it is translationally invariant, so that the distribution is the same on every node, and by symmetry it depends on $\sigma$ only. The self-consistent cavity equation reads~\cite{Gruzberg_1996}
\begin{equation}
    \tilde{P}(\sigma) = \int \de \psi^\prime e^{-J \overline{\psi} \cdot \psi^\prime - H \sigma^\prime} \tilde{P}^k (\sigma^\prime) \, .
\end{equation}
Expanding both sides in powers of the Grassmann variables and integrating them out yields
\begin{equation}  \label{eq:Ptheta_cavity}
\begin{aligned}
   & \tilde{P}(\theta) = e^{J(1 - \cosh \theta) - H} \tilde{P}^k(0) \\
   & \qquad \qquad + \int_0^\infty \! \! \de \theta^\prime L(\theta, \theta^\prime) D(\theta^\prime) \tilde{P}^k(\theta^\prime) \, , \\
   & L(\theta, \theta^\prime) = J \sinh \theta \, e^{J(1 - \cosh \theta \cosh \theta^\prime)} \, \tilde{I}_1 (J \sinh \theta \sinh \theta^\prime) \, , \\
   & D(\theta^\prime) = e^{-H \cosh \theta^\prime} \, ,
\end{aligned}
\end{equation}
where $\tilde{I}_1(x)$ is the modified Bessel function of the first kind. Evaluating this expression at $\theta=0$ gives $\tilde{P}(0) = e^{-H} \tilde{P}^k(0)$, which admits the two solutions $\tilde P(0) = e^{H /(k - 1)}$ and $\tilde P(0) = 0$; to ensure that $\tilde{P}(\theta) \equiv 1$ is a solution at $H = 0$ we select the branch $\tilde{P}(0) = e^{H /(k - 1)}$. Finally, the counterpart of the IPR within the toy model is the correlation function of Ref.~\cite{Gruzberg_1996} evaluated at zero distance, the analogue of $\eta \avg{|G_{ij}|^2}$ at $i = j$; measured at a node where all $k+1$ neighbors are present, it reads
\begin{equation} \label{eq:I2toy}
    I_2(H) = e^{- 2 H /(k - 1)} \, H \int_0^\infty \de \theta \, D(\theta) \, \tilde{P}^{k+1}(\theta) \, \sinh \theta \, .
\end{equation}
Our goal is to characterize the singular small-$H$ behavior of the fixed point $\tilde{P}(\theta)$, and hence of $I_2(H)$, as $H \to 0$ in the paramagnetic phase.

\subsection{Scaling variable and small-$H$ form of the cavity equation} \label{sec:setup}

It is convenient to introduce the variable $t$ defined through $\cosh \theta = e^t/H$, and to work with $P(t) \equiv \tilde{P}(\theta)$. At finite $H$, $t$ ranges over $[t_{\rm min},\infty)$ with $t_{\rm min}=\ln H$. Note that $P(t_{\rm min})=\tilde{P}(0)=e^{H/(k-1)}$ tends to one linearly in $H$ as $H \to 0$, while $P(t)$ decays faster than exponentially as $t \to +\infty$. The rationale for this choice is that, as we shall see, the region $t = O(1)$ is where the $H=0$ solution lives (the ``bulk''), while the region $t - t_{\rm min} = O(1)$ is where the field $H$ acts (the ``boundary layer''); the variable $t$ is the analogue of the logarithm of the Laplace variable conjugate to $\hat m$ in the Anderson problem~\cite{zirnbauer1986localization,zirnbauer1986anderson,verbaarschot1988graded,mirlin1991localization,mirlin1991universality,fyodorov1991localization,fyodorov1992novel,mirlin1994distribution}; the boundary layer at $t \approx \ln H$ thus corresponds to $\hat m \sim 1/\eta$.
To recast Eq.~\eqref{eq:Ptheta_cavity} in a form suitable for the small-$H$ limit we define
\begin{equation}
    \begin{aligned}
    t &= \ln (H \, \cosh \theta) \, , \\
        s(t) &\equiv \sqrt{1 - H^2 e^{-2 t}} \, , \\
        \sinh \theta &= \frac{e^t}{H} s(t) \, , \qquad \de \theta = \frac{\de t}{s(t)} \, , \\
        x(t,t^\prime) & \equiv J \frac{e^{t + t^\prime}}{H^2} s(t) s(t^\prime) \, , \\
        y(t,t^\prime) & \equiv J \frac{e^{t + t^\prime}}{H^2} \left( s(t) s(t^\prime) - 1 \right)\, , \\
        \tilde{I}_1(x) & \equiv \frac{e^{x + F(x)}}{\sqrt{2 \pi x}} \, ,
    \end{aligned}
\end{equation}
the last definition isolating in $F(x)$ the corrections to the large-argument asymptotics of the Bessel function, $F(x) \to 0$ as $x \to \infty$. Inserting these relations into Eq.~\eqref{eq:Ptheta_cavity}, after some algebra one obtains
\begin{subequations} 
	\begin{align}
		\label{eq:PtH}
		P(t) &= e^{J(1-e^t/H)} e^{H/(k-1)} \\
		\nonumber
		& \qquad  \,\, + \int_{t_{\rm min}}^\infty \de t' \, K_H(t,t') \, e^{-e^{t'}} P^k(t') \, , \\
		\label{eq:KH}
		K_H(t,t') &= \frac{s^{1/2}(t)\, s^{-3/2}(t')}{2K_{-1/2}(J)} \, e^{F(x(t,t'))} \, e^{\frac{t-t'}{2}+y(t,t')} \, ,
	\end{align}
\end{subequations}
where we have used $2 K_{-1/2} (J) = \sqrt{2 \pi/J} \, e^{-J}$, with $K_\nu$ the modified Bessel function of the second kind. The function $x(t,t')$ diverges as $H\to 0$ at fixed $t,t'$, so that $e^{F(x)} \to 1$, and $y(t,t') \to - J \cosh (t-t')$. In the limit $H \to 0$ at fixed $t = O(1)$, the bulk regime, the equation therefore reduces to
\begin{equation}
    P_0(t)=\int_{-\infty}^{+\infty} \de t' \, L_0(t-t') \, e^{-e^{t'}} \, P_0^k(t') \, ,
\label{eq:Pt0}
\end{equation}
with the translationally invariant kernel
\begin{equation} \label{eq:L0}
    L_0(t)=\frac{1}{2K_{-1/2}(J)}\exp \left( \frac{t}{2}-J\, \cosh t \right) \, .
\end{equation}
Here and below the subscript of $P_H(t)$ makes the $H$-dependence of the fixed-point solution explicit.

\section{Scaling analysis at small $H$ in the paramagnetic phase} \label{sec:scaling}

\begin{figure*}
    \centering
    \panel{a}{\includegraphics[width=0.325\linewidth]{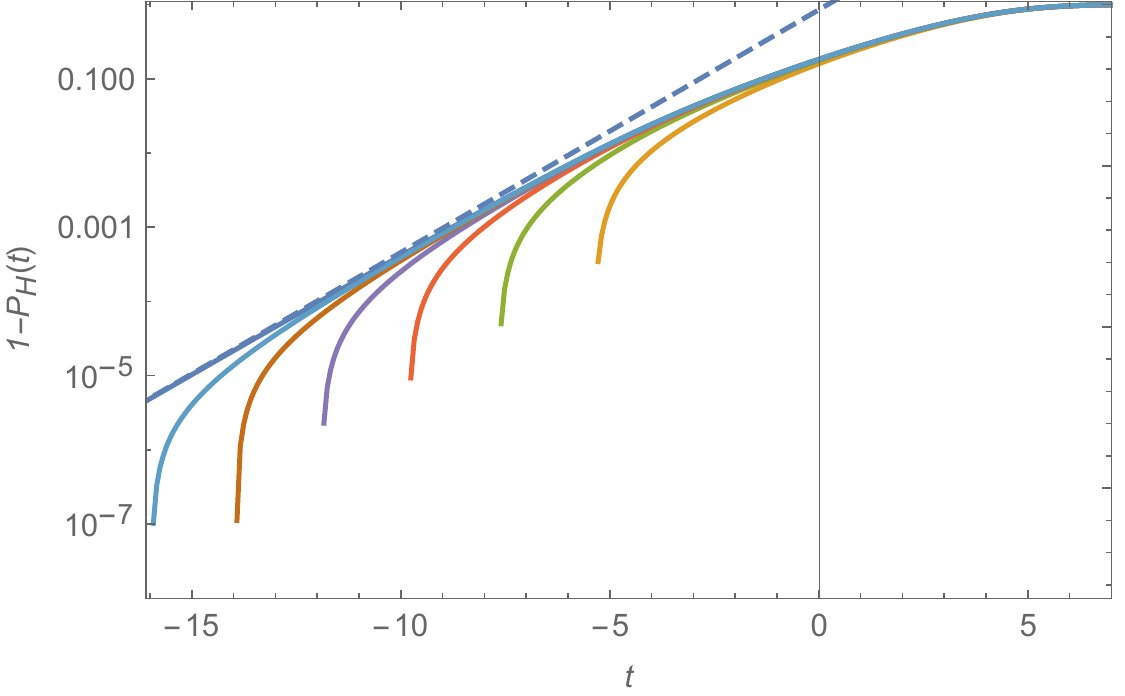}}\hfill
    \panel{b}{\includegraphics[width=0.325\linewidth]{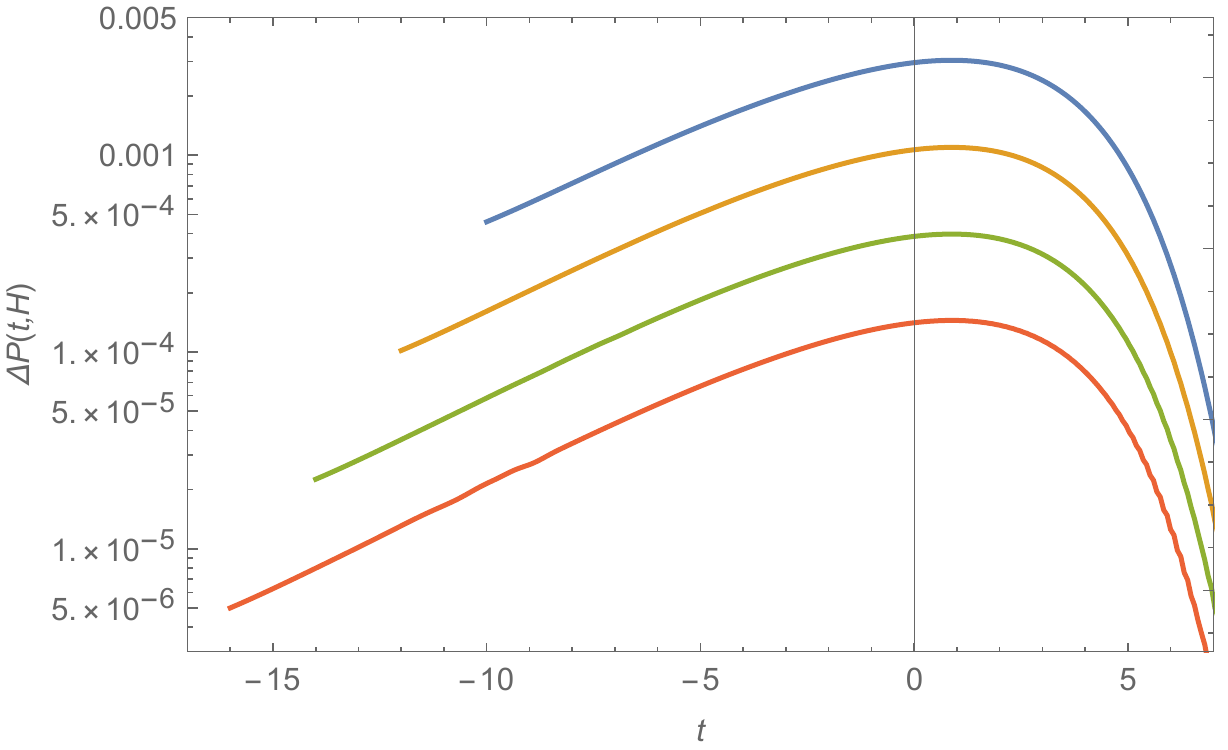}}\hfill
    \panel{c}{\includegraphics[width=0.325\linewidth]{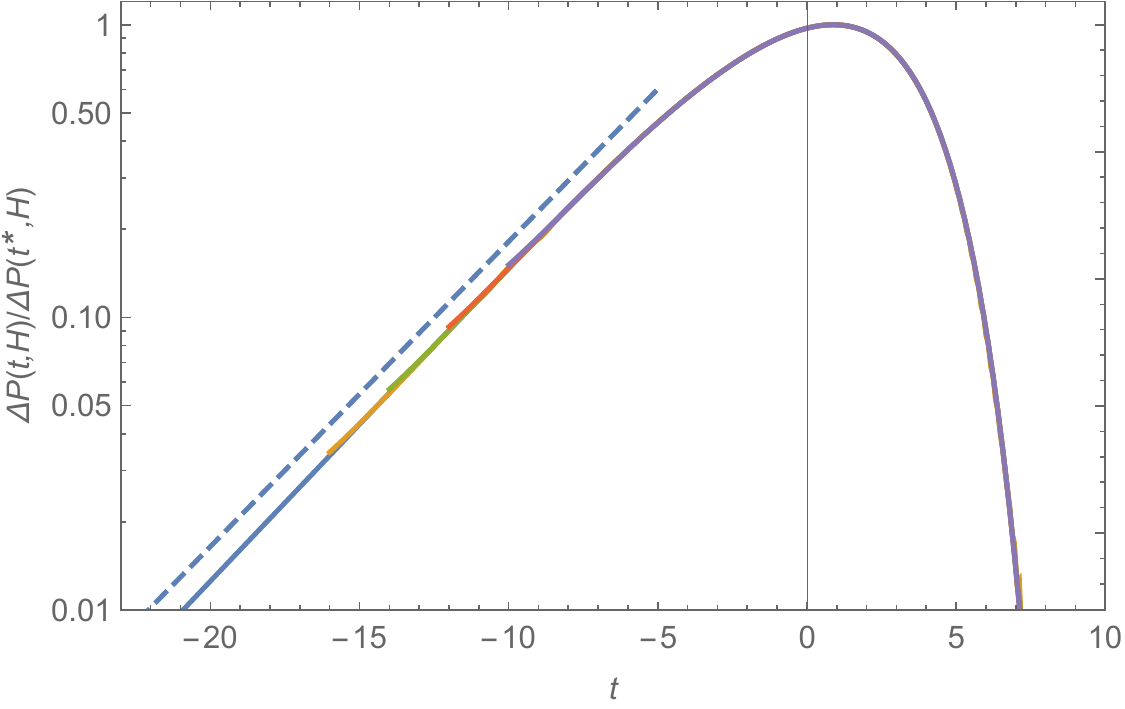}}
    \caption{(a) $\delta P_H(t)  = 1-P_H(t)$ vs. $t$ for $J=0.01$, $k=2$, and several values of $H$, corresponding to $t_{\rm min} = \ln H =-\infty, -16,-14,-12,-10,-8,-6$. The straight dashed line is the asymptote $ c\, e^{\beta\, t}$ of the $H=0$ solution ($\ln H =-\infty$). (b) $\Delta P(t,H) = P_H(t)-P_0(t)$ vs. $t$ for the same parameters and $t_{\rm min}=-16,-14,-12,-10$. (c) $\Delta P(t,H) / \Delta P(t^\star,H)$ vs. $t$, where $t^\star$ is the position of the maximum. The blue line is the scaling function $\ft(t)/\ft(t^\star)$, where $\ft(t)$ is the solution of the linearized equation~\eqref{eq:lindp} that vanishes as $e^{(1-\beta)t}$ for $t \to -\infty$; the dashed line shows $e^{(1-\beta)t}$ for reference.}
    \label{fig:1mpt}
\end{figure*}

\subsection{Bulk regime and the exponent $\beta$} \label{sec:bulk}

As $t \to - \infty$ the solution of Eq.~\eqref{eq:Pt0} must tend to one. Linearizing around $P=1$ by writing $P_0(t) = 1-\delta P_0(t)$, and noting that for $t' \to -\infty$ the factor $e^{-e^{t'}}$ can be replaced by one, one obtains an equation with a kernel that depends only on $t - t'$,
\begin{equation}
  \delta P_0(t)=  k \int_{-\infty}^{+\infty} \de t' \, L_0(t-t')  \, \delta P_0(t') \, ,
\label{eq:linsmallt}
\end{equation}
whose solutions are exponentials, $\delta P_0(t) \propto e^{\beta  t}$, provided $\beta$ satisfies
\begin{equation} \label{eq:betaJ}
\frac{1}{k} = \frac{K_{\beta-1/2}(J)}{K_{-1/2}(J)} \, .
\end{equation}
Owing to the symmetry $K_\nu = K_{-\nu}$ of the Bessel function, Eq.~\eqref{eq:betaJ} is invariant under $\beta \to 1-\beta$. It has no solution for $J>J_c$ (with $J_c \approx 0.02654$ for $k=2$), which is the ordered phase, and two symmetric solutions $(\beta,1-\beta)$ for $J<J_c$, which coalesce at $\beta = 1/2$ at the critical point. $J_c$ is indeed determined by the condition $1=k\, K_0(J_c)/K_{-1/2}(J_c)$. This is exactly the structure of the Abou-Chacra--Anderson--Thouless condition of the Anderson model discussed in Sec.~\ref{sec:linearized}, and $\beta$ plays the role of the LDoS tail exponent of the Anderson model~\cite{Gruzberg_1996}. For $J<J_c$ the physical branch is the larger root, because (i) it correctly reproduces the known behavior of the $H=0$ solution as $t\to-\infty$ in the $J\to 0$ limit and (ii) it is the one selected by the numerical solution at finite $H$ as we will see below. In what follows $\beta > 1/2$ denotes this larger root, and
\begin{equation} \label{eq:P0asympt}
    \delta P_0(t) = 1 - P_0(t) \simeq c \, e^{\beta t} \, , \qquad t \to -\infty \, ,
\end{equation}
where the prefactor $c$ is not fixed by the linearized equation and must be obtained from the numerical solution of the full Eq.~\eqref{eq:Pt0}; for $J = 0.01$ and $k=2$ we find $c \approx 0.8551$ and $\beta=0.7530$.\footnote{Determining $c$ accurately requires some care, because the asymptotic regime~\eqref{eq:P0asympt} is reached only for $t \to -\infty$, where $P_0(t)$ differs from one by an exponentially small amount: $c$ is fixed by requiring the numerical solution to join the asymptotic form $1 - c\,e^{\beta t}$ smoothly (Algorithm 2 of Appendix~\ref{app:numerics}).}
The second root $1-\beta$ plays no role in the $H=0$ solution. Since $t \to -\infty$ corresponds to the Laplace variable conjugate to $\hat m$ going to zero, i.e.\ to $\hat m \to \infty$, Eq.~\eqref{eq:P0asympt} is the counterpart of the tail~\eqref{eq:Qmhattails}.

We solve Eq.~\eqref{eq:PtH} numerically for several values of $H$; the codes are available in a public repository~\cite{coderepo}. An accurate solution is essential for the analysis that follows, since the quantities of interest, such as $\Delta P(t,H) = P_H(t) - P_0(t)$, are small differences between two solutions that must be resolved over many orders of magnitude as $H \to 0$. Figure~\ref{fig:1mpt}(a) shows $\delta P_H(t)=1-P_H(t)$ for $J=0.01$ and $k=2$, for $t_{\rm min}=-\infty$ ($H=0$) and $t_{\rm min} = -16,\ldots,-6$. The curves converge to the $H=0$ solution and deviate from it only for $t = O(t_{\rm min})$; the straight dashed line is the asymptote~\eqref{eq:P0asympt}.

\subsection{Scaling function in the bulk: $\ft(t)$} \label{sec:ftyp}

Consider now the difference between the fixed-point solutions at fixed $t$ for varying $H$, $\Delta P(t,H) \equiv P_H(t) - P_0(t) = \delta P_0(t) - \delta P_H(t)$, shown in Fig.~\ref{fig:1mpt}(b). When each curve is rescaled by its maximum value $\Delta P(t^\star, H)$, all curves collapse onto a single one [Fig.~\ref{fig:1mpt}(c)], apart from the cutoff at $t=t_{\rm min}$. This suggests that, at fixed $t=O(1)$ and for $H \to 0$,
\begin{equation} \label{eq:ansatz}
   P_H(t)-P_0(t) = \at(H) \,  \ft(t) \, ,
\end{equation}
with an amplitude $\at(H)$ which, from Fig.~\ref{fig:1mpt}(b), vanishes as a power of $H$ as $H \to 0$. Remarkably, the scaling curve vanishes as $t \to -\infty$ as $e^{(1-\beta)t}$, i.e.\ according to the {\it smaller} of the two roots of Eq.~\eqref{eq:betaJ}. To our knowledge, this is the first instance in which the subdominant solution of the linearized cavity equation acquires a direct physical meaning in this context.

The equation obeyed by $\ft$ follows from the small-$H$ expansion of Eq.~\eqref{eq:PtH} in the bulk. The boundary (source) term vanishes faster than any power of $H$ at fixed $t$, and therefore does not contribute to the algebraic small-$H$ expansion. Expanding the kernel, one finds $K_H(t,t') = L_0(t-t') + O(H^2)$; as we check self-consistently below, $\at(H)$ is much larger than $O(H^2)$, so the corrections to the kernel can be neglected. Linearizing Eq.~\eqref{eq:PtH} around $P_0(t)$ with the ansatz~\eqref{eq:ansatz} then gives
\begin{equation}
   \ft(t)= k \int_{-\infty}^{+\infty} \de t' \, L_0(t-t') \, e^{-e^{t'}} P_0^{k-1}(t') \, \ft (t') \, .
    \label{eq:lindp}
\end{equation}
In other words, $\ft$ is a zero mode of the $H=0$ cavity equation linearized around its fixed point $P_0$: the operator $1 - \mathcal{L}_0$, with $\mathcal{L}_0$ the right-hand side of Eq.~\eqref{eq:lindp}, has a non-trivial kernel. This is what makes a correction larger than the naive $O(H^2)$ possible at all. Note that Eq.~\eqref{eq:lindp} differs from Eq.~\eqref{eq:linsmallt}, which holds only for $t \to -\infty$: it describes the correction for any finite $t$, and reduces to Eq.~\eqref{eq:linsmallt} for $t,t' \to -\infty$ since $P_0(t') \to 1$ there. It therefore admits solutions behaving as $e^{\beta t}$ and $e^{(1-\beta)t}$ at $t \to -\infty$. The second one is selected by requiring that the correction vanish as $H \to 0$: the field removes a weight $\propto H^{\beta}$ in the boundary layer at $t \simeq \ln H$ (Sec.~\ref{sec:boundary}), and propagated into the bulk along $e^{\beta t}$ this perturbation would be amplified by $H^{-\beta}$, giving an $O(1)$ change of the solution, i.e.\ a different $H=0$ fixed point, which is excluded since the prefactor $c$ of Eq.~\eqref{eq:P0asympt} is fixed by the full Eq.~\eqref{eq:Pt0}; along $e^{(1-\beta)t}$ it is amplified only by $H^{-(1-\beta)}$, giving the vanishing amplitude $H^{2\beta-1}$ derived below. The numerics of Fig.~\ref{fig:1mpt}(c) confirm this selection. The zero mode is thus unique, and we fix its normalization by imposing
\begin{equation} \label{eq:ftypasympt}
    \ft(t) \simeq e^{(1-\beta)t} \, , \qquad t \to -\infty \, .
\end{equation}
The reason why the subdominant root is selected becomes clear once the boundary layer is analyzed.

\subsection{Boundary layer: the function $\fm$} \label{sec:boundary}

To determine the amplitude $\at(H)$ one must study $P_H(t)$ in the region $t=O(t_{\rm min})$, where it departs from $P_0(t)$. There $P_H(t)$ is close to one [Fig.~\ref{fig:1mpt}(a)], and Eq.~\eqref{eq:PtH} can again be linearized around $P=1$. The numerical results indicate the scaling form
\begin{equation}
    P_H (t) = 1 - \am(H) \, \fm (\Delta t)  \, , \qquad
    \Delta t \equiv t-t_{\rm min}=O(1) \, .
\label{eq:sc2}
\end{equation}
Using $t = t_{\rm min} + \Delta t$ and $t_{\rm min} = \ln H$, one finds
\begin{equation}
    \begin{aligned}
        s(t) & = \sqrt{1 - e^{-2 \Delta t}} \equiv \tilde{s}( \Delta t) \, , \\
        x(t,t^\prime) & = J e^{\Delta t + \Delta t^\prime} \tilde s(\Delta t) \tilde s( \Delta t^\prime) \equiv \tilde{x}(\Delta t, \Delta t^\prime)  \, , \\
        y(t,t^\prime) & = J e^{\Delta t + \Delta t^\prime} \left( \tilde s(\Delta t) \tilde s(\Delta t^\prime) - 1 \right) \equiv \tilde{y}(\Delta t, \Delta t^\prime)\, ,
    \end{aligned}
\end{equation}
so that, in terms of $\Delta t, \Delta t'$, the kernel of Eq.~\eqref{eq:PtH} becomes exactly $H$-independent. This is a special feature of the parametrization $t = \ln(H \cosh \theta)$, and it is what makes the boundary-layer equation tractable. The source term becomes $e^{J(1- e^{\Delta t})}$, while the factor $e^{-e^{t'}} = e^{-H e^{\Delta t'}}$ tends to one. Setting $H \to 0$ and $P \equiv 1$ in Eq.~\eqref{eq:PtH} gives the identity
\begin{widetext}
    \begin{equation} \label{eq:zeroth}
        1  = e^{J(1- e^{\Delta t})} + \frac{1}{2K_{-1/2}(J)}\int_{0}^\infty \de \Delta t' \, \tilde{s}^{\, -3/2}(\Delta t') \, \tilde{s}^{1/2} (\Delta t) \, e^{F(\tilde{x}(\Delta t,\Delta t'))} \, e^{(\Delta t-\Delta t')/2+\tilde{y}(\Delta t,\Delta t')} \, ,
    \end{equation}
which holds exactly for every $\Delta t \ge 0$, since it merely expresses the fact that $P \equiv 1$ solves the $H \to 0$ limit of the boundary-layer equation. Inserting the ansatz~\eqref{eq:sc2} into Eq.~\eqref{eq:PtH}, linearizing in $\am(H)$ and using Eq.~\eqref{eq:zeroth}, one obtains the linear equation
\begin{equation}
    \fm(\Delta t) = \frac{k}{2K_{-1/2}(J)}\int_{0}^\infty \de \Delta t' \, \tilde{s}^{\, -3/2}(\Delta t') \, \tilde{s}^{1/2} (\Delta t) \, e^{F(\tilde{x}(\Delta t,\Delta t'))} \, e^{(\Delta t-\Delta t')/2+\tilde{y}(\Delta t,\Delta t')}  \, \fm(\Delta t') \, .
\label{eq:linfmin}
\end{equation}
\end{widetext}
As $\Delta t \to \infty$ the kernel of Eq.~\eqref{eq:linfmin} reduces to that of Eq.~\eqref{eq:linsmallt}, so that $\fm(\Delta t)$ is asymptotically a combination of $e^{\beta \Delta t}$ and $e^{(1-\beta)\Delta t}$; we fix its normalization by the leading behavior
\begin{equation} \label{eq:fminasympt}
    \fm(\Delta t) \simeq e^{\beta \Delta t} \, , \qquad \Delta t \to \infty \, .
\end{equation}
At the other end, $P_H(t_{\rm min}) = e^{H/(k-1)} \approx 1 + H/(k-1)$ tends to one linearly in $H$, which, as we check self-consistently below, is faster than $\am(H)$, so that $\fm(0)=0$; this is automatically satisfied by Eq.~\eqref{eq:linfmin}, whose kernel vanishes at $\Delta t = 0$ through the factor $\tilde s^{1/2}(\Delta t)$.

\begin{figure}
    \centering
    \includegraphics[width=0.46\textwidth]{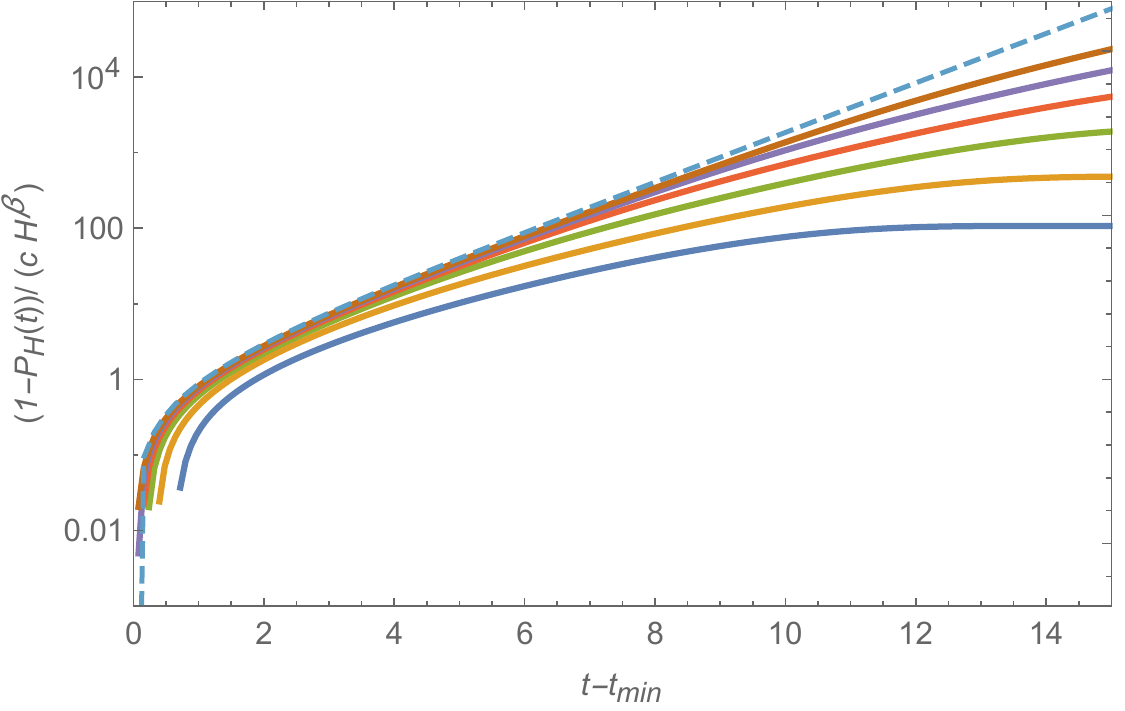}
    \caption{$[1-P_H(t)]/(c\, H^{\beta})$ vs $\Delta t = t-t_{\rm min}$ for $J=0.01$, $k=2$ and $t_{\rm min}=-16,-14,-12,-10,-8,-6$. For $H \to 0$ the curves converge to the scaling function $\fm(\Delta t)$ (dashed blue), solution of Eq.~\eqref{eq:linfmin}.}
    \label{fig:res1mPtdt}
\end{figure}

The amplitude $\am(H)$ is fixed by matching the boundary-layer solution to the bulk one in the overlap region $t_{\rm min} \ll t \ll -1$, where both Eq.~\eqref{eq:P0asympt} and Eq.~\eqref{eq:fminasympt} hold:
\begin{equation}
\begin{aligned}
     & 1- \am(H) \, e^{\beta (t-t_{\rm min})} = 1- c \, e^{\beta t} \\
     & \qquad \Longrightarrow \quad
    \am(H) = c \, e^{\beta t_{\rm min}} = c \, H^{\beta} \, .
\end{aligned}
\label{eq:amin}
\end{equation}
Figure~\ref{fig:res1mPtdt} confirms this result: the curves $[1-P_H(t)]/(c\, H^{\beta})$ converge, as $H \to 0$, to the scaling function $\fm (\Delta t)$ obtained by solving Eq.~\eqref{eq:linfmin}. It is solved with Algorithm~4 of Appendix~\ref{app:numerics}.
Note that $\am(H) \propto H^\beta$ vanishes more slowly than linearly in $H$, as anticipated. The boundary layer is the counterpart, in the Anderson problem, of the region $\hat m \sim 1/\eta$ where the regulator cuts off the power-law tail of $Q(0,\hat m)$, and Eq.~\eqref{eq:amin} is the statement that the weight of this region scales as $\eta^\beta$ [cf.\ Eq.~\eqref{eq:Qmhattails}].

\subsection{Matching and the correction exponent $\kappa=2\beta-1$} \label{sec:matching}

We can now determine $\at(H)$ by matching the $t \to - \infty$ behavior of the bulk ansatz~\eqref{eq:ansatz} with the $\Delta t \to \infty$ behavior of the boundary-layer ansatz~\eqref{eq:sc2}. The key observation is that, beyond its leading behavior~\eqref{eq:fminasympt}, $\fm(\Delta t)$ contains a {\it subleading} divergent term proportional to the second root,
\begin{equation} \label{eq:fminsub}
    \fm(\Delta t) = e^{\beta \Delta t} - c_3 \, e^{(1-\beta)\Delta t} + \ldots \, , \qquad \Delta t \gg 1 \, .
\end{equation}
Such a term is allowed because both $e^{\beta \Delta t}$ and $e^{(1 - \beta) \Delta t}$ solve the asymptotic linearized equation~\eqref{eq:linsmallt}; its coefficient $c_3$ is however not free, since it is determined by the full boundary-layer equation~\eqref{eq:linfmin}, whose solution is unique once normalized by Eq.~\eqref{eq:fminasympt}. Solving Eq.~\eqref{eq:linfmin} numerically for $J=0.01$ and $k=2$ we find $c_3 \approx 0.969181$ ($c_3$ depends on $J$, and we argue in Sec.~\ref{sec:nearcritical} that it tends to one as $J \to J_c$). The subleading term is precisely what determines $\at(H)$. In the overlap region $t_{\rm min} \ll t \ll -1$ the leading terms cancel between $\am(H) \fm(\Delta t)$ and $c \, e^{\beta t}$, by construction of Eq.~\eqref{eq:amin}, and what is left is
\begin{equation}
\begin{aligned}
     P_H(t)-P_0(t) &= -  \am(H) \, \fm (\Delta t) + c \, e^{\beta t} \\ &=    c_3 \, \am(H)\, e^{(1-\beta)(t-t_{\rm min})} \\ &= \at(H ) \, e^{(1-\beta)t} \, ,
     \end{aligned}
\end{equation}
where the last equality is the $t \to -\infty$ limit of the bulk ansatz~\eqref{eq:ansatz} with the normalization~\eqref{eq:ftypasympt}. This explains why the bulk correction is carried by the subdominant root, and gives
\begin{equation} \label{eq:atyp}
    \at(H) = c_3 \, \am(H) \, e^{-(1-\beta) t_{\rm min}}= c_3 \, c \, H^{2\beta-1} \, .
\end{equation}
Since the IPR proxy~\eqref{eq:I2toy} is an integral of $P_H(t)$ dominated by the bulk region $t = O(1)$, Eqs.~\eqref{eq:ansatz} and~\eqref{eq:atyp} imply
\begin{equation} \label{eq:toyresult}
    I_2(H) -I_2(0) \propto H^{2\beta -1} \, ,
\end{equation}
i.e.\ $\kappa = 2\beta - 1$, in agreement with the small-$\eta$ corrections to the IPR of the Anderson model, Eq.~\eqref{eq:kappa}. The mechanism can be summarized as follows: the field $H$ removes a weight $\propto H^\beta$ from the tail of the distribution; this perturbation does not propagate into the bulk along the dominant mode $e^{\beta t}$, which is used up to match the unperturbed solution, but along the subdominant one, $e^{(1-\beta)t}$, and in doing so it is amplified by the factor $e^{-(1-\beta)t_{\rm min}} = H^{-(1-\beta)}$ between the boundary layer and the bulk. The product $H^{\beta} \cdot H^{-(1-\beta)} = H^{2\beta - 1}$ is the anomalous correction.

\begin{figure}
    \centering
    \includegraphics[width=0.46\textwidth]{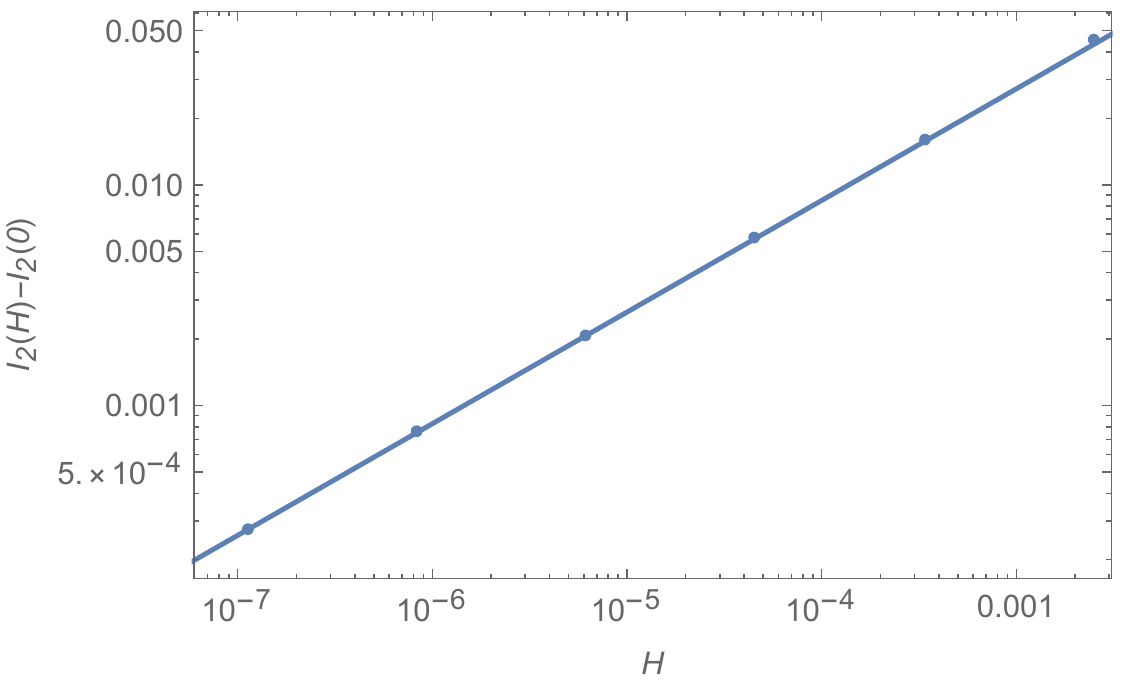}
    \caption{IPR difference $I_2(H)-I_2(0)$ as a function of $H=e^{t_{\rm min}}$, for $J = 0.01$ and $k=2$. The solid line is the parameter-free prediction, Eq.~\eqref{eq:I2scal}.}
    \label{fig:DI2vsH}
\end{figure}

To demonstrate the aforementioned behavior let us write the expression of $I_2(H)$ in terms of the variable $t$: Eq.~\eqref{eq:I2toy} yields
\begin{equation}
    I_2(H)= e^{-2H/(k-1)} \int_{t_{\rm min}}^\infty\, e^{-e^t} \, e^t \, P_H^{k+1}(t) \, \de t \ .
\end{equation}
The prefactor $e^{-2H/(k-1)} = 1 + O(H)$ only produces analytic corrections, subdominant with respect to $H^{2\beta - 1}$, and we neglect it in the following.
In the limit $H \to 0$ the difference with $I_2(0)$ is controlled by the bulk region $t=O(1)$, and recalling that $P_H(t)\approx P_0(t)+c\, c_3\, H^{2\beta-1}\ft(t)$ we obtain
\begin{equation}
\begin{aligned}
   I_2(H)-I_2(0) \approx {} & H^{2\beta-1} c \, c_3 \\
   & \times (k+1) \int_{-\infty}^{+\infty} e^{-e^t} \, e^t \, P_0^k(t) \,\ft(t)\, \de t \, .
\end{aligned}
\label{eq:I2scal}
\end{equation}
The above prediction~\eqref{eq:I2scal} can be checked quantitatively, indeed all its ingredients can be computed, in particular from the numerical solution of the functions $\fm(\Delta t)$, $\ft(t)$ and $P_0(t)$ we have
\begin{equation}
  (k+1)  \int_{-\infty}^{+\infty} \, e^{-e^t} \, e^t \, P_0^k(t) \,\ft(t)\, \de t=1.08772 \, , 
\end{equation}
\begin{equation}
    c=0.8551 \, , \qquad c_3=0.969181 \, .
\end{equation}
Figure~\ref{fig:DI2vsH} shows $I_2(H)-I_2(0)$ as a function of $H$, together with the asymptotic behavior~\eqref{eq:I2scal}.
The agreement is excellent, both in the exponent and in the prefactor. The exponent $2\beta - 1$ vanishes at the critical point, implying a slower than power-law convergence of the IPR as $H \to 0$ at criticality, which we analyze in Sec.~\ref{sec:critical}.

\subsection{Summary of the scaling regimes} \label{sec:summary}

For convenience we collect the behavior of $P_H(t)$ in the different regimes. In the bulk, $t=O(1)$,
\begin{equation}
    P_H(t)= P_0(t) + c_3 \, c \, \ft(t) \, H^{2\beta-1}  \, ,
\label{eq:1nonc}
\end{equation}
where $\ft(t)$ is determined by Eq.~\eqref{eq:lindp} with the normalization~\eqref{eq:ftypasympt}. In the intermediate region $(t-t_{\rm min})/|t_{\rm min}|=O(1)$,
\begin{equation}
   P_H(t) = 1 - c \, e^{\beta t} \left[ 1 - c_3  \,e^{(1-2\beta) (t-t_{\rm min})} \right] \, ,
\label{eq:2nonc}
\end{equation}
and in the boundary layer $\Delta t=O(1)$,
\begin{equation}
   P_H(t) =1- c \, H^{\beta} \fm(\Delta t)  \, ,
\label{eq:3nonc}
\end{equation}
where $\fm(\Delta t)$ obeys Eq.~\eqref{eq:linfmin}, satisfies $\fm(0)=0$, and behaves as in Eq.~\eqref{eq:fminsub} for $\Delta t \to \infty$. The constants $c$ and $c_3$ are determined numerically from Eqs.~\eqref{eq:Pt0} and~\eqref{eq:linfmin}; for $J = 0.01$ and $k=2$, $c \approx 0.8551$ and $c_3 \approx 0.969181$. As we shall see in Secs.~\ref{sec:anderson} and~\ref{sec:numerics}, the bulk regime~\eqref{eq:1nonc} is precisely the scaling regime found for the Anderson model on the BL, both for the finite-$\eta$ corrections to the distribution $\overline{Q}_\eta(\hat m)$ and for the finite-$N$ corrections to the distribution of the eigenstates' amplitudes.

\subsection{Eigenstate correlations in the toy model: complex field} \label{sec:complexH}

In the Anderson model the finite-$\eta$ corrections are fixed, through the spectral identity~\eqref{eq:spectral}, by the correlation function $N^2\mathcal{C}(\omega)$ of two eigenstates at energy distance $\omega$. The toy model has no eigenstates, since $H_{\rm hs}$ is not a quantum Hamiltonian, but the counterpart of $\mathcal{C}(\omega)$ can be defined, and computed, by promoting $H$ to a complex variable, $H = \eta - {\rm i}\omega/2 = |H|\, e^{-{\rm i}\chi}$, with $\eta > 0$, i.e.\ $|\chi| < \pi/2$. The field enters Eq.~\eqref{eq:Ptheta_cavity} only through $D(\theta)$, the source term and the boundary value $\tilde P(0) = e^{H/(k-1)}$, and the integrals over $\theta$ converge for ${\rm Re}\, H > 0$. The fixed point continued from real $H$, and with it $I_2(H)$ of Eq.~\eqref{eq:I2toy}, is therefore analytic in the right half plane, provided it meets no singularity, of which we see no sign; $I_2(H)$ is real for real $H$, and tends to one for $H \to \infty$, where the measure $H D(\theta) \sinh\theta \, \de\theta$ concentrates at $\theta = 0$ and the factor $e^{-2H/(k-1)}$ compensates $e^{-H}\tilde P^{k+1}(0)$. By the Schwarz integral formula for the half plane, $I_2(H)/H$ then admits the spectral representation
\begin{equation} \label{eq:spectraltoy}
\begin{aligned}
    \frac{I_2(H)}{H} &= \int_{-\infty}^{\infty} \de\Omega \, \frac{\mathcal{S}(\Omega)}{H + {\rm i}\Omega/2} \, , \\
    \mathcal{S}(\omega) &= \frac{1}{2\pi} \lim_{\eta \to 0^+} {\rm Re} \, \frac{I_2(\eta - {\rm i}\omega/2)}{\eta - {\rm i}\omega/2} \, ,
\end{aligned}
\end{equation}
where $\mathcal{S}(\omega) = I_2(0)\,\delta(\omega) + \mathcal{S}_{\rm reg}(\omega)$ is real and even, and $\int \de\omega\, \mathcal{S}(\omega) = \lim_{H\to\infty} I_2(H) = 1$.

For real $H = \eta$, Eq.~\eqref{eq:spectraltoy} reads $I_2(\eta) - I_2(0) = \int \de\omega \, \mathcal{S}_{\rm reg}(\omega)\, 4\eta^2/(\omega^2 + 4\eta^2)$, which is Eq.~\eqref{eq:spectral}, while $\int \de\omega\, \mathcal{S}_{\rm reg}(\omega) = 1 - I_2(0)$ is the sum rule~\eqref{eq:sumrule} on the infinite BL: $\mathcal{S}_{\rm reg}(\omega)$ is the analogue of $\avg{\rho}N^2\mathcal{C}(\omega)$, and $I_2(0)\,\delta(\omega)$ that of the diagonal terms. In the Anderson model, indeed, the same continuation of $I_2(\eta)/\eta = \avg{|G_{ii}|^2}/(\pi\avg{\rho})$ replaces $G_{ii}({\rm i}\eta)\, G_{ii}(-{\rm i}\eta)$ with $G_{ii}(\omega/2 + {\rm i}\eta)\, G_{ii}(-\omega/2 - {\rm i}\eta)$, whose average over energy obeys Eq.~\eqref{eq:spectraltoy} with $\mathcal{S}(\omega) = I_2^\infty \delta(\omega) + \avg{\rho} N^2\mathcal{C}(\omega)$, up to analytic terms.\footnote{The analogy has two limits. First, $\mathcal{S} \ge 0$ by construction in the Anderson model, where it is a sum over pairs of eigenstates, whereas in the toy model, whose Hamiltonian~\eqref{eq:Hhs} is the energy function of a classical spin model rather than a quantum Hamiltonian with eigenstates, the positivity of $\mathcal{S}$, equivalent to ${\rm Re}\,[I_2(H)/H] \ge 0$ for ${\rm Re}\,H > 0$, must be checked. Second, in the Anderson model the two Green's functions at energies $\pm\omega/2$ are no longer complex conjugate, and must be followed through a joint recursion at the two energies, as in Eq.~\eqref{eq:OpsPD}: the continuation cannot be carried out on the single-energy distribution $P_\eta(g,\hat g)$, and a frequency is not simply a complex $\eta$. In the supersymmetric NL$\sigma$M, instead, $\eta$ and $\omega$ enter only through the complex coefficient of the symmetry-breaking term~\cite{efetov1983supersymmetry,efetov1987density,mirlin2000statistics}, as $H$ does here, so that the continuation discussed below should apply there as well.}

Let $c_H \equiv c\, c_3\, (k+1) \int \de t \, e^{-e^t} e^t P_0^k(t)\, \ft(t)$ be the prefactor of Eq.~\eqref{eq:I2scal}, $I_2(H) - I_2(0) \simeq c_H H^{2\beta-1}$ ($c_H = 0.9014$ for $J = 0.01$). Assuming that the continued fixed point meets no singularity, this law can be continued by a rotation in the complex plane, $H = |H|\,e^{-{\rm i}\chi}$, with the same $c_H$ up to the imaginary axis, where $[I_2(H) - I_2(0)]/H \simeq c_H H^{2\beta - 2}$, with $H^{2\beta-2} = (|\omega|/2)^{-2(1-\beta)}\, e^{{\rm i}\pi(1-\beta)\,{\rm sgn}\,\omega}$; Eq.~\eqref{eq:spectraltoy} then gives
\begin{equation} \label{eq:Stoy}
    \mathcal{S}_{\rm reg}(\omega) \simeq \frac{c_H}{2\pi} \, \cos[\pi(1-\beta)] \left( \frac{|\omega|}{2} \right)^{-2(1-\beta)} , \qquad \omega \to 0 \, ,
\end{equation}
i.e.\ $\mu = 2(1-\beta)$, with an amplitude that contains no free parameter. Since $\cos[\pi(1-\beta)] = \sin(\pi\kappa/2)$, Eq.~\eqref{eq:Stoy} is equivalent to Eq.~\eqref{eq:kappamu}, with $c_\eta \to c_H$ and $\avg{\rho}B|\omega|^{-\mu} \to \mathcal{S}_{\rm reg}(\omega)$, as it must be, since at real $H$ Eq.~\eqref{eq:spectraltoy} has the form of Eq.~\eqref{eq:spectral}. The continuation, however, computes the correlation function itself, analytically and with no assumption on its sign or monotonicity: it is the counterpart, for the toy model, of the direct numerical computations of $N^2\mathcal{C}(\omega)$ in Sec.~\ref{sec:corrtests}. We have verified numerically, by solving the cavity equation at complex $H$, that the rotation is allowed and that Eq.~\eqref{eq:Stoy} holds, with no adjustable parameter (Appendix~\ref{app:complexH} and Fig.~\ref{fig:Stoy}).

The rotation works because each step of Secs.~\ref{sec:bulk}--\ref{sec:matching} goes through at complex $H$. In the real variable $t = \ln(|H|\cosh\theta)$ the kernel of Eq.~\eqref{eq:PtH} stays real, $K_{|H|}(t,t')$, and only the factor $e^{-e^{t'}}$ becomes $\exp(-e^{t' - {\rm i}\chi})$. (i)~In the bulk the equation reduces to $P(t) = \int \de t' \, L_0(t-t') \exp(-e^{t'-{\rm i}\chi}) P^k(t')$, which is solved by $P_0(t - {\rm i}\chi)$, as follows by shifting the integration contour by ${\rm i}\chi$ [$L_0(u)$ is analytic and decays for ${\rm Re}\,u \to \pm\infty$ in the strip $|{\rm Im}\,u| < \pi/2$, where $\exp(-e^{u})$ also decays for ${\rm Re}\,u \to +\infty$]: the bulk solution is the $H = 0$ one evaluated at the complex argument $\ln(H\cosh\theta)$, and the zero mode of Eq.~\eqref{eq:lindp} is likewise $\ft(t - {\rm i}\chi)$. (ii)~In the boundary layer, $\Delta t = t - \ln|H| = O(1)$, one has $|H\cosh\theta| \to 0$: as for real $H$, the source term and the boundary value lose their dependence on $H$, so that Eq.~\eqref{eq:linfmin}, $\fm$ and $c_3$ are unchanged. (iii)~The matching is the same algebra as in Eqs.~\eqref{eq:amin} and~\eqref{eq:atyp}, with $H\cosh\theta$ in place of $e^{t}$ on the bulk side; it gives $\am(H) = c\,H^{\beta}$ and $\at(H) = c\,c_3\,H^{2\beta-1}$ on the principal branch and, after shifting back the contour in the integral of Eq.~\eqref{eq:I2scal}, $I_2(H) - I_2(0) \simeq c_H H^{2\beta-1}$ for all $|\chi| < \pi/2$. At $|\chi| = \pi/2$ the contour shift is no longer allowed, and Eq.~\eqref{eq:Stoy} follows as a boundary value.\footnote{This assumes that the limit $|\chi| \to \pi/2$ is uniform. This assumption, the existence of the continued fixed point and the positivity of $\mathcal{S}_{\rm reg}$ beyond small $\omega$ are supported by the numerics of Appendix~\ref{app:complexH}, but not proven; at small $\omega$, positivity follows from Eq.~\eqref{eq:Stoy}, since $\cos[\pi(1-\beta)] > 0$ for $\beta > 1/2$.} The amplitude of Eq.~\eqref{eq:Stoy} vanishes as $\beta \to 1/2$, and the critical point requires a separate treatment, to which we now turn.

\section{Corrections at and near the critical point} \label{sec:critical}

\subsection{Logarithmic corrections at $J = J_c$}

\begin{figure}
    \centering
    \includegraphics[width=0.46\textwidth]{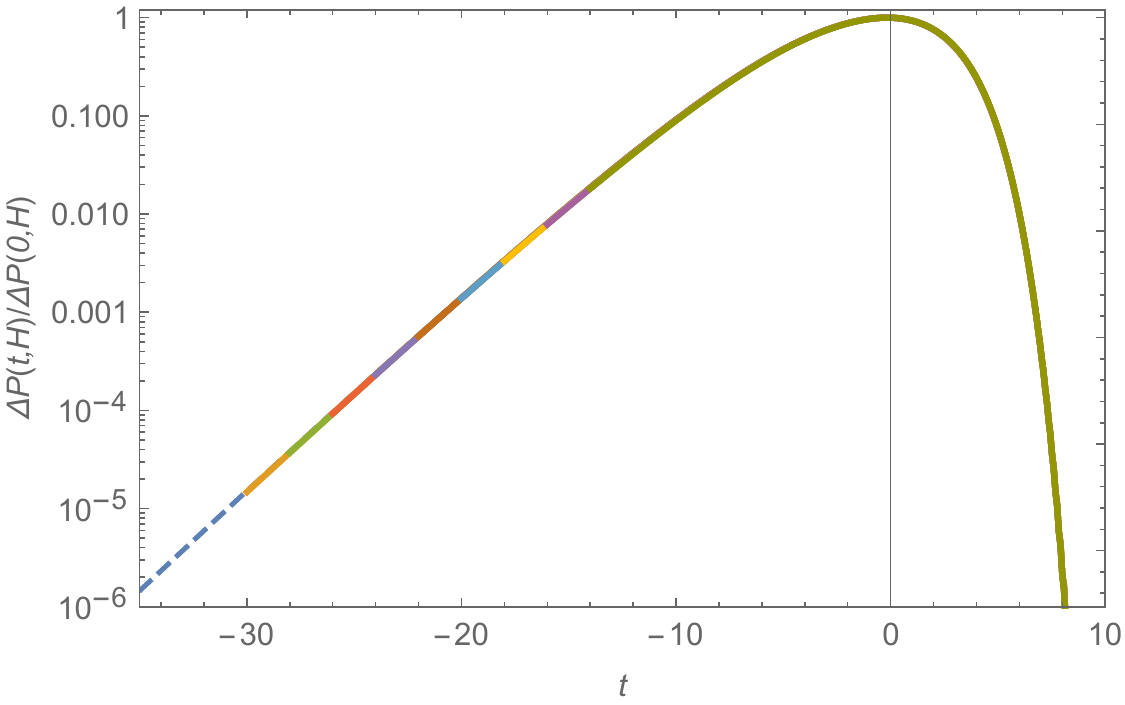}
    \caption{$\Delta P(t,H)/\Delta P(0,H)$ vs. $t$ for $-t_{\rm min}=30,28,26,24,22,20,18,16,14$ at the critical point $J=J_c$ ($k=2$). The rescaled curves collapse onto the rescaled scaling function $\ft(t)/\ft(0)$ (dashed).}
    \label{fig:DpDP0crit}
\end{figure}

The critical point is characterized by $\beta=1/2$, i.e.\ $2\beta-1=0$: the IPR converges to its $H=0$ value more slowly than any power law. Figure~\ref{fig:DpDP0crit} shows the quantity $\Delta P(t,H) \equiv P_H(t) - P_0(t)$ as obtained from the numerical solutions of Eq.~\eqref{eq:PtH} at $J = J_c$ for a sequence of small values of $H$. All curves collapse onto a single master curve when divided by $\Delta P(0,H)$, so that the factorized form~\eqref{eq:ansatz} still holds, with $\ft(t)$ obeying Eq.~\eqref{eq:lindp} at $J = J_c$. The asymptotic behavior of the master curve for $t \ll 0$, however, is now
\begin{equation}
    \ft(t) \sim -t\, e^{t/2} \, , \qquad t \to -\infty \, .
\end{equation}
Indeed, while away from criticality the linearized equation~\eqref{eq:linsmallt} admits the two solutions $e^{\beta t}$ and $e^{(1-\beta)t}$, at the critical point, where the two roots coalesce, it admits the two solutions $e^{t/2}$ and $-t\, e^{t/2}$.
This can be seen by plugging $e^{\beta t}$ into Eq.~\eqref{eq:linsmallt} and differentiating with respect to $\beta$: the term coming from the kernel $L_0(t-t')$ is proportional to $\de J/\de\beta$ that vanishes at the critical point, therefore the derivative of $e^{\beta t}$ with respect to $\beta$ is a solution as well at $J=J_c$.
It follows that for $t_{\rm min} \ll t \ll 0$ the fixed-point solution reads
\begin{equation}
\begin{aligned}
   P_H(t) &= P_0(t) +  \at(H) \,  \ft(t) \\
   &= 1 - c \, e^{t/2} - \at(H)\,  t\, e^{t/2} \, ,
\end{aligned}
\label{eq:sc1}
\end{equation}
where $c$ is obtained from the solution of Eq.~\eqref{eq:Pt0} at $J=J_c$.

\begin{figure*}
    \centering
    \panel{a}{\includegraphics[width=0.325\linewidth]{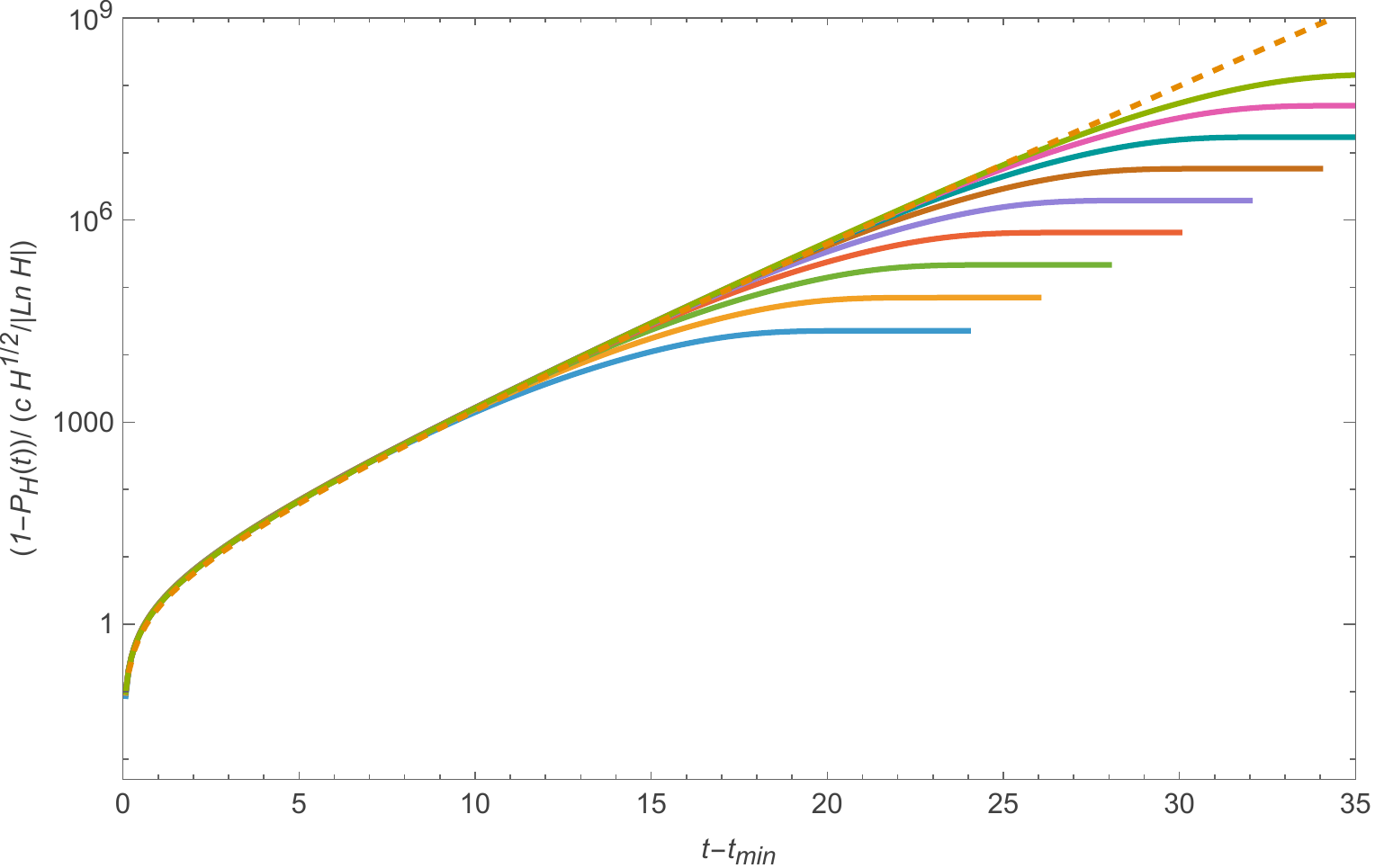}}\hfill
    \panel{b}{\includegraphics[width=0.325\linewidth]{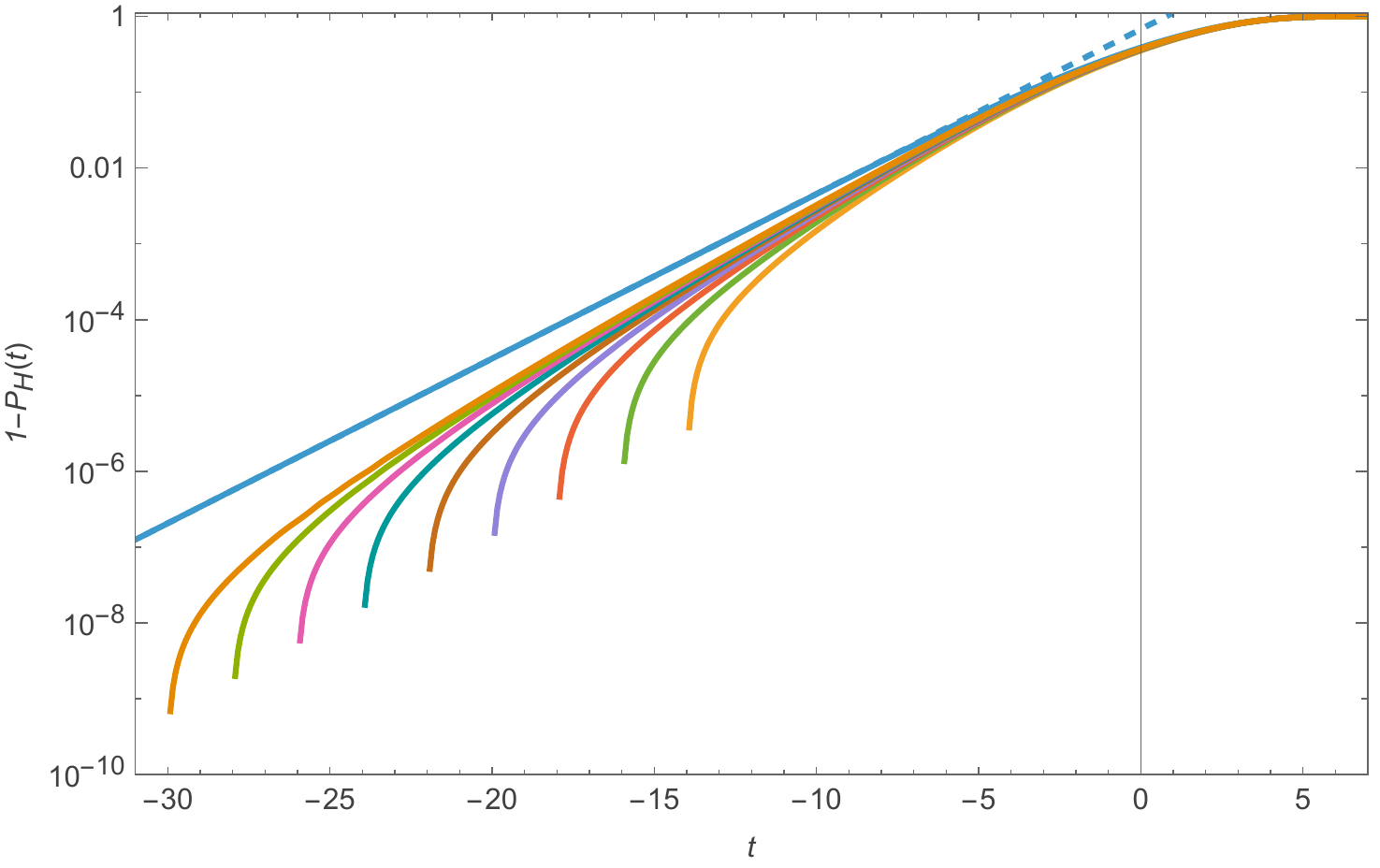}}\hfill
    \panel{c}{\includegraphics[width=0.325\linewidth]{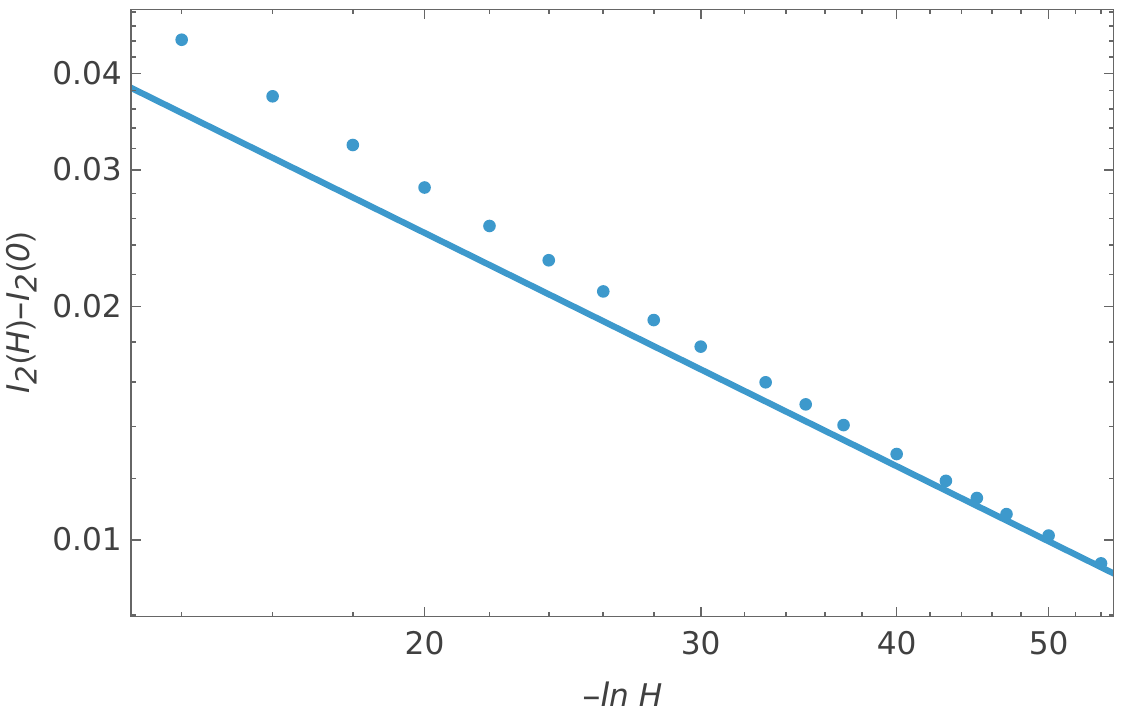}}
    \caption{Critical point $J = J_c$, $k=2$. (a) $[1-P_H(t)]/\am(H)$, with $\am=c \, H^{1/2}/ |\ln H|$ [Eq.~\eqref{eq:amin_crit}], vs. $\Delta t$ for $-t_{\rm min}=30,28,\ldots,14$. The curves collapse onto the scaling function $\fm(\Delta t)$ (dashed line), solution of Eq.~\eqref{eq:linfmin} at $J=J_c$. (b) $1-P_H(t)$ vs $t$ for the same values of $t_{\rm min}$. The continuous blue line is the $H=0$ solution of Eq.~\eqref{eq:Pt0} at $J_c$, which vanishes as $c\, e^{t/2}$ (dashed line). (c) $I_2(H) - I_2(0)$ vs. $- \ln H=-t_{\rm min}$. The straight line is the asymptotic parameter-free prediction~\eqref{eq:I2scalcrit}, obtained from the numerical solution of Eq.~\eqref{eq:lindp} at $J_c$, for $-t_{\rm min}$ up to 54 ($H=3.5\, \times \, 10^{-24}$). }
    \label{fig:fmincrit}
\end{figure*}

As in the non-critical case, $\at(H)$ follows from matching with the boundary layer, where the ansatz~\eqref{eq:sc2} still holds with $\fm$ obeying Eq.~\eqref{eq:linfmin} at $J_c$, $\fm(0)=0$, and $\fm(\Delta t)$ solving the shift-invariant equation~\eqref{eq:linsmallt} for $\Delta t \gg 1$. Requiring Eq.~\eqref{eq:sc1} for $t \ll 0$ to match Eq.~\eqref{eq:sc2} for $1 \ll \Delta t \ll |t_{\rm min}|$ gives
\begin{equation} \label{eq:matching_critical}
\begin{aligned}
    \am(H)\, \fm (\Delta t) &=  c \, e^{t/2} + \at(H)\,  t\, e^{t/2} \\
    &= e^{t_{\rm min}/2} e^{\Delta t/2} \big[ c + \at(H) \, t_{\rm min} \\
    & \qquad \qquad \qquad + \at(H)\,  \Delta t  \big] \, .
\end{aligned}
\end{equation}
For the right-hand side to be compatible with $\fm(0) = 0$, the constant term in the square bracket must vanish, $c+ \at(H) t_{\rm min}=0$,\footnote{Strictly speaking, the condition $\fm(0)=0$ does not fix the constant term in the large-$\Delta t$ asymptotics of $\fm$: generically $\fm(\Delta t) \simeq (\Delta t + A_c)\, e^{\Delta t/2}$, with a constant $A_c$ determined by the full Eq.~\eqref{eq:linfmin} at $J_c$, as $c_3$ is in the non-critical case. The matching then gives $\at(H) = c/(|\ln H| + A_c)$ and $\am(H) = c\,H^{1/2}/(|\ln H| + A_c)$. The leading $1/|\ln H|$ behavior, and hence Eqs.~\eqref{eq:atyp_crit} and~\eqref{eq:amin_crit} to leading order, are unchanged, while $A_c$ produces corrections of relative order $1/|\ln H|$, which contribute to the slow approach to the asymptotic law visible in Fig.~\ref{fig:fmincrit}(c).} i.e.
\begin{equation} \label{eq:atyp_crit}
    \at(H) = \frac{c}{|t_{\rm min}|} = \frac{c}{|\ln H|} \, .
\end{equation}
With this value, the two terms of Eq.~\eqref{eq:sc1} are of the same order in the region $t = O(t_{\rm min})$ and cancel at $t = t_{\rm min}$, consistently with $P_H(t_{\rm min}) = 1$ up to terms decaying faster than $e^{t_{\rm min}/2}$. Inserting Eq.~\eqref{eq:atyp_crit} back into Eq.~\eqref{eq:matching_critical} one finds $\am(H)\, \fm (\Delta t) = (c/|t_{\rm min}|) \, e^{t_{\rm min}/2} \, \Delta t \, e^{\Delta t/2}$, which is satisfied provided that $\fm (\Delta t) \sim \Delta t \, e^{\Delta t/2}$ for $\Delta t \to \infty$ and
\begin{equation} \label{eq:amin_crit}
    \am(H)= c \, \frac{H^{1/2}}{|\ln H|} \, .
\end{equation}

The scaling regimes at criticality are therefore: in the bulk, $t=O(1)$,
\begin{equation}
    P_H(t)= P_0(t)+ \frac{c}{|\ln H|} \ft(t) \, ,
\label{eq:1c}
\end{equation}
with $\ft(t)\sim -t\, e^{t/2}$ for $t \to -\infty$; in the intermediate region $(t-t_{\rm min})/|t_{\rm min}|=O(1)$,
\begin{equation}
   P_H(t) = 1 - c \, e^{t/2} \left( 1-\frac{t}{t_{\rm min}} \right) \, ;
\label{eq:2c}
\end{equation}
and in the boundary layer $\Delta t=O(1)$,
\begin{equation}
    P_H(t) = 1 -  c \, \frac{H^{1/2}}{|\ln H|}\, \fm (\Delta t) \, .
\label{eq:3c}
\end{equation}
Since the IPR proxy~\eqref{eq:I2toy} is dominated by the bulk region, the correction to the IPR at the critical point vanishes logarithmically,
\begin{equation} \label{eq:logcorr}
    I_2(H) -I_2^{\rm crit} (0) \propto \frac{1}{| \ln H |} \, ,
\end{equation}
consistent with the vanishing of $2\beta - 1$ as $J \to J_c$ and with the logarithmically slow finite-size and finite-$\eta$ corrections to the IPR observed numerically at the Anderson transition on the BL~\cite{tikhonov2019statistics}. 
More precisely, we have
\begin{equation}
\begin{aligned}
   I_2(H)-I_2^{\rm crit}(0) \approx {} & \frac{c}{|\ln H|} \\
   & \times (k+1) \int_{-\infty}^{+\infty} e^{-e^t} \, e^t \, P_0^k(t) \,\ft(t)\, \de t \, .
\end{aligned}
\label{eq:I2scalcrit}
\end{equation}
From the scaling function $\ft(t)$ and $P_0(t)$ computed numerically at $J=J_c$, $k=2$, we have
\begin{equation}
    (k+1)\int_{-\infty}^{+\infty} \, e^{-e^t} \, e^t \, P_0^k(t) \,\ft(t)\, \de t=0.7369\ ,
\end{equation}
while $c = 0.67573$.

We note that in Ref.~\cite{tikhonov2019statistics} the critical correction was fitted, from the solution of the self-consistency equations at imaginary frequency, to an effective form $(\ln 1/\eta)^{-\zeta}$ (the exponent $\zeta$ is denoted $\mu$ there and in Refs.~\cite{chen2024describing,garcia2026connecting}; it should not be confused with the exponent $\mu = 2(1-\beta)$ of the eigenstate correlations of Sec.~\ref{sec:realspace}), with $\zeta \simeq 1/2$ for the Anderson model and $\zeta \simeq 3/2$ for the present toy model, whereas our analysis predicts a pure logarithm, $\zeta = 1$. The difference is most likely a preasymptotic effect. Equation~\eqref{eq:logcorr} is the leading term of an expansion in powers of $1/|\ln H|$, and the corrections to it decay only logarithmically as well: as Fig.~\ref{fig:fmincrit}(c) shows, the $1/|\ln H|$ law is approached slowly, and one must reach $|\ln H| \gtrsim 50$, i.e.\ $H \lesssim 10^{-22}$, before the asymptotic behavior sets in. Over the ranges accessible in practice ($\ln N \lesssim 12$--$14$ for the largest exact diagonalizations, $|\ln\eta| \lesssim 20$--$30$ in population dynamics), the correction is still far from its asymptotic form, and a power-law fit in $\ln N$ or $\ln(1/\eta)$ necessarily returns an effective exponent $\zeta \neq 1$. Since a finite graph acts on the bulk of the distribution as a regulator $\eta_{\rm eff} \sim 1/N$ (Sec.~\ref{sec:Tu}), we correspondingly expect a logarithmically slow finite-size correction at the critical point, $\propto 1/\ln N$ asymptotically. In the toy model the asymptotic regime can be reached because the cavity equation can be solved at arbitrarily small $H$; this however requires a dedicated numerical treatment of the region $t \le 0$, where the deficit $1 - P_H(t) \sim e^{t/2}$ becomes exponentially small as $H \to 0$ (Algorithm~5 of Appendix~\ref{app:numerics}). 

Figure~\ref{fig:fmincrit} tests these predictions. Panel (a) demonstrates the scaling~\eqref{eq:3c} in the boundary layer; panel (b) shows that, in agreement with Eq.~\eqref{eq:2c}, the deviations from the $H=0$ solution in the region $t=O(t_{\rm min})$ are much larger than in the non-critical case [cf.\ Fig.~\ref{fig:1mpt}(a)]; panel (c) displays the slow, logarithmic convergence of the IPR, Eq.~\eqref{eq:I2scalcrit}. Note that at the critical point corrections to the leading small-$H$ behavior are expected to be large and indeed the asymptotic regime is reached for exceedingly small values of $H=O(10^{-24})$ corresponding to $-\ln H \approx 54$, compared to the non-critical case where the asymptotic regime is reached already at $H=O(10^{-3})$ (see Fig.~\ref{fig:DI2vsH}).
Computing numerically $P_H(t)$ at such small values of $H$ is challenging; the algorithm we developed to do so is Algorithm~5 of Appendix~\ref{app:numerics}.

\subsection{Crossover close to the critical point} \label{sec:nearcritical}

It is instructive to see how the power-law regime of Sec.~\ref{sec:scaling} crosses over to the logarithmic one as $J \to J_c$. At fixed small $H$, there is always a value of $J$ close enough to $J_c$ that $\delta \beta \equiv \beta-1/2$ is so small that the factor $e^{-2 \, \delta \beta \, (t-t_{\rm min})}$ in Eq.~\eqref{eq:2nonc} is $O(1)$ for $t=O(t_{\rm min})$; when this happens the physics must be the same as at criticality, and Eq.~\eqref{eq:1nonc} must match Eq.~\eqref{eq:1c}, and Eq.~\eqref{eq:3nonc} must match Eq.~\eqref{eq:3c}. For small $|J-J_c|$ one can argue that $\ft^{J}(t) \propto \ft^{J_c}(t)$ up to $|t| = O(\delta \beta^{-1})$, where the two asymptotic behaviors $e^{(1-\beta)t}$ and $-t e^{t/2}$ depart from each other; given the normalization~\eqref{eq:ftypasympt}, this implies $\ft^{J}(t) = O(\delta \beta)\, \ft^{J_c}(t)$ at any finite $t$. The bulk shift~\eqref{eq:1nonc} is then $H^{2 \delta \beta} \, O(\delta \beta)$, which for $|\ln H| = O(\delta \beta^{-1})$ is $O(\delta\beta) = O(1/|\ln H|)$, the same order as in Eq.~\eqref{eq:1c}. The crossover between the two regimes thus takes place at $|\ln H| \sim 1/(\beta - 1/2) \sim (J_c - J)^{-1/2}$ [$(W - W_c)^{-1/2}$ in the Anderson model]. The same reasoning applied to Eqs.~\eqref{eq:3nonc} and~\eqref{eq:3c}, i.e.\ $\fm^{J}(\Delta t) = O(\delta \beta)\, \fm^{J_c}(\Delta t)$ for $\Delta t \ll \delta \beta^{-1}$, requires the leading and subleading terms of Eq.~\eqref{eq:fminsub} to cancel at leading order, which implies that $c_3 \to 1$ as $J \to J_c$. These statements can be made uniform across the crossover. Like $\fm$, the bulk zero mode contains the second root as a subleading term, $\ft^{J}(t) \simeq e^{(1-\beta)t} - b_3\, e^{\beta t}$ for $t \ll -1$, and $\ft^{J}(t) = O(\delta\beta)\, \ft^{J_c}(t)$ requires $b_3 \to 1$ as $J \to J_c$, so that $\ft^{J}(t) \simeq 2\,\delta\beta\, (-t)\, e^{t/2}$ for $1 \ll |t| \ll \delta\beta^{-1}$. Keeping both roots in the bulk and in the boundary layer, the matching of the coefficients of $e^{\beta t}$ and $e^{(1-\beta)t}$ in the overlap region $t_{\rm min} \ll t \ll -1$ gives $\am(H)\, e^{-\beta t_{\rm min}} = c + b_3\, \at(H)$ and $\at(H) = c_3\, \am(H)\, e^{-(1-\beta) t_{\rm min}}$, i.e.
\begin{equation} \label{eq:atypuniform}
    \at(H) = \frac{c \, c_3 \, H^{2\beta-1}}{1 - b_3 \, c_3 \, H^{2\beta-1}} \, .
\end{equation}
For $|\ln H| \gg \delta\beta^{-1}$ this reduces to Eq.~\eqref{eq:atyp}, while for $1 \ll |\ln H| \ll \delta\beta^{-1}$, with $b_3, c_3 \to 1$ and $1 - H^{2\delta\beta} \simeq 2\,\delta\beta\, |\ln H|$, it gives $\at(H)\, \ft^{J}(t) \simeq (c/|\ln H|)\, \ft^{J_c}(t)$, i.e.\ the critical bulk correction~\eqref{eq:1c}.

We can thus obtain a scaling form describing $I_2(H)-I_2(0)$ in the critical region, where $\kappa = 2\beta - 1$ is small:
\begin{equation} \label{eq:I2crossover}
\begin{aligned}
    I_2(H)-I_2(0) & \simeq \frac{C}{|\ln H|}\, f\big(\kappa\, |\ln H|\big) \, , \\
    f(x) & = \frac{x\, e^{-x}}{1-e^{-x}} = \frac{x}{e^{x}-1} \, ,
\end{aligned}
\end{equation}
where
\begin{equation} \label{eq:Ccrit}
    C = c\,(k+1) \int_{-\infty}^{+\infty} e^{-e^t}\, e^t\, P_0^k(t)\, \ft^{J_c}(t)\, \de t \, ,
\end{equation}
i.e.\ $C = 0.7369\, c \approx 0.50$ for $k = 2$. Since $f(0) = 1$, Eq.~\eqref{eq:I2crossover} gives back the critical result~\eqref{eq:I2scalcrit} for $\kappa |\ln H| \ll 1$, while for $\kappa |\ln H| \gg 1$, where $f(x) \simeq x\, e^{-x}$, it gives
\begin{equation} \label{eq:I2kappasmall}
    I_2(H)-I_2(0) \simeq C\, \kappa\, H^{\kappa} \, .
\end{equation}
This also follows directly from Eq.~\eqref{eq:I2scal}, using $c_3 \simeq 1$ and $\ft^{J}(t) \simeq \kappa\, \ft^{J_c}(t)$, valid for $\kappa \to 0$: the prefactor of the $H^\kappa$ correction vanishes linearly in $\kappa$ at the critical point. The crossover between Eqs.~\eqref{eq:I2scalcrit} and~\eqref{eq:I2kappasmall} takes place at $\kappa |\ln H| \sim 1$. Equation~\eqref{eq:I2crossover} is the finite-size-scaling form of the IPR in the toy model; its counterpart for the Anderson model is stated and tested in Sec.~\ref{sec:FSS}.

The structure at play --- a linear recursion whose two roots merge at the transition, cut off by a boundary at $t_{\rm min}$ --- is that of traveling waves on trees~\cite{derrida1988polymers}, which also describe the Anderson transition on the Cayley tree~\cite{monthus2009anderson}, and in which a cutoff likewise produces slow, logarithmic corrections~\cite{brunet1997shift}. Translated to the Anderson model, these considerations imply that the power law $\eta^{2\beta - 1}$ is only visible for $|\ln \eta| \gg (W-W_c)^{-1/2}$, and that at fixed small $\eta$ the corrections cross over to the logarithmic critical behavior in a window $W - W_c \lesssim 1/\ln^2(1/\eta)$ around the transition.

Finally, the continuation to complex $H$ of Sec.~\ref{sec:complexH} goes through at $J = J_c$ as well, since none of its steps requires $\beta \neq 1/2$; we have not carried it out numerically, as it would require Algorithm~5 of Appendix~\ref{app:numerics} at complex $H$ and $|\ln H| \gtrsim 50$. Continuing Eq.~\eqref{eq:I2scalcrit}, $C/|\ln H| \to C/\ln(1/H)$, with $\ln(1/H) = \ln(2/|\omega|) + {\rm i}(\pi/2)\,{\rm sgn}\,\omega$ at $H = 0^+ - {\rm i}\omega/2$, we expect at the critical point
\begin{equation} \label{eq:Scrit}
    \mathcal{S}_{\rm reg}(\omega) \simeq \frac{C}{2|\omega| \left[ \ln^2(2/|\omega|) + \pi^2/4 \right]} \simeq \frac{C}{2|\omega| \ln^2(1/|\omega|)} \, ,
\end{equation}
with $C \approx 0.50$ given by Eq.~\eqref{eq:Ccrit} and relative corrections of order $1/\ln(1/|\omega|)$: this is $\mu = 1$, with a logarithmic factor that keeps $\mathcal{S}_{\rm reg}$ integrable [its weight at $|\omega| < \eta$, $C/\ln(1/\eta)$, is the correction~\eqref{eq:I2scalcrit}], i.e.\ the toy-model counterpart of the form $N^2\mathcal{C}(\omega) \propto 1/[\omega \ln^2(1/\omega)]$ anticipated in Sec.~\ref{sec:proliferation}. The crossover form~\eqref{eq:I2crossover}, which reads $I_2(H) - I_2(0) \simeq C\kappa H^\kappa/(1 - H^\kappa)$, continues in the same way, and interpolates between Eq.~\eqref{eq:Scrit} for $\kappa \ln(1/|\omega|) \ll 1$ and Eq.~\eqref{eq:Stoy} with $c_H = C\kappa$, whose amplitude $\propto \kappa \sin(\pi\kappa/2)$ vanishes as $\kappa^2$, for $\kappa \ln(1/|\omega|) \gg 1$.

\section{The Anderson model: the same mechanism under a few assumptions} \label{sec:anderson}

We now show that the mechanism identified in the toy model --- a zero mode of the linearized $\eta=0$ recursion, decaying with the subdominant root, whose amplitude is fixed by matching to a boundary layer --- is also at work in the cavity equations of the Anderson model, and reproduces $\kappa = 2 \beta - 1$. The derivation requires a few explicit assumptions, which we state as we go; they are the natural counterparts of properties that are established in the toy model, and we test their consequences numerically in Sec.~\ref{sec:numerics}.

\subsection{Zero mode of the linearized cavity equation} \label{sec:zeromode}

Writing $G_{i \to j} = g_{i \to j} + {\rm i}\eta \hat g_{i \to j}$, the exact finite-$\eta$ cavity equations~\eqref{eq:cavity_a} read
\begin{subequations} \label{eq:gghateta}
\begin{align}
    g_{i \to j} & = \frac{a_i}{a_i^2 + \eta^2 b_i^2} \, , \qquad
    \hat{g}_{i \to j} = \frac{b_i}{a_i^2 + \eta^2 b_i^2} \, , \\
    a_i & \equiv \epsilon_i - \!\!\! \sum_{m \in \partial i \setminus j} \!\!\! g_{m \to i} \, , \qquad
    b_i \equiv 1 + \!\!\! \sum_{m \in \partial i \setminus j} \!\!\! \hat g_{m \to i} \, , \label{eq:ab_def}
\end{align}
\end{subequations}
and the corresponding exact distributional equation for the joint distribution $P_\eta(g,\hat g)$ of the cavity Green's functions is
\begin{widetext}
\begin{equation} \label{eq:Peta}
    P_\eta(g,\hat{g}) = \int \de \epsilon \, p(\epsilon) \left[ \prod_{i=1}^k  \de g_i \, \de \hat{g}_i P_\eta(g_i,\hat{g}_i) \right]
    \delta \left(g - \frac{a}{a^2 + \eta^2 b^2}\right)
    \delta \left(\hat{g} - \frac{b}{a^2 + \eta^2 b^2}  \right)\, ,
    \qquad a = \epsilon - \sum_{i=1}^k g_i \, , \quad b = 1 + \sum_{i=1}^k \hat g_i \, .
\end{equation}
The regulator enters Eq.~\eqref{eq:Peta} only through $\eta^2$. Expanding around the $\eta = 0$ solution $P_0$, the correction $\delta P_\eta \equiv P_\eta - P_0$ obeys
\begin{equation} \label{eq:deltaP}
\begin{aligned}
    \delta P_\eta(g,\hat{g}) = k \int \de \epsilon \, p(\epsilon) & \left[ \prod_{i=2}^k  \de g_i \, \de \hat{g}_i P_0(g_i,\hat{g}_i) \right] \de g_1 \, \de \hat{g}_1 \, \delta P_\eta(g_1,\hat{g}_1) \\
    & \times \delta \left(g - \frac{1}{a}\right) \delta \left(\hat{g} - g^2 b \right) + O(\eta^2)
    \equiv \left( \mathcal{L}_0 \, \delta P_\eta \right)(g,\hat g) + O(\eta^2) \, ,
\end{aligned}
\end{equation}
\end{widetext}
where $\mathcal{L}_0$ is the $\eta = 0$ recursion linearized around its fixed point. If $1 - \mathcal{L}_0$ were invertible, Eq.~\eqref{eq:deltaP} would imply $\delta P_\eta = O(\eta^2)$, in blatant contradiction with the numerical results of Sec.~\ref{sec:evidence}. The resolution is the same as in the toy model, Eq.~\eqref{eq:lindp}: $\mathcal{L}_0$ has a zero mode, i.e.\ a non-trivial solution $\ft(g,\hat g)$ of
\begin{equation} \label{eq:ftypAL}
\begin{aligned}
    \ft(g,\hat{g}) = k \int \de \epsilon \, p(\epsilon) & \left[ \prod_{i=2}^k  \de g_i \, \de \hat{g}_i P_0(g_i,\hat{g}_i) \right] \\
    & \times \de g_1 \, \de \hat{g}_1 \, \ft(g_1,\hat{g}_1) \\
    & \times \delta \left(g - \frac{1}{a}\right) \delta \left(\hat{g} - g^2 b \right) \, ,
\end{aligned}
\end{equation}
and the leading correction is
\begin{equation} \label{eq:deltaPzm}
    \delta P_\eta(g,\hat{g}) = c_{\rm typ}(\eta)\, \ft(g,\hat{g}) + O(\eta^2) \, ,
\end{equation}
with an amplitude $c_{\rm typ}(\eta)$ that is not determined by Eq.~\eqref{eq:deltaP} and must be fixed, as in the toy model, by matching to the region $\hat g \sim 1/\eta$ where the expansion in $\eta^2$ breaks down. Note that integrating Eq.~\eqref{eq:ftypAL} over $(g,\hat g)$ gives $(1-k) \int \ft = 0$: the zero mode has zero integral, as required by the normalization of $P_\eta$. More precisely, integrating Eq.~\eqref{eq:ftypAL} over $\hat g$ only shows that the $g$-marginal of $\ft$ obeys the linearization of the closed recursion for the real parts; assuming that the latter has no zero mode, as is natural since $P_0(g)$ is a regular and stable fixed point, this marginal vanishes, so that the density of states is not affected by the zero mode. This supports the assumption, made in Sec.~\ref{sec:linearized} to derive Eq.~\eqref{eq:I2deltaQ}, that no singular correction arises in the average density of states.

The existence of the zero mode is our first assumption; it is supported by the following considerations. The large-$\hat g$ structure of Eq.~\eqref{eq:ftypAL} is the same as that determining the tail exponent of $P_0$: looking for a solution of the form $\ft(g,\hat g) \simeq f_\gamma(g) \hat g^{-1-\gamma}$ at large $\hat g$, and using the fact that a large $\hat g$ is produced by a large $\hat g_1$ amplified by $g^2$, one finds that $f_\gamma$ must satisfy Eq.~\eqref{eq:fbeta} with $\beta \to \gamma$, i.e.\ $\lambda(\gamma) = 1$. Thanks to the symmetry $\lambda(\gamma) = \lambda(1-\gamma)$~\cite{anderson1973selfconsistent,abou1974self,mirlin1991localization,kravtsov2018non,parisi2019anderson}, this equation has the two solutions $\gamma = \beta$ and $\gamma = 1 - \beta$, so that, exactly as in the toy model, the linearized recursion admits two admissible power-law behaviors at large $\hat g$, while regularity at small $\hat g$ imposes one condition: generically, a one-dimensional space of solutions survives. [In the single-neighbor approximation used to extract the tail, we have assumed that a large $\hat g$ comes from a large $\hat g_1$ in the factor $\delta P_\eta(g_1,\hat g_1)$ and not from one of the $P_0(g_i,\hat g_i)$ factors; this is consistent, because $P_0 \propto \hat g^{-1-\beta} \ll \ft \propto \hat g^{-(2-\beta)}$ at large $\hat g$ as soon as $\beta > 1/2$.] As in the toy model, the requirement that the correction vanish as $\eta \to 0$ selects the second root: the regulator removes a weight $\propto \eta^{\beta}$ at $\hat m \sim 1/\eta$, which, propagated into the bulk along the $\hat g^{-1-\beta}$ mode, would produce an $O(1)$ change of the fixed point (Sec.~\ref{sec:ftyp}). We therefore assume that the zero mode has a non-vanishing component along the second root, so that, since $2 - \beta < 1 + \beta$, it decays at large $\hat g$ as
\begin{equation} \label{eq:ftypALtail}
    \ft(g,\hat{g}) \simeq f_{1-\beta}(g) \, \hat{g}^{-(2-\beta)} \, , \qquad \hat g \to \infty \, ,
\end{equation}
where $f_{1-\beta}$ is the eigenfunction of Eq.~\eqref{eq:fbeta} at the second root. This is our second assumption, and it is the one with the most directly testable consequences: it predicts that the correction to the distribution of $\hat m$ at $m=0$ decays, in the window $1 \ll \hat m \ll 1/\eta$, as $\hat m^{-(2-\beta)}$, i.e.\ {\it more slowly} than the distribution itself.

\subsection{Boundary layer and matching} \label{sec:boundaryAL}

For $\hat g = O(1/\eta)$ the term $\eta^2 b^2$ in the denominators of Eq.~\eqref{eq:gghateta} can no longer be neglected and the expansion~\eqref{eq:deltaP} breaks down. On these scales, however, $P_\eta(g,\hat g)$ is small, of order $\eta^{1+\beta}$ as we shall see, and Eq.~\eqref{eq:Peta} can again be linearized, now in $P_\eta$ itself: a large $b$ comes from a single large $\hat g_1$ (third assumption, the counterpart of the single-neighbor approximation above), so that $b \simeq \hat g_1$ and the other $k-1$ neighbors only contribute to $a$ through their real parts. In terms of the rescaled variable $\tilde g = \eta \hat g$, the solution takes the scaling form
\begin{equation} \label{eq:flargeansatz}
    P_{\eta}(g,\hat{g}) \simeq c_{\rm large}(\eta) \, f_{\rm large}(g, \eta \hat{g}) \, ,
\end{equation}
where $f_{\rm large}(g,\tilde g)$ obeys the $\eta$-independent linear equation
\begin{widetext}
\begin{equation} \label{eq:flarge}
        f_{\rm large}(g,\tilde{g}) = k \int \de \epsilon \, p(\epsilon) \left[ \prod_{i=2}^k  \de g_i \,  P_0(g_i) \right] \de g_1 \, \de \tilde g_1 \,
        \delta \left(g - \frac{a}{a^2 + \tilde{g}_1^2 }\right)
        \delta \left(\tilde{g} - \frac{\tilde{g}_1}{a^2 + \tilde{g}_1^2 }\right) f_{\rm large}(g_1,\tilde{g}_1) \, ,
        \qquad a = \epsilon - \sum_{i=1}^k g_i \, .
\end{equation}
\end{widetext}
This is the counterpart of the boundary-layer equation~\eqref{eq:linfmin} of the toy model, and $\tilde g = \eta \hat g$ plays the role of $e^{-\Delta t}$: as there, the rescaling by $\eta$ makes the kernel independent of the regulator. At small $\tilde g$ (i.e.\ $\hat g \ll 1/\eta$) the term $\tilde g_1^2$ in the denominators can be neglected, Eq.~\eqref{eq:flarge} reduces to the $\eta = 0$ tail equation, and $f_{\rm large}$ is a combination of the two admissible power laws,
\begin{equation} \label{eq:flargesmall}
\begin{aligned}
      f_{\rm large}(g,\tilde{g}) = f_\beta(g) \, \tilde{g}^{-1-\beta} & + c_{\rm sub} \, f_{1-\beta}(g) \, \tilde{g}^{-(2-\beta)} + \ldots \, , \\
      & \tilde g \ll 1 \, ,
\end{aligned}
\end{equation}
where the first term fixes the normalization of $f_{\rm large}$ and the coefficient $c_{\rm sub}$ of the subleading term is determined by the solution of the full Eq.~\eqref{eq:flarge}, exactly as $c_3$ in Eq.~\eqref{eq:fminsub}.

The matching proceeds in two steps, in the overlap region $1 \ll \hat g \ll 1/\eta$ where both descriptions hold. The leading term of Eq.~\eqref{eq:flargesmall} must match the unperturbed tail $P_0(g,\hat g) \simeq f_\beta(g) \hat g^{-1-\beta}$,
\begin{equation}
\begin{aligned}
    & f_\beta(g) \, \hat{g}^{-1-\beta} = c_{\rm large}(\eta) \, f_\beta(g) \, (\eta \hat g)^{-1-\beta} \\
    & \qquad \Longrightarrow \quad c_{\rm large}(\eta) = \eta^{1+\beta} \, ,
\end{aligned}
\end{equation}
the counterpart of $\am(H) = c H^\beta$ in Eq.~\eqref{eq:amin} (the extra power of $\eta$ comes from the fact that $P_\eta$ is a density in $\hat g$ while $P(t)$ is not). The subleading term must match the zero-mode correction~\eqref{eq:deltaPzm} with the tail~\eqref{eq:ftypALtail},
\begin{equation}
    c_{\rm typ}(\eta) \, f_{1-\beta}(g) \, \hat g^{-(2-\beta)} = c_{\rm sub} \, c_{\rm large}(\eta) \, f_{1-\beta}(g) \, (\eta \hat g)^{-(2-\beta)} ,
\end{equation}
which gives the final result
\begin{equation} \label{eq:ctyp}
    c_{\rm typ}(\eta) = c_{\rm sub} \, \eta^{1+\beta} \eta^{-(2-\beta)} = c_{\rm sub} \, \eta^{2\beta-1} \, ,
\end{equation}
the exact counterpart of Eq.~\eqref{eq:atyp}. Since the correction to $Q_\eta(0,\hat m)$, and through Eq.~\eqref{eq:I2deltaQ} the correction to the IPR, are linear functionals of $\delta P_\eta$ in the bulk region $\hat g = O(1)$, Eqs.~\eqref{eq:deltaPzm} and~\eqref{eq:ctyp} imply
\begin{equation} \label{eq:distscalingQ}
\begin{aligned}
    \overline{Q}_\eta(\hat{m}) - \overline{Q}_{0^+}(\hat{m}) &= \eta^{2\beta - 1} \, g_\eta(\hat{m}) \, , \\
    I_2(\eta) - I_2^\infty &= \eta^{2\beta-1} \int_0^\infty g_\eta(\hat m) \, \hat m^{-1} \, \de \hat m \, ,
\end{aligned}
\end{equation}
where $g_\eta(\hat m)$ is the image of the zero mode $\ft$ on the local distribution at $m=0$, with the same tail $g_\eta(\hat m) \propto \hat m^{-(2-\beta)}$ for $1 \ll \hat m \ll 1/\eta$. This is Eq.~\eqref{eq:kappa}. The structure of the argument is identical to that of the toy model: the regulator removes a weight $\eta^\beta$ from the tail of the distribution [the total weight of $P_0$ beyond $\hat g \sim 1/\eta$, cf.\ Eq.~\eqref{eq:Qmhattails}]; this perturbation is transmitted to the bulk along the subdominant root, which amplifies it by $\eta^{-(1-\beta)}$ between $\hat g \sim 1/\eta$ and $\hat g = O(1)$; and the product $\eta^{\beta - (1-\beta)} = \eta^{2\beta - 1}$ is the anomalous correction. What the toy model adds is a derivation in which each of these steps is made explicit and checked against the exact numerical solution of the cavity equation.

\subsection{The far tail $\hat m \gg 1/\eta$ and the exponent $\beta - 3$} \label{sec:fartail}

The matching argument only used the behavior of $f_{\rm large}(g,\tilde{g})$ at small $\tilde{g}$. We now show that its behavior at large $\tilde{g}$, i.e.\ the shape of $P_\eta(g,\hat g)$ and of $\overline{Q}_\eta(\hat m)$ for $\hat g, \hat m \gg 1/\eta$, is fixed by the $\eta=0$ solution alone, and that it produces a new power law, $\hat m^{\beta-3}$, which is clearly visible in the numerical data of Sec.~\ref{sec:numerics}. The key observation is that the cavity recursion~\eqref{eq:gghateta} is a complex inversion, $G = g + {\rm i}\eta\hat g = 1/(a - {\rm i}\eta b)$. Let $\mathcal{R}_\eta(a,b)$ denote the joint distribution of the variables $(a,b)$ of Eq.~\eqref{eq:ab_def} induced by $P_\eta$; it is the cavity counterpart of $Q_\eta(m,\hat m)$, with $k$ instead of $k+1$ neighbors, so that in the $\eta \to 0^+$ limit and at $a=0$ it has the same power-law tail as Eq.~\eqref{eq:Qmhattails},
\begin{equation} \label{eq:Rtail}
    \mathcal{R}_{0^+}(0,b) \simeq C_{\mathcal{R}} \, b^{-1-\beta} \, , \qquad b \gg 1 \, ,
\end{equation}
with $C_{\mathcal{R}} = k \int \de \epsilon \, p(\epsilon) [ \prod_{i=2}^k \de g_i \, P_0(g_i) ] \de g_1 \, f_\beta(g_1) \, \delta( \epsilon - \sum_{i=1}^k g_i )$, and it is a smooth function of $a$ around $a=0$, since $\epsilon$ is uniformly distributed. The map $z = a - {\rm i}\eta b \mapsto w = 1/z = g + {\rm i} \eta \hat{g}$ is conformal with Jacobian $|\de w / \de z|^2 = |w|^4$, so that densities transform as $p_w(w) = |w|^{-4} p_z(z)$. Since the density of $(a, \eta b)$ is $\eta^{-1} \mathcal{R}_\eta(a,b)$, and the inverse map is $a = g/(g^2 + \eta^2 \hat{g}^2)$, $b = \hat{g}/(g^2 + \eta^2 \hat{g}^2)$, one obtains the exact identity
\begin{equation} \label{eq:PfromR}
    P_\eta(g,\hat{g}) = \frac{1}{\left( g^2 + \eta^2 \hat{g}^2 \right)^2} \,
    \mathcal{R}_\eta \! \left( \frac{g}{g^2 + \eta^2 \hat{g}^2} \, , \, \frac{\hat{g}}{g^2 + \eta^2 \hat{g}^2} \right) \, ,
\end{equation}
valid for all $(g,\hat{g})$ and all $\eta$ [at $\eta = 0$ it reduces to $P_0(g,\hat{g}) = g^{-4} \mathcal{R}_0(1/g, \hat{g}/g^2)$, the integrated form of Eq.~\eqref{eq:linearized}]. Equation~\eqref{eq:PfromR} is not closed, since $\mathcal{R}_\eta$ is a functional of $P_\eta$, but it becomes predictive whenever the arguments of $\mathcal{R}_\eta$ fall in a region where $\mathcal{R}_\eta \simeq \mathcal{R}_{0^+}$ is known, namely $b \ll 1/\eta$.

Consider a fixed, non-resonant real part $g = O(1)$. Equation~\eqref{eq:PfromR} displays two regimes separated by $\eta \hat{g} \sim |g|$, i.e.\ by $\hat{g} \sim 1/\eta$. (i) For $\eta \hat{g} \ll |g|$, $a \simeq 1/g = O(1)$ and $b \simeq \hat{g}/g^2 \gg 1$: a large $\hat{g}$ is produced by a large $b$ amplified through a typical factor $g^2$. Inserting the tail~\eqref{eq:Rtail} one recovers the $\eta=0$ result, $P_\eta(g,\hat{g}) \simeq f_\beta(g) \, \hat{g}^{-1-\beta}$ with $f_\beta(g) = |g|^{2\beta - 2} r_\beta(1/g)$, where $r_\beta(a)$ is the coefficient of $b^{-1-\beta}$ in $\mathcal{R}_{0^+}(a,b)$ [$r_\beta(0) = C_{\mathcal{R}}$]. This is the propagation regime, which produces the leading small-$\tilde{g}$ behavior~\eqref{eq:flargesmall}. (ii) For $\eta \hat{g} \gg |g|$, the arguments of $\mathcal{R}_\eta$ become
\begin{equation} \label{eq:preimage}
    a \simeq \frac{g}{\eta^2 \hat{g}^2} \to 0 \, , \qquad b \simeq \frac{1}{\eta^2 \hat{g}} \, ,
\end{equation}
and the prefactor is $(\eta \hat{g})^{-4}$. In this regime the recursion acts as an inversion rather than as an amplification: a value $\hat{g} \gg 1/\eta$ requires the site to be essentially exactly at resonance, $|a| \lesssim |g| / (\eta \hat{g})^2$, so that the magnitude of $G$ is capped by the imaginary part of its denominator, $\eta \hat{g} \simeq 1/(\eta b)$; the value of $b$ needed to produce a given $\hat g$ is {\it smaller} the larger $\hat g$ is. The smoothness of $\mathcal{R}$ in $a$ makes the result independent of $g$ for $|g| \ll \eta \hat{g}$,
\begin{equation} \label{eq:fartail}
\begin{aligned}
    P_\eta(g,\hat{g}) &\simeq \frac{1}{\eta^4 \hat{g}^4} \, \mathcal{R}_{0^+} \! \left( 0 , \frac{1}{\eta^2 \hat{g}} \right) \, , \\
    & \qquad \hat{g} \gg \eta^{-1} \, , \quad |g| \ll \eta \hat{g} \, ,
\end{aligned}
\end{equation}
where replacing $\mathcal{R}_\eta$ by $\mathcal{R}_{0^+}$ is legitimate precisely because $b = 1/(\eta^2 \hat{g}) \ll 1/\eta$ as soon as $\hat{g} \gg 1/\eta$: the far tail of $P_\eta$ is fed by the part of the input distribution that has not yet been affected by the regulator. As long as $b \gg 1$, i.e.\ for $\eta^{-1} \ll \hat{g} \ll \eta^{-2}$, one can use the tail~\eqref{eq:Rtail}, and
\begin{equation} \label{eq:betaminus3}
\begin{aligned}
    P_\eta(g,\hat{g}) &\simeq \frac{C_{\mathcal{R}}}{\eta^4 \hat{g}^4} \left( \eta^2 \hat{g} \right)^{1+\beta}
    = C_{\mathcal{R}} \, \eta^{1+\beta} \left( \eta \hat{g} \right)^{\beta - 3} \, , \\
    & \qquad \qquad \eta^{-1} \ll \hat{g} \ll \eta^{-2} \, .
\end{aligned}
\end{equation}
The exponent $\beta - 3 = -4 + (1+\beta)$ is the sum of the Jacobian of the inversion and of the tail exponent of the input distribution evaluated at the inverted argument; more generally, a term $\propto b^{-1-\gamma}$ in $\mathcal{R}_{0^+}(0,b)$ is mapped onto a term $\propto \hat{g}^{\gamma - 3}$ in the far tail. In the language of Eq.~\eqref{eq:flargeansatz}, Eq.~\eqref{eq:betaminus3} reads
\begin{equation} \label{eq:flarge_large}
    f_{\rm large}(g,\tilde{g}) \simeq C_{\mathcal{R}} \, \tilde{g}^{\,\beta - 3} \, , \qquad \tilde{g} \gg 1 \, ,
\end{equation}
with the {\it same} amplitude $c_{\rm large}(\eta) = \eta^{1+\beta}$ found from the matching at small $\tilde{g}$, as it should: the two regimes are the two limits of the single crossover function $f_{\rm large}$, which is of order one at $\tilde{g} = O(1)$. Unlike the subleading coefficient $c_{\rm sub}$ of Eq.~\eqref{eq:flargesmall}, which requires solving the full boundary-layer equation because it is produced by the self-consistent region $b \sim 1/\eta$, the large-$\tilde g$ behavior is inherited from the $\eta=0$ solution through the inversion. It can also be read off directly from Eq.~\eqref{eq:flarge}: for $\tilde g \gg 1$ the two $\delta$-functions force $\tilde g_1 \simeq 1/\tilde g \ll 1$ and $a \simeq g/\tilde g^2 \to 0$, so that only the leading small-$\tilde g_1$ term of $f_{\rm large}$ itself contributes, with the Jacobian $\tilde g^{-4}$. Applied to the subleading term of Eq.~\eqref{eq:flargesmall}, the same reasoning produces a correction $\propto \tilde g^{(1-\beta)-3}= \tilde g^{-2-\beta}$ to Eq.~\eqref{eq:flarge_large}, subleading at large $\tilde g$ precisely because $\beta > 1/2$. Finally, Eq.~\eqref{eq:fartail} also describes the ultimate cutoff: when $\hat g$ approaches $\eta^{-2}$ the argument $b = 1/(\eta^2 \hat g)$ enters the bulk of $\mathcal{R}_{0^+}(0,b)$, the power law~\eqref{eq:betaminus3} bends over following the shape of $\mathcal{R}_{0^+}(0,b)$ at $b = O(1)$, and $P_\eta$ vanishes identically for $\hat g > \eta^{-2}$ since $b \geq 1$ (the bound $\im G \le 1/\eta$). Equation~\eqref{eq:fartail} thus provides a parameter-free prediction for the entire far tail of the finite-$\eta$ distribution in terms of the $\eta=0^+$ distribution $\mathcal{R}_{0^+}(0,b)$, which is directly accessible from the linearized population dynamics.

The translation to the local quantity $\overline{Q}_\eta(\hat m)$ is immediate. At $m=0$ the constraint $\epsilon = \sum_{i=1}^{k+1} g_i$, with $|\epsilon| \le W/2$, strongly suppresses large values of the $g_i$ (a large $g_i$ would have to be compensated by another one of opposite sign, at the price of a second factor from the $g^{-2}$ tail of $P_0(g)$), so that the last-step resonance mechanism is switched off and a value $\hat m \gg 1/\eta$ of $\hat m = 1 + \sum_{i=1}^{k+1} \hat g_i$ must come from a single neighbor with $\hat g_1 \simeq \hat m$ and $g_1 = O(1) \ll \eta \hat m$, exactly the regime (ii) in which Eq.~\eqref{eq:fartail} is independent of $g_1$. Using $\int \de \epsilon \, p(\epsilon) \, \de g_1 \, \delta(\epsilon - \sum_{i=1}^{k+1} g_i) = 1$ for any value of the remaining $g_i$'s, one finds
\begin{equation} \label{eq:Qfartail}
\begin{aligned}
    Q_\eta(0,\hat{m}) &\simeq \frac{k+1}{\eta^4 \hat{m}^4} \, \mathcal{R}_{0^+} \! \left( 0, \frac{1}{\eta^2 \hat{m}} \right) \\
    &\simeq (k+1) \, C_{\mathcal{R}} \, \eta^{1+\beta} \left( \eta \hat{m} \right)^{\beta-3} \, , \\
    & \qquad \qquad \eta^{-1} \ll \hat{m} \ll \eta^{-2} \, .
\end{aligned}
\end{equation}
The full structure of the distribution at $m=0$ is therefore
\begin{equation} \label{eq:Qregimes}
    \overline{Q}_\eta (\hat{m}) \propto
    \left\{ \begin{array}{@{}l@{\;\;}l@{}}
        \hat{m}^{-1 - \beta} \, [1 + O((\eta \hat m)^{2\beta - 1})] & 1 \ll \hat{m} \ll \eta^{-1} ,\\[2pt]
        \eta^{1+\beta} (\eta \hat{m})^{\beta -3} & \eta^{-1} \ll \hat{m} \ll \eta^{-2} ,
    \end{array} \right.
\end{equation}
with the two expressions of the same order $\eta^{1+\beta}$ at the crossover $\hat m \sim 1/\eta$, and the relative correction $(\eta \hat m)^{2\beta-1}$ in the first line, coming from the zero mode, being $O(1)$ precisely at the crossover. We stress that the conditioning on $m = 0$ is essential: the unconditioned marginal distribution of $\hat g$ is dominated, at large $\hat g$, by the last-step resonances $|g| \gg 1$ of regime (i), and since $f_\beta(g) \simeq |g|^{2\beta - 2} C_{\mathcal{R}}$ is not integrable at large $g$ for $\beta > 1/2$, integrating $f_\beta(g) \hat g^{-1-\beta}$ over $|g| \lesssim \sqrt{\hat g}$ turns the exponent $1+\beta$ into the familiar $3/2$ of the LDoS distribution in the localized phase, $P(\im G) \sim \sqrt{\eta}/(\im G)^{3/2}$~\cite{mirlin1991localization,mirlin1994distribution}. The exponents $1+\beta$ and $\beta - 3$ are instead the signature of the propagation along the tree and of its inverted image, respectively, and are only visible once the trivial resonance of the last site is removed, as in $Q_\eta(0,\hat m)$ or in the amplitude distribution $T(u)$.

\subsection{At criticality} \label{sec:criticalAL}

At the critical point, where $\beta=1/2$, one can argue that $\ft(g,\hat{g})$ has a logarithmic tail:
\begin{equation} \label{eq:ftypALtail-crit}
    \ft(g,\hat{g}) \simeq f_{1/2} (g) \, \frac{\ln \hat{g}}{\hat{g}^{3/2}} \, , \qquad \hat g \to \infty \, .
\end{equation}
To show this, we recall that the asymptotic equation for $P_0(g,\hat{g})$ at large $\hat{g}$ admits the solution $P_0(g,\hat{g}) \approx f_\beta(g)/\hat{g}^{1+\beta}$. Differentiating the whole cavity equation for $P_0(g,\hat{g})$ with respect to $\beta$, we obtain a source term proportional to $\de W/\de\beta$ and a shift term proportional to $\de P_0(g,\hat{g})/\de\beta$. However, at the critical point $\beta=1/2$ we have $\de W/\de\beta=0$ according to Eq.~\eqref{eq:beta}: as a consequence, $\de P_0(g,\hat{g})/\de\beta$ obeys the same equation~\eqref{eq:ftypAL} as $\ft(g,\hat{g})$. Note that $\de P_0/\de\beta$ is finite and nonzero at $\beta = 1/2$ even though $\de W/\de\beta = 0$ there, because $P_0$ depends on $W$ through $\sqrt{W - W_c} \propto \beta - 1/2$~\cite{rizzo2024localized}, so that the vanishing of $\de W/\de\beta$ is compensated by the divergence of $\de P_0/\de W$. On the other hand, the leading behavior at large $\hat{g}$ of the derivative of $P_0(g,\hat{g}) = f_\beta(g)/\hat{g}^{1+\beta}$ with respect to $\beta$ is
\begin{equation}
    \frac{\de P_0(g,\hat{g})}{\de\beta} \simeq -f_{\beta} (g) \, \frac{\ln \hat{g}}{\hat{g}^{\beta+1}} \, , \qquad \hat g \to \infty \, ,
\end{equation}
and, according to the above argument, at $\beta=1/2$ this coincides, up to a normalization constant, with the behavior of $\ft(g,\hat{g})$, which gives Eq.~\eqref{eq:ftypALtail-crit}.
At large $\hat{g}$ and $\beta=1/2$ we can thus write
\begin{equation}
    P_\eta(g,\hat{g}) \approx f_{1/2}(g) \, \frac{1}{\hat{g}^{3/2}} + c_{\rm typ}(\eta) \, f_{1/2}(g) \, \frac{\ln \hat{g}}{\hat{g}^{3/2}} \, .
\end{equation}
Following the arguments of the previous sections, the leading and subleading contributions become of the same order, and cancel, for $\hat{g}=O(\eta^{-1})$, which implies
\begin{equation}
    c_{\rm typ}(\eta) \propto -\frac{1}{|\ln \eta|} \, ,
\end{equation}
leading to
\begin{equation}
    I_2(\eta) - I_2^\infty \propto \frac{1}{|\ln \eta|} \qquad \textrm{for} \ W=W_c \, .
\end{equation}

\section{Numerical tests on the full distributions} \label{sec:numerics}

The analysis of the previous sections makes predictions that go well beyond the exponent of the average IPR, and concern the full probability distributions. In this section we test them against the numerical solution of the cavity equations of the Anderson model at finite $\eta$, and then show that the same scaling function describes, in the appropriate variables, the finite-size corrections to the distribution of the eigenstates' amplitudes.

\subsection{The distribution $\overline{Q}_\eta(\hat m)$ at finite $\eta$} \label{sec:Qeta}

\begin{figure*}
\centering
\panel{a}{\includegraphics[width=0.325\textwidth]{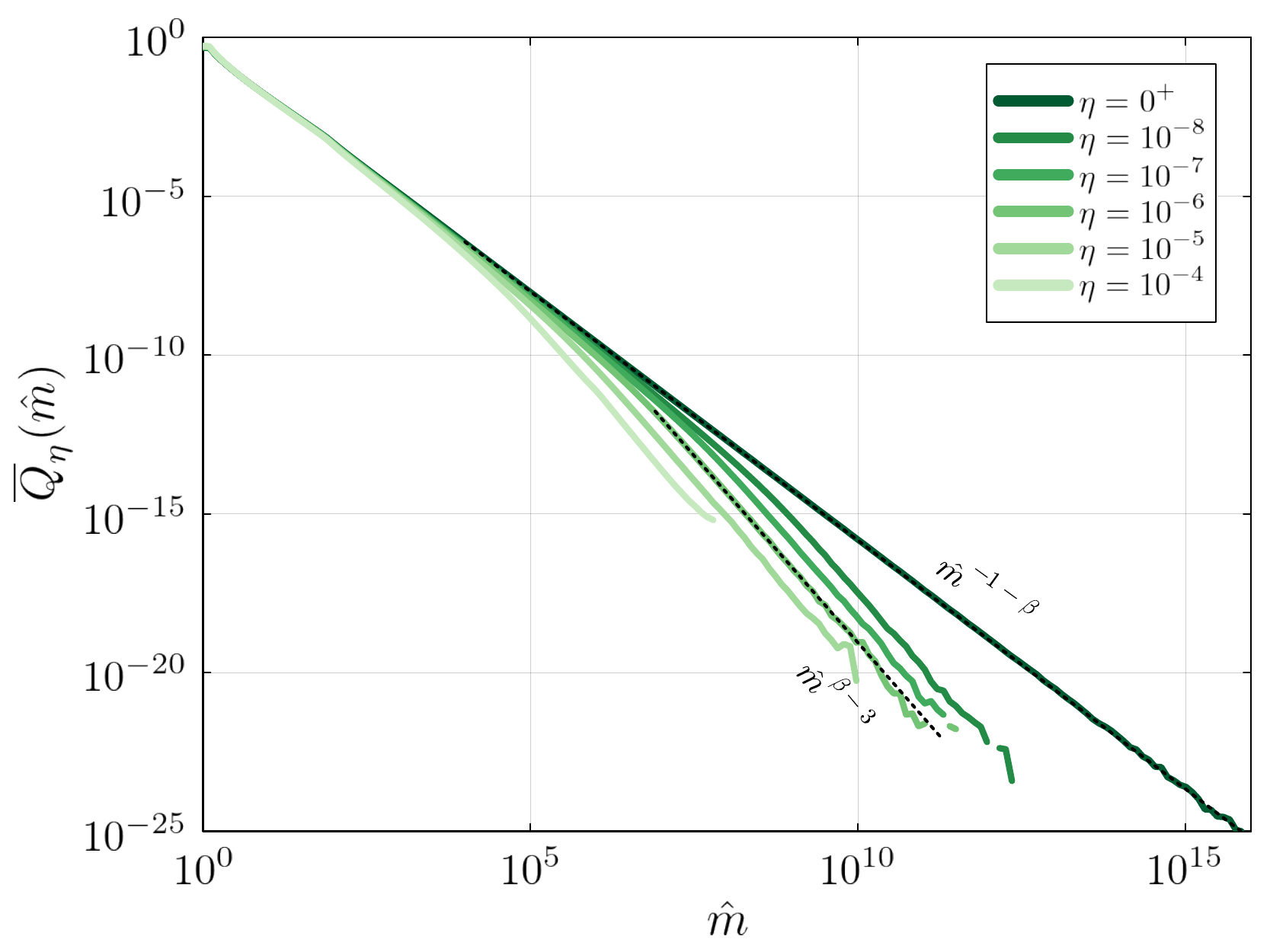}} \hspace{0.01\textwidth}
\panel{b}{\includegraphics[width=0.325\textwidth]{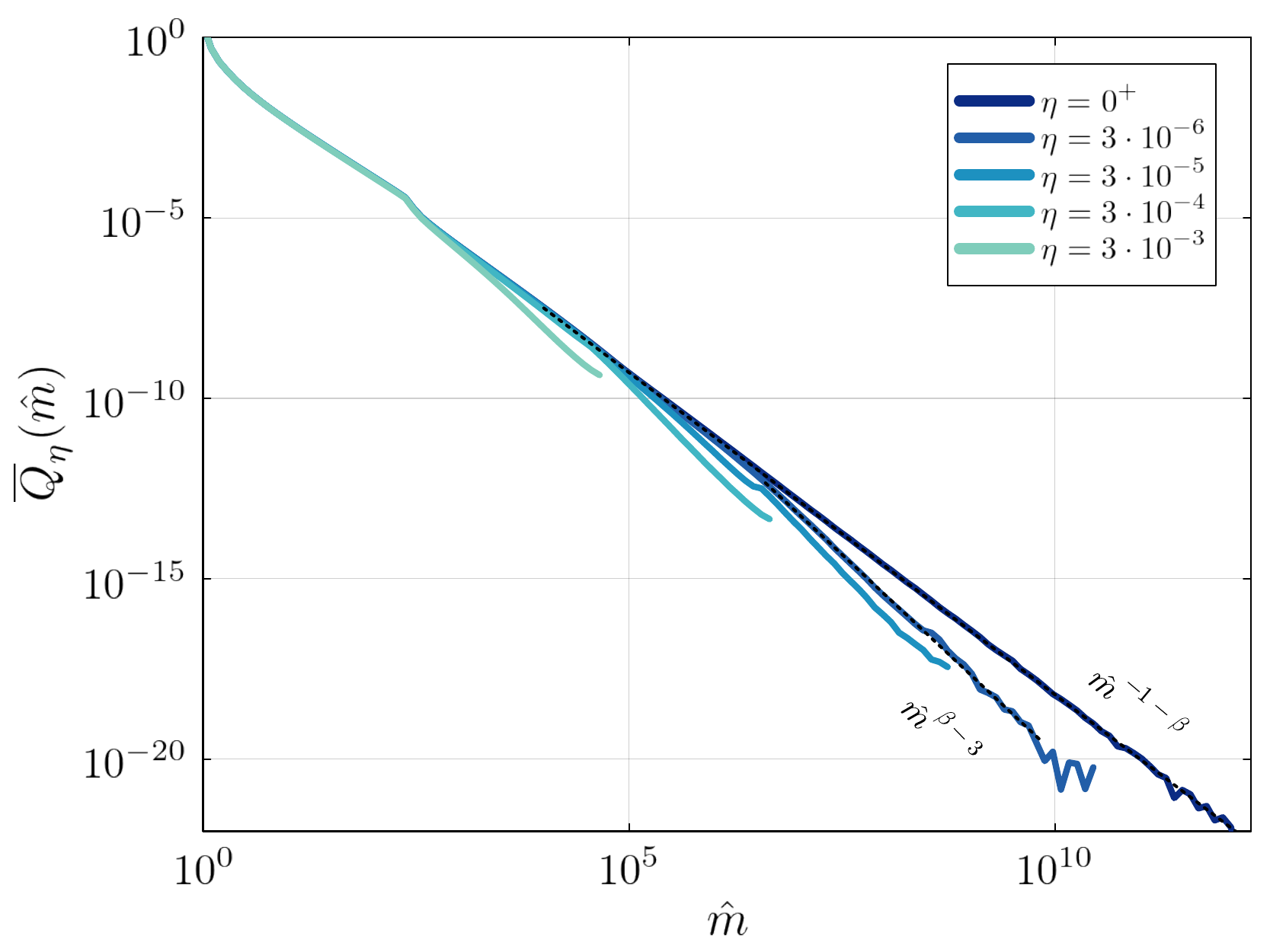}} \\[3mm]
\panel{c}{\includegraphics[width=0.325\textwidth]{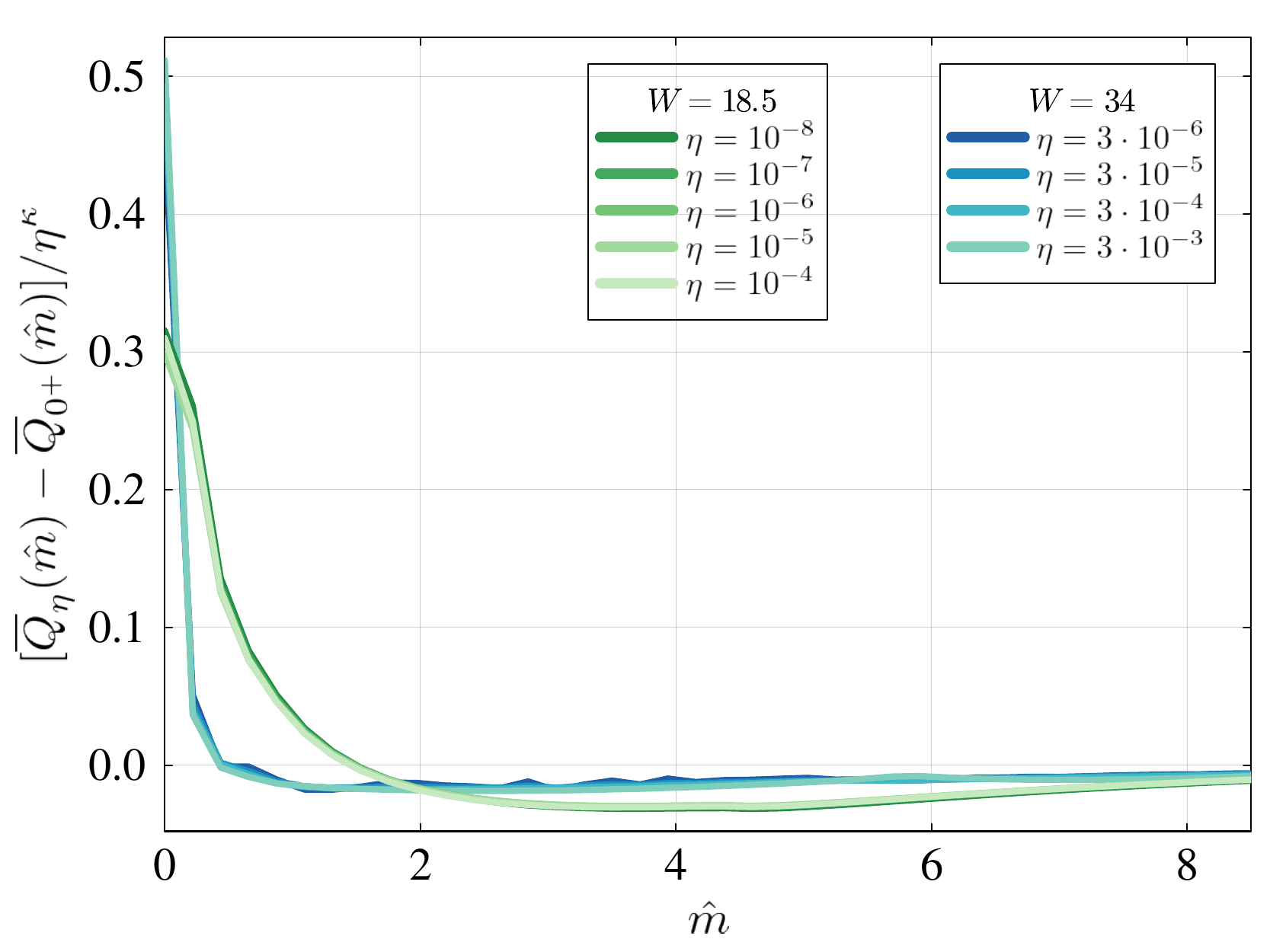}} \hfill
\panel{d}{\includegraphics[width=0.325\textwidth]{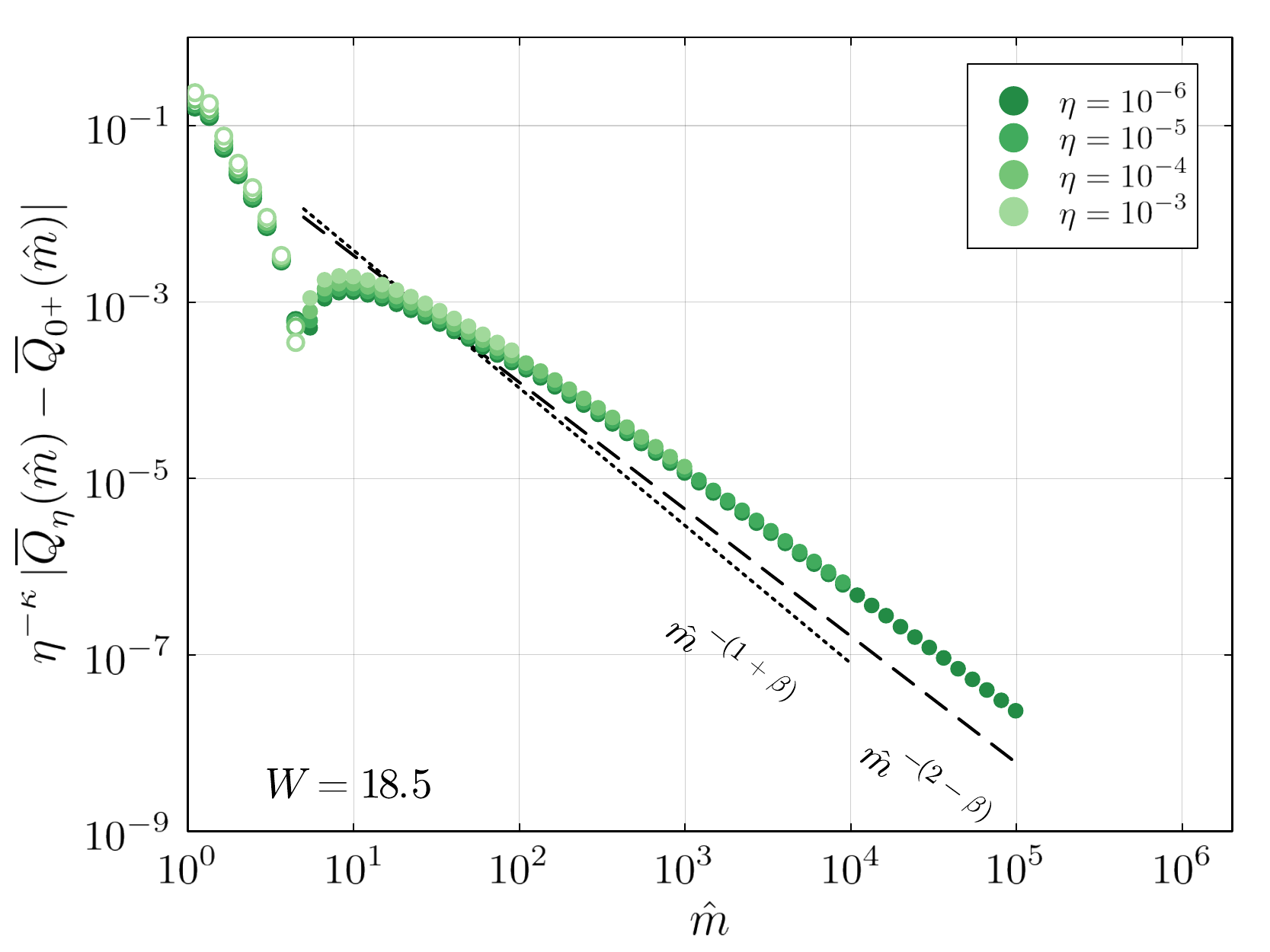}} \hfill
\panel{e}{\includegraphics[width=0.325\textwidth]{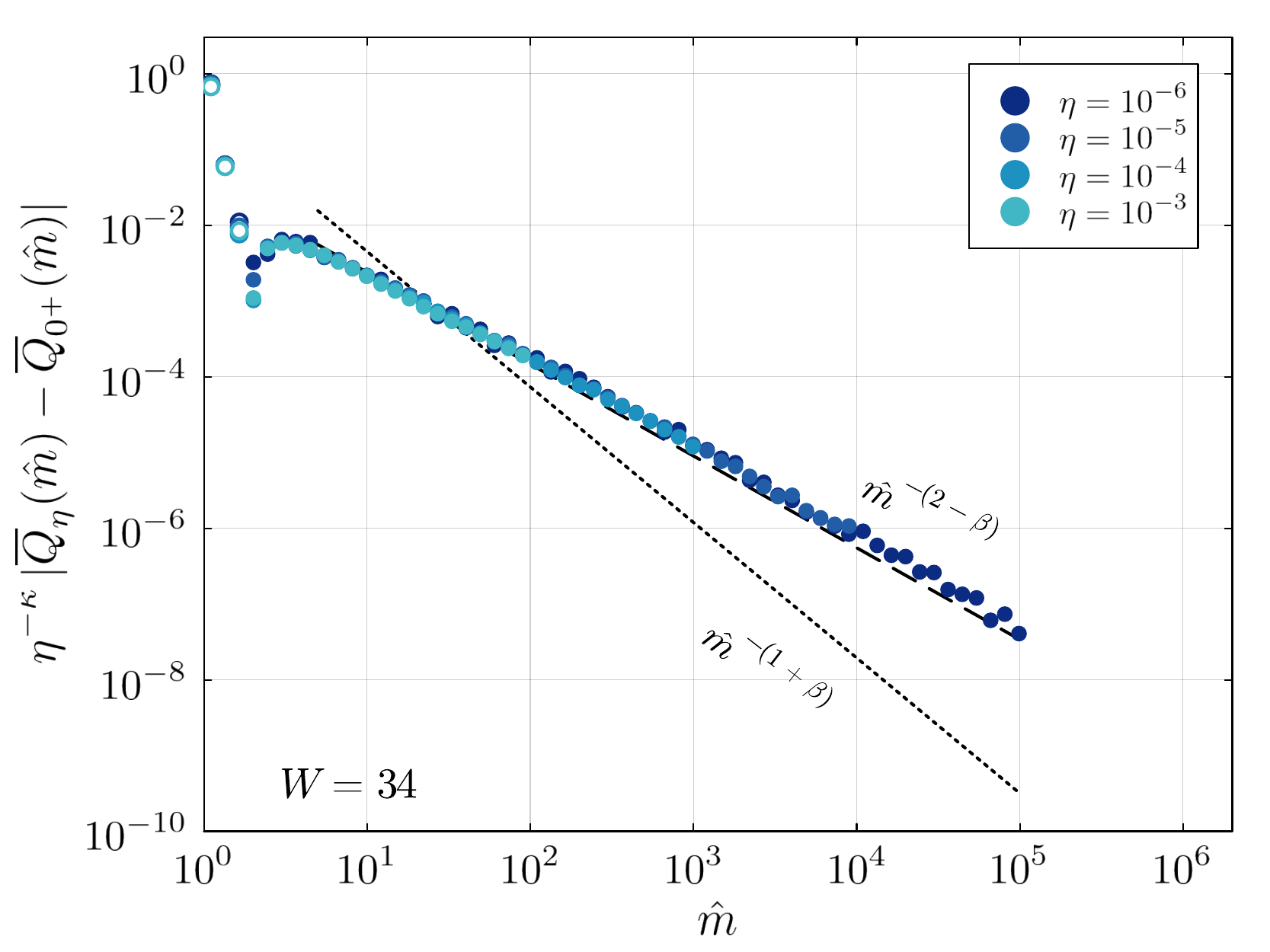}}
\caption{Top row: normalized distributions $\overline{Q}_\eta (\hat{m})$ of the variable $\hat m$ at $m=0$, from the population-dynamics solution of the cavity equations at finite $\eta$, for (a) $W=18.5$ and (b) $W=34$ and several values of the regulator $\eta$. The two power-law regimes of Eq.~\eqref{eq:Qregimes}, $\hat m^{-1-\beta}$ for $\hat m \ll 1/\eta$ and $\hat m^{\beta-3}$ for $\hat m \gg 1/\eta$, are indicated. Bottom row: the scaling function $g_\eta$, with $\kappa = 2\beta - 1$ in all panels, where $\beta = 0.57 \pm 0.01$ ($\kappa = 0.14 \pm 0.02$) for $W = 18.5$ and $\beta = 0.78 \pm 0.01$ ($\kappa = 0.56 \pm 0.02$) for $W = 34$, from the tail of $\overline{Q}_{0^+}$; the error bar covers the fits of the independent runs of (c) and of (d), (e), and each rescaling uses the value within it that best fits the corresponding run (the values used are $\kappa = 0.152$ and $0.546$ in (c), $0.122$ in (d) and $0.576$ in (e)). (c) Rescaled difference $[\overline{Q}_\eta (\hat{m}) - \overline{Q}_{0^+} ( \hat{m})]/\eta^{\kappa}$ in the bulk region $\hat{m} = O(1)$, for the same sets of curves as in (a) and (b) (green: $W=18.5$; blue: $W=34$). (d), (e) Its tail: $\eta^{-\kappa} |\overline{Q}_\eta (\hat{m}) - \overline{Q}_{0^+} ( \hat{m})|$ on logarithmic scales, in the window $1 \ll \hat m \ll 1/\eta$ (each curve is shown up to $\hat m = 0.1/\eta$), for $W=18.5$ (d) and $W=34$ (e) and $\eta = 10^{-3}, 10^{-4}, 10^{-5}, 10^{-6}$ (independent runs with populations of $2^{23}$ elements, $\sim 2.5 \times 10^9$ conditioned samples per value of $\eta$). Filled symbols: $g_\eta < 0$; open symbols: $g_\eta > 0$ (the zero crossing of $g_\eta$ is at $\hat m \approx 4.6$ for $W = 18.5$ and $\hat m \approx 1.8$ for $W=34$). The dashed line is the predicted tail of the zero mode, $\propto \hat m^{-(2-\beta)}$ [Eq.~\eqref{eq:ftypALtail}]; the dotted line is the decay $\propto \hat m^{-(1+\beta)}$ of $\overline{Q}_{0^+}$ itself, shown for comparison.
}
\label{fig:deltaQ}
\end{figure*}

We solve Eq.~\eqref{eq:cavity_a} at finite $\eta$ by population dynamics (with populations of up to $10^8$ elements) and sample the distribution $Q_\eta(0,\hat m)$ of $\hat m = 1 + \sum_{m \in \partial i} \hat g_{m \to i}$ at $m = 0$ by the conditioning technique of Ref.~\cite{rizzo2024localized}: since $m = \epsilon_i - \sum_{l \in \partial i} g_{l \to i}$ and $\epsilon_i$ is uniformly distributed, the density at $m=0$ is $1/W$ whenever $|\sum_m g_{m \to i}| < W/2$ and zero otherwise, so that $Q_\eta(0,\hat m)$ is obtained by histogramming $\hat m$ over the $(k+1)$-tuples of population elements satisfying this constraint. The $\eta \to 0^+$ distribution $Q_{0^+}(0,\hat m)$ is obtained in the same way from the linearized equations~\eqref{eq:linearized}, and both are normalized by $\int_0^\infty Q_{0^+}(0,\hat m) \de \hat m$ to give $\overline{Q}_\eta$ and $\overline{Q}_{0^+}$.

Panels (a) and (b) of Fig.~\ref{fig:deltaQ} show $\overline{Q}_\eta(\hat m)$ for $W = 18.5$, close to the transition ($\beta = 0.57 \pm 0.01$), and $W = 34$, deep in the localized phase ($\beta = 0.78 \pm 0.01$), for several values of $\eta$. The two regimes predicted by Eq.~\eqref{eq:Qregimes} are clearly visible: the $\eta=0^+$ power law $\hat m^{-1-\beta}$ for $\hat m \ll 1/\eta$, and the steeper law $\hat m^{\beta - 3}$ beyond the crossover at $\hat m \sim 1/\eta$, which moves to larger $\hat m$ as $\eta$ is decreased. The latter is the inverted image of the former discussed in Sec.~\ref{sec:fartail}; note that for $\beta \to 1/2$ the two exponents, $-3/2$ and $-5/2$, differ by exactly one, while for $\beta \to 1$ they tend to $-2$ on both sides.

Panel (c) is the central test of the scaling picture. It shows the difference $\overline{Q}_\eta(\hat m) - \overline{Q}_{0^+}(\hat m)$ in the bulk region $\hat m = O(1)$, rescaled by $\eta^{\kappa}$ with $\kappa = 2\beta - 1$ taken from the tail exponent of $\overline{Q}_{0^+}$ at the same $W$: for both disorder strengths, all curves collapse onto a single master function $g_\eta(\hat m)$, over several orders of magnitude in $\eta$.\footnote{At $W = 18.5$, close to the transition, $\kappa \approx 0.14$ is small and the factor $\eta^{\kappa}$ varies only by a factor of order two over the range of $\eta$ shown, so that the collapse tests the factorized form rather than the precise value of $\kappa$, which is tested sharply at $W = 34$.} This is the direct counterpart of Fig.~\ref{fig:1mpt}(c) for the toy model, and it confirms Eq.~\eqref{eq:distscalingQ}: the entire correction to the distribution in the bulk factorizes into the amplitude $\eta^{2\beta-1}$ times an $\eta$-independent function. Equally good collapses are obtained at other values of $W$ with the corresponding $\kappa(W)$. Since $I_2(\eta) - I_2^\infty$ is, by Eq.~\eqref{eq:I2deltaQ}, the moment $\int g_\eta \hat m^{-1} \de \hat m$ of this very function, the collapse is a considerably more stringent test than the exponent match of Fig.~\ref{fig:kappa}: it shows that the mechanism operates at the level of the full statistics of the LDoS, not only of its first inverse moment. The master function $g_\eta(\hat m)$ changes sign as a function of $\hat m$, as it must since it integrates to zero; the positive sign of $c_\eta = \int g_\eta \hat m^{-1} \de \hat m$ reflects the fact that the finite-$\eta$ distribution has more weight at small $\hat m$ (large LDoS) than the asymptotic one. Our second assumption in Sec.~\ref{sec:zeromode} further predicts that $g_\eta(\hat m)$ decays as $\hat m^{-(2-\beta)}$ in the window $1 \ll \hat m \ll 1/\eta$, i.e.\ more slowly than $\overline{Q}_{0^+}$ itself. This is tested in panels (d) and (e) of Fig.~\ref{fig:deltaQ}, which show the same rescaled difference on logarithmic scales, in the tail region, for $W = 18.5$ and $W = 34$ and four values of $\eta$ spanning three decades. The collapse persists over the entire window $1 \ll \hat m \ll 1/\eta$, which extends over five decades in $\hat m$ for the smallest $\eta$, and the master function follows the predicted power law $\hat m^{-(2-\beta)}$ (dashed line), clearly distinguishable from the $\hat m^{-(1+\beta)}$ decay of $\overline{Q}_{0^+}$ (dotted line): the correction is negative throughout the tail and decays with the exponent $2-\beta$ associated with the subdominant root $1-\beta$ of the Abou-Chacra--Anderson--Thouless condition, not with the exponent $1+\beta$ of the distribution itself. At $W=34$ the two candidate exponents, $1.22$ and $1.78$, are far apart and the data leave no doubt; at $W = 18.5$ they are close ($1.43$ and $1.57$) and, this close to $W_c$, the effective exponent of the master function drifts slowly towards its asymptotic value ($1.23$, $1.33$ and $1.37$ over the three successive decades above $\hat m = 100$), as expected from the logarithmic crossover of Sec.~\ref{sec:nearcritical}, since $|\ln \eta|$ is not large compared to $(\beta - 1/2)^{-1} \approx 14$; the collapse in $\eta$, however, is as good as at $W = 34$. The relative correction $\eta^\kappa g_\eta / \overline{Q}_{0^+} \propto (\eta \hat m)^{2\beta-1}$ thus grows along the tail and becomes of order one at the crossover $\hat m \sim 1/\eta$, as anticipated in Eq.~\eqref{eq:Qregimes}, where the curves of Fig.~\ref{fig:deltaQ}(d,e) depart from the master function and the boundary-layer description of Secs.~\ref{sec:boundaryAL} and~\ref{sec:fartail} takes over. The same power law appears, in the variable $u = 1/\hat m$, in the finite-size corrections discussed next.

\subsection{The same scaling function for the finite-$N$ corrections to $T_N(u)$} \label{sec:Tu}

\begin{figure*}
\panel{a}{\includegraphics[width=0.325\textwidth]{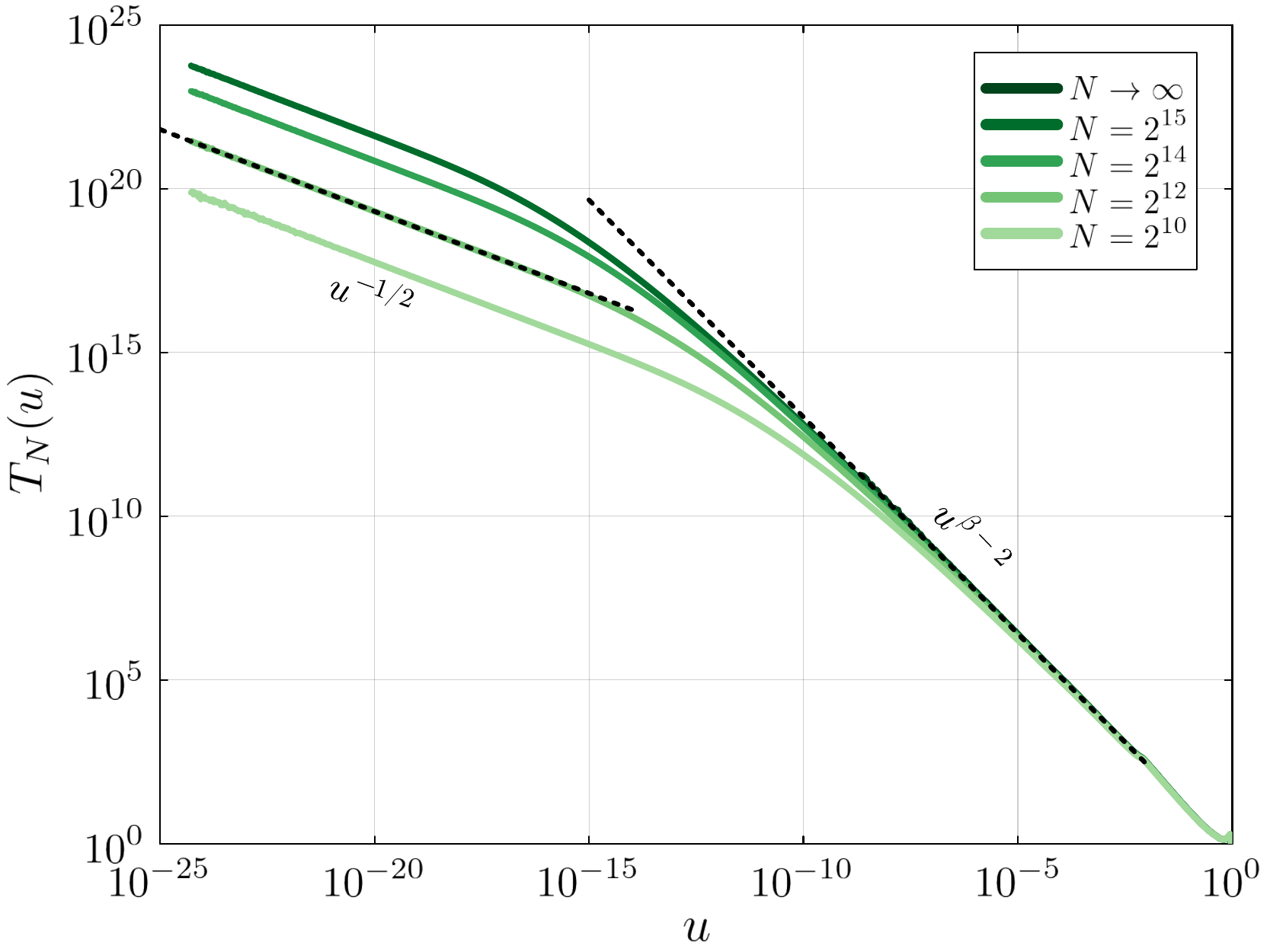}}\hfill
\panel{b}{\includegraphics[width=0.325\textwidth]{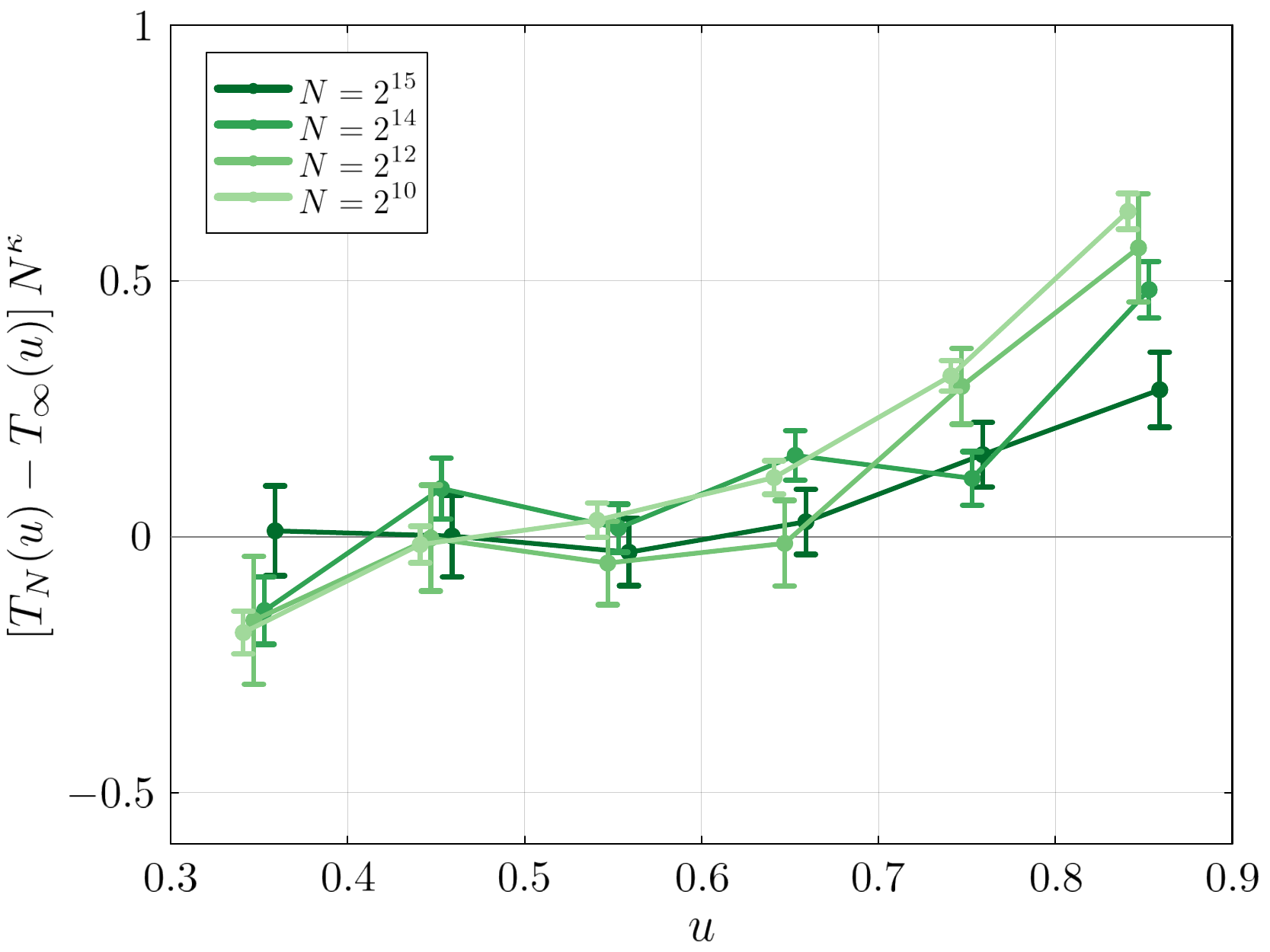}}\hfill
\panel{c}{\includegraphics[width=0.325\textwidth]{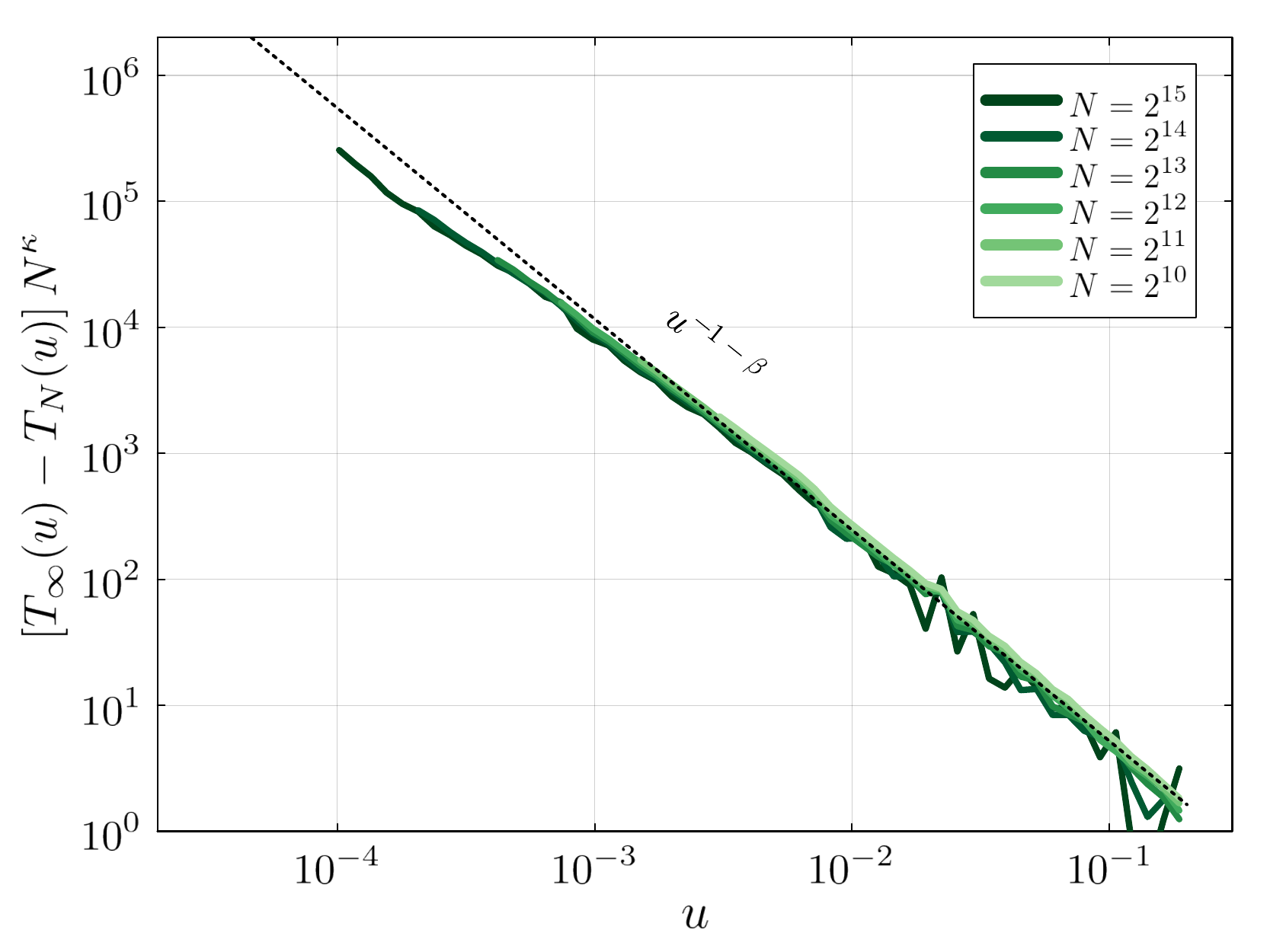}}
\caption{(a) Distribution $T_N(u)$ (normalized to $N$) of the squared amplitudes $u = |\psi_\alpha(i)|^2$ of the eigenstates closest to $E=0$, from exact diagonalization of the Anderson model on random regular graphs with $N = 2^n$ nodes, for several $n$ (see legend) and $W = 22$; the $N \to \infty$ distribution $T_\infty(u)$ is obtained from the cavity solution through Eq.~\eqref{eq:Tuinf}. (b) Rescaled difference $[T_N(u) - T_\infty(u)]\, N^{\kappa}$ in the bulk region $u = O(1)$, on linear bins of width $0.1$, with $\kappa = 2\beta - 1 \approx 0.35$ at $W=22$. The data include $1.6 \times 10^6$, $5.2 \times 10^5$, $4.5 \times 10^6$ and $4.4 \times 10^6$ eigenstates for $N = 2^{10}$, $2^{12}$, $2^{14}$ and $2^{15}$ respectively; the error bars are statistical, and the points are slightly displaced horizontally for legibility. (c) $[T_\infty(u) - T_N(u)]\, N^{\kappa}$ on logarithmic scales for $3/N \le u \le 0.2$, where $T_N < T_\infty$, and $N = 2^{10}, \ldots, 2^{15}$. The dotted line is the power law $u^{-1-\beta}$ predicted by Eq.~\eqref{eq:gNtail}, with a fitted amplitude, $0.11\, u^{-1-\beta}$, the image under $u = 1/\hat m$ of the tail $\hat m^{-(2-\beta)}$ of the zero mode. In all panels $T_\infty$ is the population-dynamics solution, which satisfies the sum rule $\int u\, T_\infty(u)\, \de u = 1$.}
\label{fig:deltaT}
\end{figure*}

The derivation of Sec.~\ref{sec:anderson} has a corollary that reaches beyond the finite-$\eta$ problem. The bulk correction is a zero mode of the linearized $\eta = 0$, $N = \infty$ recursion; nothing in Eq.~\eqref{eq:ftypAL} refers to the regulator. Any weak perturbation of the cavity recursion that is injected in the far tail of the distribution and vanishes in the bulk --- a finite $\eta$, but equally well the loop-induced corrections present on a finite graph, which the companion Letter~\cite{letter} argues to excite the same zero mode as a regulator of order $1/N$ --- must relax, in the bulk, onto the same zero mode; the nature of the perturbation can only enter through the amplitude, which is fixed in the boundary layer where the two problems differ. The finite-size corrections to the full distributions should therefore be described by the {\it same} scaling function as the finite-$\eta$ ones, once expressed in corresponding variables.

The natural observable at finite $N$ is the distribution $T_N(u)$ of the squared amplitudes $u = |\psi_\alpha(i)|^2$ of the eigenstates at $E=0$, defined with the normalization $\int_0^1 T_N(u) \de u = N$ [its second moment is $I_2(N)$], whose $N \to \infty$ limit is related to $\overline{Q}_{0^+}$ by Eq.~\eqref{eq:Tuinf}. The dictionary between the two problems is thus $u = 1/\hat m$, with $T(u) \, \de u \leftrightarrow \overline{Q}(\hat m) \, \hat m \, \de \hat m$: the amplitude of a resonant eigenstate on a site is $u = 1/\hat m$, i.e.\ its LDoS in units of $1/\eta$ (see Sec.~\ref{sec:linearized}). Writing, in analogy with Eq.~\eqref{eq:distscalingQ},
\begin{equation} \label{eq:distscalingT}
    T_N(u) - T_\infty(u) = N^{-\kappa} \, g_N(u) \, , \qquad \kappa = 2\beta - 1 \, ,
\end{equation}
the zero-mode argument predicts
\begin{equation} \label{eq:gNgeta}
    g_N(u) = A \, u^{-3} \, g_\eta(1/u) \, ,
\end{equation}
and the constant $A$ is not free: since $c_\eta = \int_0^\infty g_\eta(\hat m) \hat m^{-1} \de \hat m$ and $c_N = \int_0^1 g_N(u) u^2 \de u$ are the amplitudes of the corrections to the IPR, the change of variables in Eq.~\eqref{eq:gNgeta} gives exactly $A = c_N / c_\eta$. Equivalently, in the bulk a finite system of size $N$ behaves as the infinite BL at an effective regulator $\eta_{\rm eff} = A^{1/\kappa}/N$.
Two immediate consequences of Eq.~\eqref{eq:gNgeta} are that $T_N(u) - T_\infty(u)$, rescaled by $N^{\kappa}$ with the same $\kappa = 2\beta - 1$, must collapse onto a single master curve in the region $u = O(1)$, and that this master curve must inherit the tail of the zero mode: $g_\eta(\hat m) \propto \hat m^{-(2-\beta)}$ for $1 \ll \hat m \ll 1/\eta$ translates into
\begin{equation} \label{eq:gNtail}
    g_N(u) \propto u^{-3} \, u^{2-\beta} = u^{-1-\beta} \, , \qquad N^{-1} \ll u \ll 1 \, ,
\end{equation}
the boundary layer $\hat m \sim 1/\eta$ being mapped onto $u \sim \eta_{\rm eff} \sim 1/N$, i.e.\ onto the scale of the ergodic amplitude $|\psi|^2 \sim 1/N$. Note that the relative correction $N^{-\kappa} g_N(u)/T_\infty(u) \propto (N u)^{1 - 2\beta}$ becomes of order one precisely there, exactly as $(\eta \hat m)^{2\beta - 1}$ does at $\hat m \sim 1/\eta$ in Eq.~\eqref{eq:Qregimes}.

Figure~\ref{fig:deltaT} tests these predictions on exact-diagonalization data for the Anderson model on random regular graphs of $N = 2^n$ sites at $W = 22$ (see also Ref.~\cite{letter}). Figure~\ref{fig:deltaT}(a) shows $T_N(u)$ for several sizes together with the asymptotic $T_\infty(u)$ from the cavity solution: at finite $N$ the power law $T_\infty \propto u^{-(2-\beta)}$ is cut off at very small amplitudes, $u \lesssim u_N \propto N^{-1/(1-\beta)}$, where it is replaced by an integrable $u^{-1/2}$ behavior~\cite{mirlin1994distribution,rizzo2024localized}; this cutoff, however, produces corrections to the IPR of order $N^{-(1+\beta)/(1-\beta)}$, far too small and of the wrong sign to account for the observed ones, which must therefore come from the region $u = O(1)$. Figure~\ref{fig:deltaT}(b) shows $[T_N(u) - T_\infty(u)]\, N^{\kappa}$ in this region, with $\kappa = 2\beta - 1 \approx 0.35$ at $W=22$: the curves collapse onto a single master function within the statistical errors, which are large at the largest sizes since each eigenstate contributes only a few sites with $u = O(1)$; this is the finite-$N$ counterpart of Fig.~\ref{fig:deltaQ}(c), and it confirms Eq.~\eqref{eq:distscalingT}. Figure~\ref{fig:deltaT}(c) shows the same rescaled difference, with the opposite sign, for $3/N \le u \le 0.2$, where $T_N < T_\infty$: the collapse persists, and for $u \gtrsim 10^{-3}$ the master curve follows the power law $u^{-1-\beta}$ predicted by Eq.~\eqref{eq:gNtail} at all sizes, while at the largest sizes the curves start to depart from it below $u \approx 10^{-3}$. In the regime where it is clean, $10^{-3} \lesssim u \ll 1$, this observation is, to our mind, the most direct evidence for the subdominant root: the finite-size correction to the amplitude distribution decays, at small $u$, with the exponent $1 + \beta$ that corresponds to $1-\beta$ in the language of the Abou-Chacra--Anderson--Thouless condition, and not with the exponent $2-\beta$ of the distribution itself. For $u \lesssim 1/N$ the deficit is no longer given by the zero mode: we expect that, at the smallest amplitudes $u \sim u_N \propto N^{-1/(1-\beta)}$, the cutoff scale of the tree law first predicted in Ref.~\cite{mirlin1994distribution} (see also Ref.~\cite{rizzo2024localized}), $T_N/T_\infty$ becomes a function of the single scaling variable $z = N u^{1-\beta}$. The structure of this crossover will be discussed elsewhere. 
\begin{figure*}
\panel{a}{\includegraphics[width=0.46\textwidth]{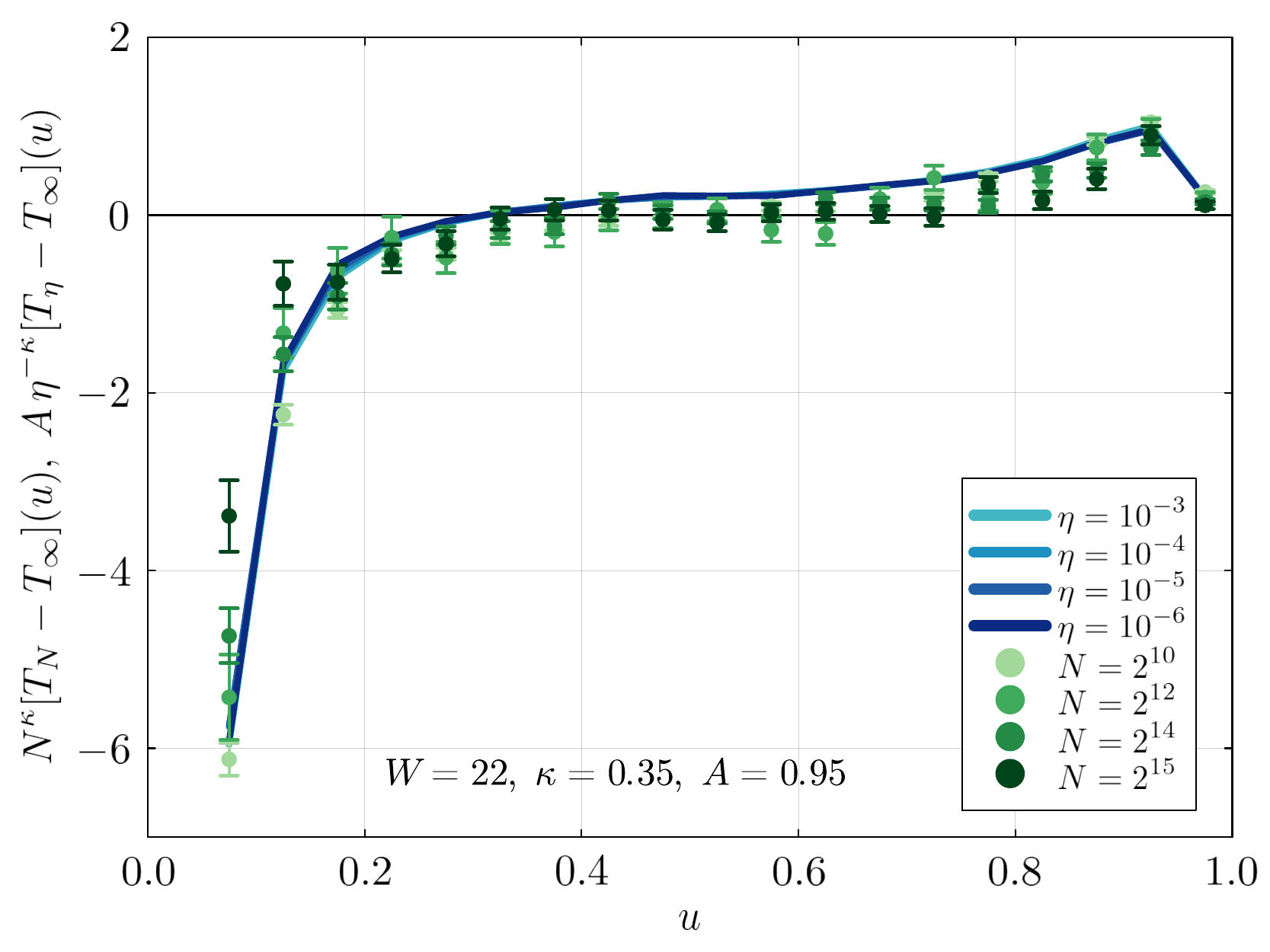}} \hfill
\panel{b}{\includegraphics[width=0.46\textwidth]{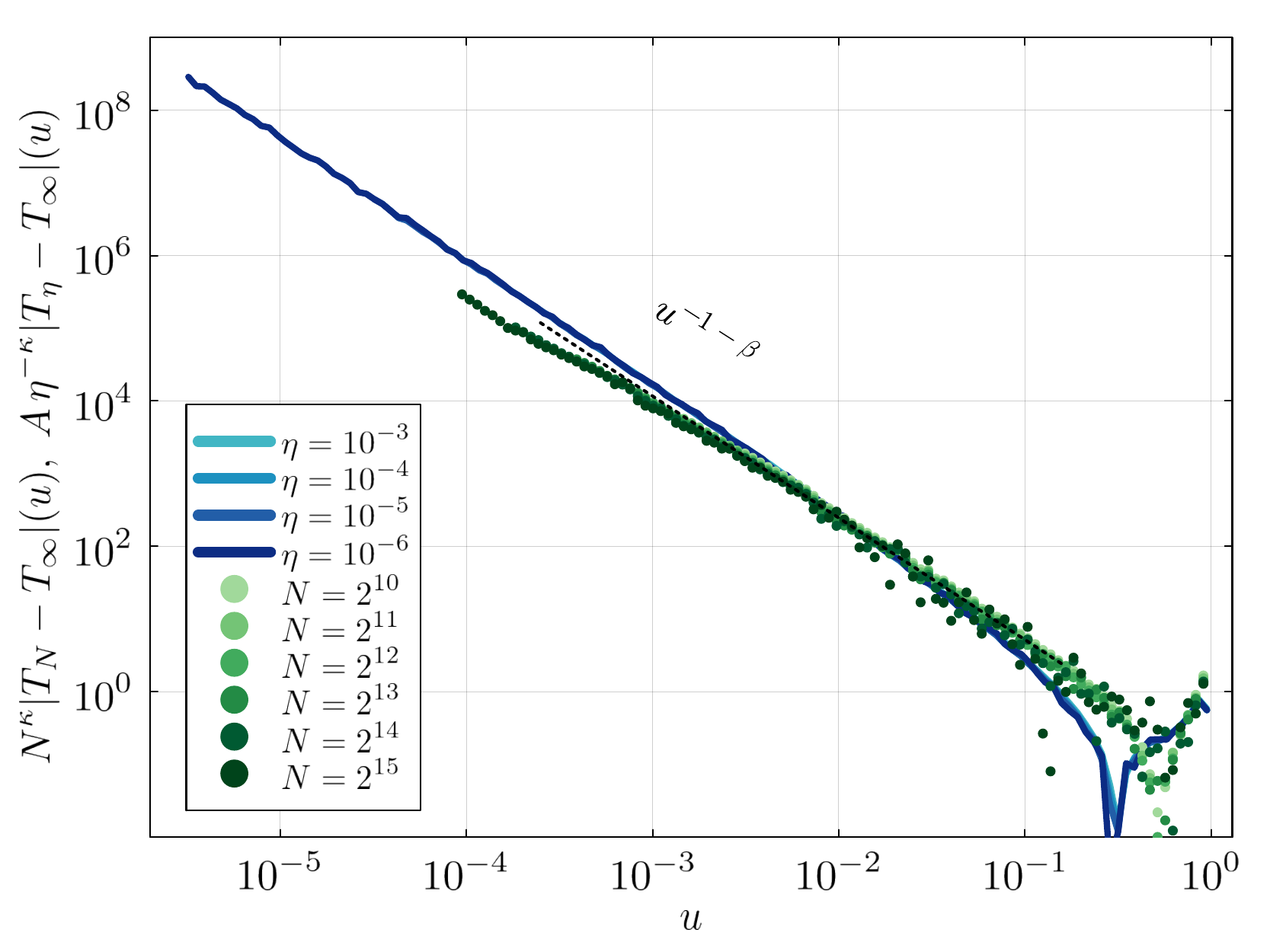}}
\caption{Direct comparison of the two scaling functions at $W = 22$ ($\beta \approx 0.675$, $\kappa \approx 0.35$). Symbols: finite-size master curve $N^{\kappa} [T_N(u) - T_\infty(u)]$ from exact diagonalization, the same data as in Fig.~\ref{fig:deltaT}: $N = 2^{10}, 2^{12}, 2^{14}, 2^{15}$ in (a) and $N = 2^{10}, \ldots, 2^{15}$ in (b). Lines: finite-regulator master curve $A \, \eta^{-\kappa} [T_\eta(u) - T_\infty(u)]$, with $T_\eta(u) \equiv u^{-3} \overline{Q}_\eta(1/u)$, from the population-dynamics solution of the cavity equations, $\eta = 10^{-3}, \ldots, 10^{-6}$. (a) Bulk region, linear bins of width $0.05$. (b) Absolute value on logarithmic scales, in the regions $1/N \ll u \ll 1$ and $\eta \ll u \ll 1$ respectively, where both master curves are negative and follow $u^{-1-\beta}$ (dotted line: $0.11\, u^{-1-\beta}$); the dip at $u \approx 0.3$ is the zero crossing [see (a)]. The curves are drawn with $A = 0.95$; fits on the flank $\max(3 \times 10^{-3}, 3/N) < u < 0.08$ give $A$ independent of $\eta$ and close to one, decreasing slowly with $N$ from $1.15$ ($N = 2^{10}$) to $0.85$ ($N = 2^{15}$).}
\label{fig:overlay}
\end{figure*}

The test of Eq.~\eqref{eq:gNgeta} can be completed by a direct overlay of the two master curves, which requires data at the same $W$ for the two problems. Figure~\ref{fig:overlay} shows it at $W = 22$, where we have solved the cavity equations at finite $\eta$ as in Sec.~\ref{sec:Qeta}; the exact-diagonalization data are those of Fig.~\ref{fig:deltaT}, $N = 2^{10}$ to $2^{15}$. The finite-regulator master curves, obtained from $\eta = 10^{-3}$ to $10^{-6}$, collapse onto each other to within the line width, both in the bulk and along the $u^{-1-\beta}$ flank. The finite-size ones, much noisier because each eigenstate contributes only a few sites with $u = O(1)$, have the same shape --- the negative flank $\propto u^{-1-\beta}$, a zero crossing and a positive bulk at larger $u$ --- and along the flank they coincide with the former once the latter is multiplied by a constant $A$, independent of $\eta$ and close to one, which however decreases slowly with $N$, from $A \approx 1.15$ at $N = 2^{10}$ to $A \approx 0.85$ at $N = 2^{15}$ (roughly as $N^{-0.09}$). In the bulk the finite-size curves follow the finite-$\eta$ ones, including the zero crossing at $u \approx 0.3$, within the statistical errors except at the smallest values of $u$, where the largest sizes lie above them. The slow drift of $A$, like that of the effective amplitude of the IPR correction discussed below and the excess of $\kappa_N$ over $\kappa_\eta$ close to $W_c$ (Sec.~\ref{sec:FSS}), points to slowly decaying subleading finite-size corrections. The two scaling functions are therefore really the same function in the appropriate variables, as Eq.~\eqref{eq:gNgeta} asserts: the finite-size and the finite-regulator corrections to the full distributions are the same zero mode of the linearized recursion, excited with different amplitudes, and in the bulk a system of $N$ sites is equivalent to the infinite BL at $\eta_{\rm eff} = A^{1/\kappa}/N$, of order $1/N$. The amplitudes of the IPR corrections are consistent with this value through $A = c_N/c_\eta$: $c_\eta \approx 0.15$ from the finite-$\eta$ populations, with $[I_2(\eta)-I_2^\infty]/\eta^{\kappa}$ essentially constant over three decades in $\eta$, and $c_N = 0.15 \pm 0.02$ from the exact-diagonalization IPR of Ref.~\cite{letter} fitted with the same $\kappa$ for $N \geq 8192$, give $c_N/c_\eta = 1.0 \pm 0.15$, consistent with $A$; the effective amplitude $[I_2(N) - I_2^\infty]\,N^{\kappa}$ itself decreases slowly, from $0.19$ at $N = 2^{10}$ to about $0.15$ at $N \gtrsim 10^4$, in parallel with $A(N)$.

\section{Finite-size scaling of the IPR} \label{sec:FSS}

The results of the previous sections fix the finite-size-scaling form of the IPR in the localized phase, including its crossover to the critical point, with no free parameter in the scaling variable. Since the finite-size scaling of the IPR on random regular graphs has been the subject of a long debate~\cite{tikhonov2019statistics,garcia2022critical,10.21468/SciPostPhys.15.2.045,garcia2017scaling,chen2023quantum,chen2024describing,vanoni2023renormalization,garcia2026connecting}, we state this form explicitly, test it on the finite-$\eta$ and finite-$N$ data, and discuss its consequences for previous finite-size analyses.

\subsection{The scaling form}

In the toy model, Eq.~\eqref{eq:I2crossover} of Sec.~\ref{sec:nearcritical} describes the whole crossover from the power law $H^{2\beta-1}$ to the critical $1/|\ln H|$ decay as a function of the single variable $x = (2\beta - 1)\,|\ln H|$, with an amplitude $C$ that is computed, not fitted.

In the Anderson model the same structure is expected with $L = \ln(1/\eta)$ at finite $\eta$, and $L = \ln N$ at finite $N$, since a graph of $N$ sites acts on the bulk as a regulator $\eta_{\rm eff} \propto 1/N$ (Sec.~\ref{sec:Tu}):
\begin{equation} \label{eq:FSS}
    I_2 - I_2^\infty \simeq \frac{1}{L} \, \Phi(\kappa L) \, , \qquad \Phi(x) = \Phi_0\, f(a x) = \frac{\Phi_0\, a\, x}{e^{a x} - 1} \, ,
\end{equation}
with $\kappa = 2\beta - 1$. For $x \gg 1$, i.e.\ $L \gg 1/\kappa$, Eq.~\eqref{eq:FSS} reduces to $I_2 - I_2^\infty \simeq \Phi_0\, a\kappa\, e^{-a\kappa L}$, the power laws of Sec.~\ref{sec:evidence}: this is the asymptotic regime, which moves to ever larger sizes (smaller $\eta$) as $W \to W_c$, and the prefactor of the power law, $\Phi_0\, a\kappa$, vanishes $\propto 2\beta - 1$, as in Eq.~\eqref{eq:I2kappasmall}. For $x \ll 1$ the correction becomes $\Phi_0/L$, with a finite amplitude. The amplitude $\Phi_0$, the counterpart of $C$, is not universal, and the theory predicts $a = 1$, as required by the asymptotic exponent $\kappa = 2\beta - 1$. Away from $W_c$, $b_3, c_3 \neq 1$ replace $e^{x} - 1$ by $e^{x} - b_3 c_3$ in Eq.~\eqref{eq:atypuniform} and make the amplitude depend on $W$, and a constant factor in $\eta_{\rm eff} \propto 1/N$ only shifts $L$ by a constant, a correction of relative order $1/L$, so that none of these effects changes $a$. In the fits below we nevertheless keep $a$ as an effective parameter, which absorbs the $W$-dependence of the amplitude and the subleading corrections at finite $L$; the data confirm $a = 1$ once these effects are separated. Combined with the square-root singularity $I_2^\infty \simeq I_2^{\infty,{\rm crit}} + c_{I_2}\,(W - W_c)^{1/2}$ of the asymptotic value (Sec.~\ref{sec:linearized}), and with $\kappa \propto (W - W_c)^{1/2}$, Eq.~\eqref{eq:FSS} gives close to $W_c$
\begin{equation} \label{eq:FSSlinear}
\begin{aligned}
    I_2(N, W) - I_2^{\infty,{\rm crit}} & \simeq \frac{1}{\ln N}\, \mathcal{F}\big(\kappa \ln N\big) \, , \\
    \mathcal{F}(x) & \simeq \frac{c_{I_2}}{2c_\beta}\, x + \Phi(x) \, ,
\end{aligned}
\end{equation}
where we used $\kappa \simeq 2c_\beta\,(W - W_c)^{1/2}$ close to $W_c$, Eq.~\eqref{eq:beta}. This has the form of the ``linear'' single-parameter scaling ansatz of Refs.~\cite{garcia2017scaling,garcia2022critical}, with scaling variable $\ln N/\xi_\kappa$, but with a scaling length $\xi_\kappa = 1/\kappa \propto (W - W_c)^{-1/2}$, i.e.\ $\nu = 1/2$, rather than their $\nu = 1$ (as discussed in Sec.~\ref{sec:previous}, this collapse in one variable does not make the underlying flow one-parameter); and the critical value is approached as $1/\ln N$, i.e.\ with $\zeta = 1$, where $\zeta$ is the exponent of the critical correction $(\ln N)^{-\zeta}$ introduced in Sec.~\ref{sec:critical} (denoted $\mu$ in Refs.~\cite{tikhonov2019statistics,chen2024describing,garcia2026connecting}, in the critical-localization scenario of the latter two), rather than with the effective $\zeta \simeq 1/2$ suggested by the fits of Ref.~\cite{tikhonov2019statistics}, which we have argued in Sec.~\ref{sec:critical} to be preasymptotic. All the ingredients of the scaling form except the amplitude $\Phi_0$ are known independently: $\beta$, which fixes the scaling variable, from the tail of the LDoS distribution, and $I_2^\infty$ from the $\eta \to 0^+$ cavity solution.

\begin{figure*}
\panel{a}{\includegraphics[width=0.325\textwidth]{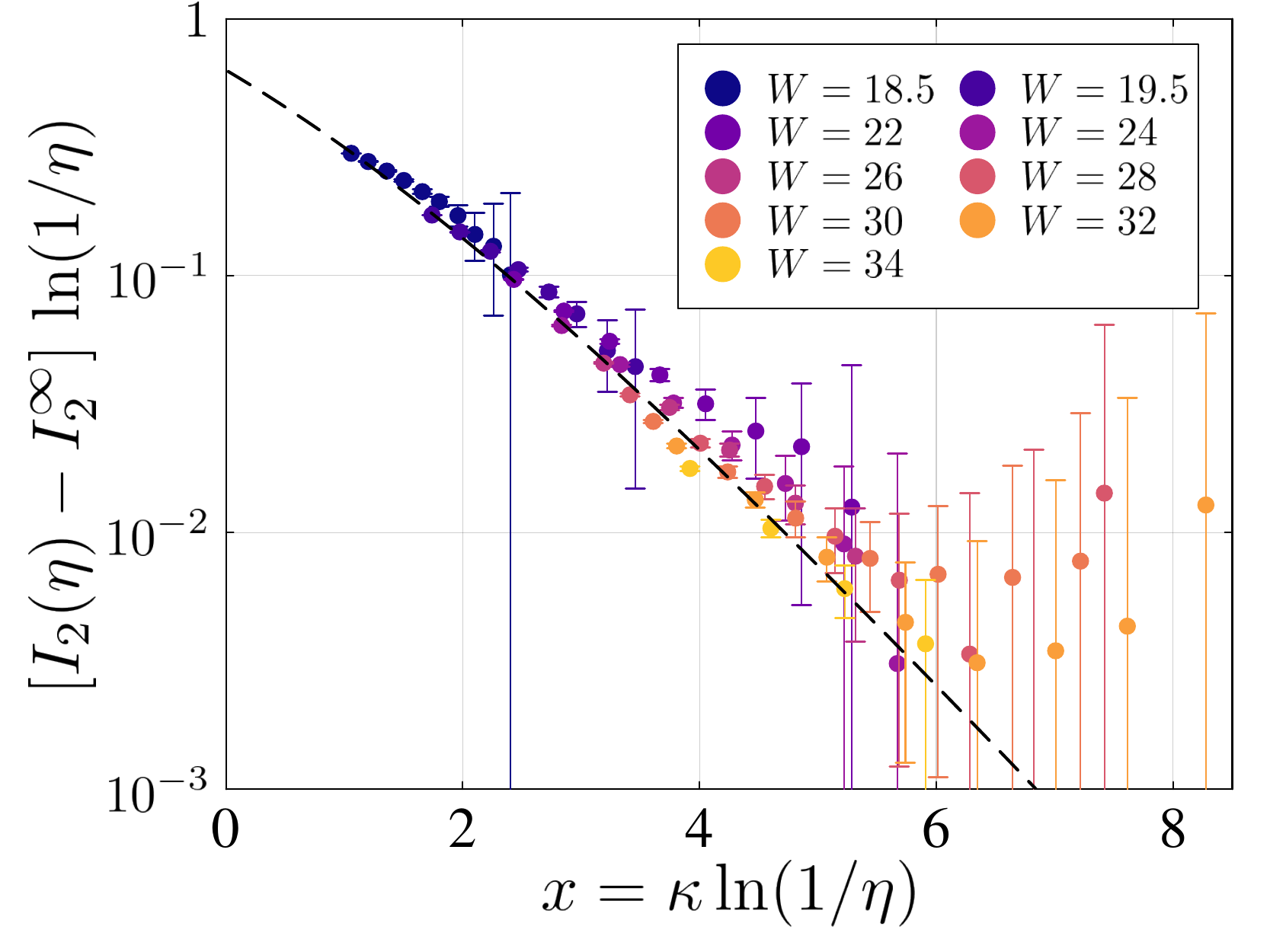}}\hfill
\panel{b}{\includegraphics[width=0.325\textwidth]{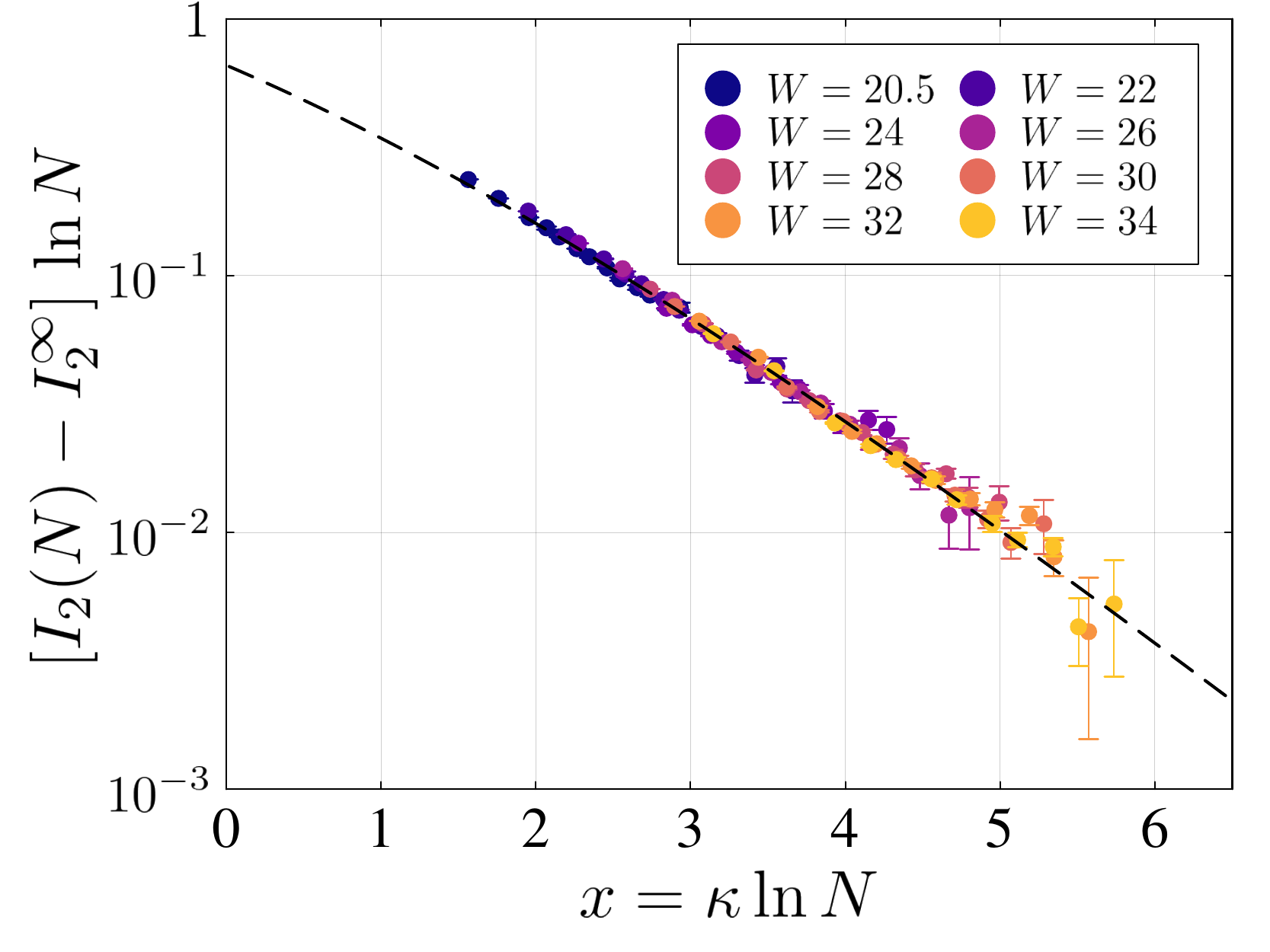}}\hfill
\panel{c}{\includegraphics[width=0.325\textwidth]{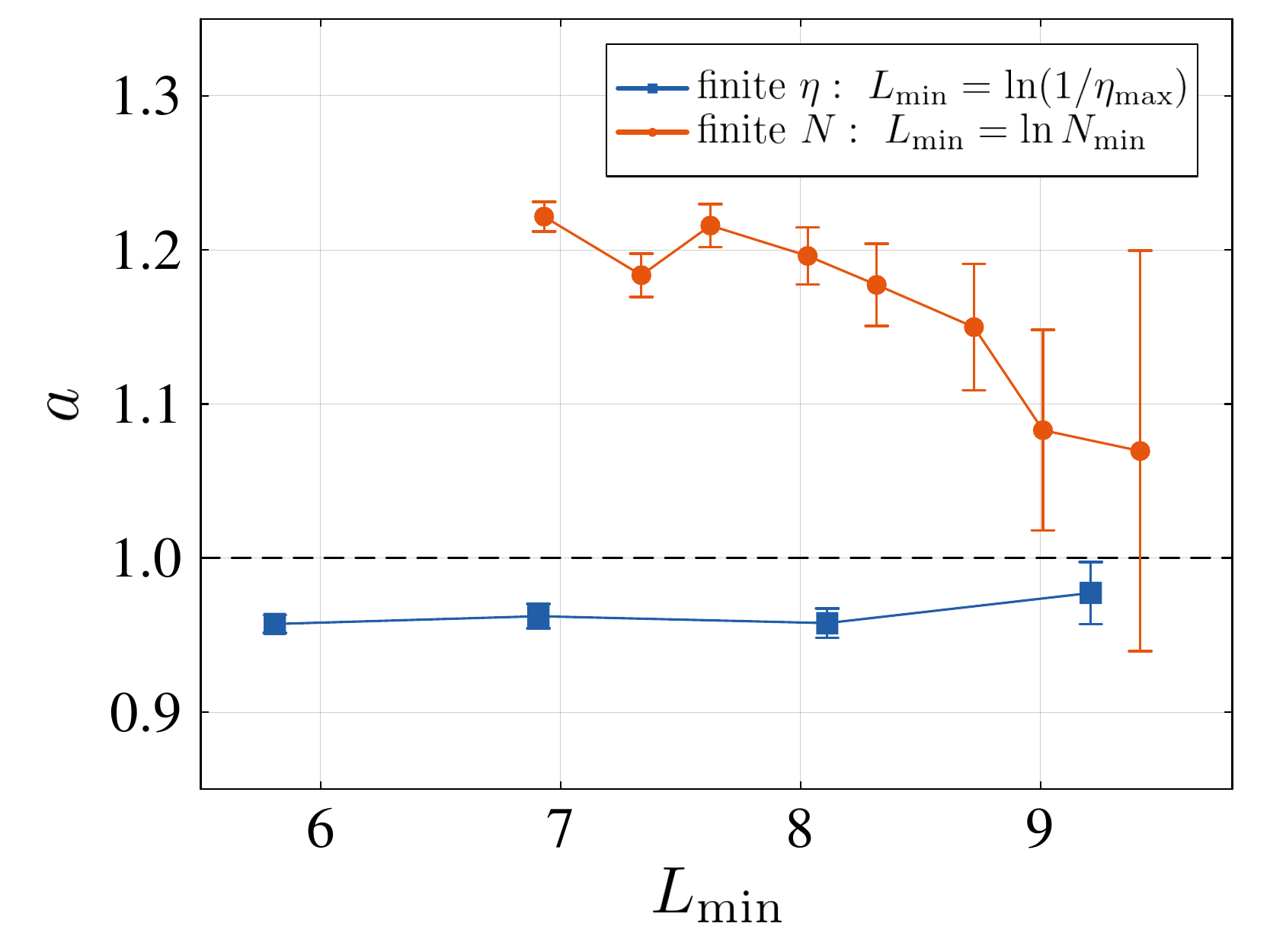}}
\caption{Test of the finite-size-scaling form~\eqref{eq:FSS}, with $\kappa = 2\beta - 1$ from the tail of the LDoS distribution and $I_2^\infty$ from the $\eta \to 0^+$ cavity solution, so that nothing is adjusted in the scaling variable. (a) Finite $\eta$, population dynamics on the infinite BL, $\eta \le 10^{-3}$, $W = 18.5$ to $34$: $[I_2(\eta) - I_2^\infty]\ln(1/\eta)$ against $x = \kappa \ln(1/\eta)$. (b) Finite $N$, exact diagonalization of random regular graphs with $N = 2^8$ to $2^{15}$, $W = 20.5$ to $34$: $[I_2(N) - I_2^\infty] \ln N$ against $x = \kappa \ln N$. Dashed lines: $\Phi(x)$ of Eq.~\eqref{eq:FSS} with a common $\Phi_0$ and $a$ for all $W$, fitted on $\eta \le 10^{-3}$ [(a): $\Phi_0 = 0.63$, $a = 1.26$] and on $N \ge 2^{11}$ [(b): $\Phi_0 = 0.66$, $a = 1.19$]. (c) Shape parameter $a$ from a fit with a common $a$ for all $W$ and an amplitude free at each $W$, against the cutoff $L_{\min}$: finite $\eta$, with $\eta \le \eta_{\max}$ and $L_{\min} = \ln(1/\eta_{\max})$ (squares), and finite $N$, with $N \ge N_{\min}$ and $L_{\min} = \ln N_{\min}$ (circles); the dashed line is the leading-order value $a = 1$.}
\label{fig:FSS}
\end{figure*}

\begin{figure}[t]
\centering
\includegraphics[width=0.46\textwidth]{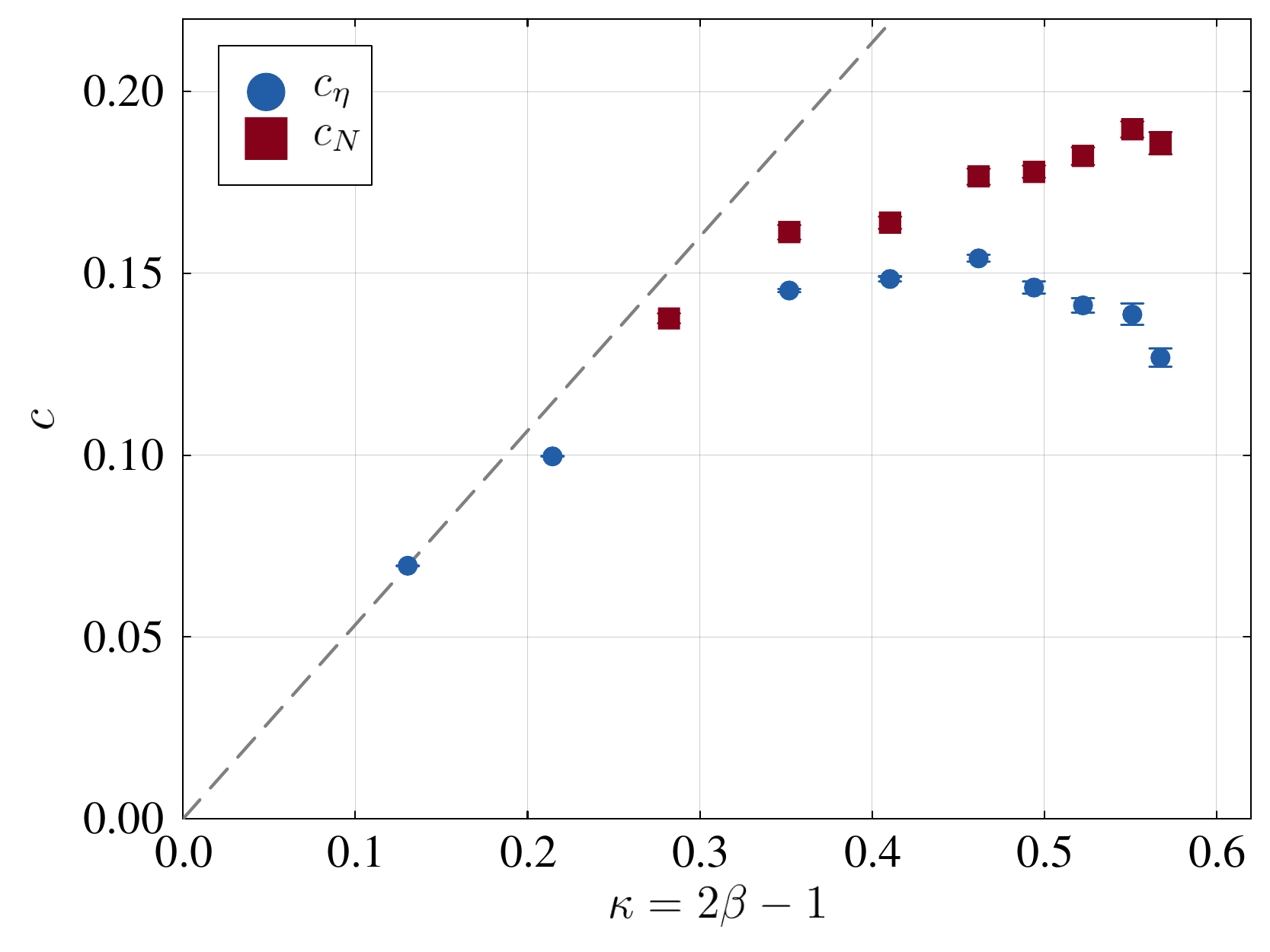}
\caption{Prefactors $c_\eta$ (circles) and $c_N$ (squares) of the asymptotic power laws $I_2(\eta) - I_2^\infty \simeq c_\eta\, \eta^{\kappa}$ and $I_2(N) - I_2^\infty \simeq c_N N^{-\kappa}$, against $\kappa = 2\beta - 1$. They are obtained as $c = \Phi_0 \kappa$ from fits of the scaling form~\eqref{eq:FSS} with $a = 1$ and an amplitude $\Phi_0$ free at each $W$, on the data of Fig.~\ref{fig:FSS} with $\eta \le 10^{-3}$ (population dynamics, $W = 18.5$ to $34$) and $N \ge 2^{11}$ (exact diagonalization, $W = 20.5$ to $34$). The dashed line, $\propto \kappa$, is a guide to the eye.}
\label{fig:cprefactor}
\end{figure}

\subsection{Test on the finite-$\eta$ and finite-$N$ data}

Figure~\ref{fig:FSS} tests Eq.~\eqref{eq:FSS} on the population-dynamics data of Fig.~\ref{fig:I2eta}, for $\eta \le 10^{-3}$ [$\ln(1/\eta) \approx 7$ to $19$, which brings $x$ down to $\approx 1$ at $W = 18.5$], and on the exact-diagonalization data of Ref.~\cite{letter}, $N = 2^8$ to $2^{15}$ ($\ln N \le 10.4$). In both cases the data for all disorder strengths collapse onto a single curve as functions of $x = \kappa L$ [Figs.~\ref{fig:FSS}(a) and (b)], with $\zeta = 1$: the scaling variable is $(2\beta - 1)\ln N$, as predicted.\footnote{A generic collapse test, in which $[I_2 - I_2^\infty]\,L^{\zeta}$ is fitted by a smooth function of the scaling variable, confirms this assignment. With a scaling variable $\propto (W - W_c)^{1/2} L$ (either $\kappa L$ or $(W - W_c)^{1/2} L$) the best collapse is obtained for $\zeta$ between $0.5$ and $1.5$, compatible with $\zeta = 1$ for both data sets. With $\nu = 1$, i.e.\ a scaling variable $\propto (W - W_c)\,L$, a collapse of comparable quality requires $\zeta \approx 1.5$ for the finite-$\eta$ data and $\zeta \approx 2.5$ for the finite-$N$ data, which are inconsistent with each other and with the $1/\ln N$ behavior at the critical point.}

The dashed lines in Figs.~\ref{fig:FSS}(a) and (b) are the scaling function~\eqref{eq:FSS} with a single amplitude $\Phi_0$ and a single shape parameter $a$ for all disorder strengths. The fitted shape parameter, $a = 1.26$ at finite $\eta$ and $a = 1.19$ at finite $N$, differs from its leading-order value $a = 1$. The reason is that the amplitude is not exactly the same at all $W$: fitted separately at each $W$ with $a = 1$, it decreases smoothly from $\Phi_0 = 0.53$ at $W = 18.5$ to $0.22$ at $W = 34$ at finite $\eta$, a dependence allowed away from $W_c$, where $b_3, c_3 \neq 1$. The corresponding prefactors $c = \Phi_0 \kappa$ of the asymptotic power laws, $c_\eta$ at finite $\eta$ and $c_N$ at finite $N$, are shown in Fig.~\ref{fig:cprefactor}: close to $W_c$, where $\Phi_0$ tends to a finite value, they vanish linearly in $\kappa$, as predicted by Eq.~\eqref{eq:I2kappasmall}, while deeper in the localized phase the decrease of $\Phi_0$ makes them depend only weakly on $W$. Since larger values of $W$ correspond to larger values of $x$, a common amplitude makes the collapsed curve look steeper, and the fit compensates with $a > 1$. A fit with a common shape parameter $a$ for all $W$ but an amplitude free at each $W$ separates the two effects [Fig.~\ref{fig:FSS}(c)]. Since Eq.~\eqref{eq:FSS} describes the asymptotic regime, up to corrections that decay as $L$ grows, we repeat this fit while progressively excluding the largest values of $\eta$ or the smallest sizes, i.e.\ restricting the data to $L \ge L_{\min}$ with $L_{\min} = \ln(1/\eta_{\max})$ or $\ln N_{\min}$: a value of $a$ that does not depend on $L_{\min}$ is free of such corrections, while a drift with $L_{\min}$ reveals them. At finite $\eta$ the fit gives $a$ between $0.96$ and $0.98$ for every cutoff $\eta_{\max}$ between $3 \times 10^{-3}$ and $10^{-4}$: the shape of the scaling function is the predicted one, with no visible drift. At finite $N$ it gives $a = 1.22$ when all sizes $N \ge 2^{10}$ are included, and $a$ decreases towards one, down to $1.07 \pm 0.13$, as the smallest sizes are progressively excluded.

The scaling form also explains why the pure power-law fits of Figs.~\ref{fig:I2eta} and~\ref{fig:kappa} give exponents systematically larger than $2\beta - 1$ close to $W_c$, at finite $\eta$ and, even more, at finite $N$. Close to $W_c$, $\kappa$ is small, and the accessible values of $L$ do not reach the regime $x = \kappa L \gg 1$ where Eq.~\eqref{eq:FSS} reduces to a pure power law. For $x = O(1)$ the term $-1$ in the denominator of $\Phi(x)$ cannot be neglected, and the local decay rate $-{\rm d}\ln[I_2 - I_2^\infty]/{\rm d}L$ is $a\kappa/(1 - e^{-a\kappa L})$, larger than $\kappa$. A power-law fit over this crossover region therefore returns an apparent exponent larger than $2\beta - 1$, while Eq.~\eqref{eq:FSS}, with $\kappa = 2\beta - 1$, describes the data.

Finally, the different behavior of $a$ at finite $\eta$ and at finite $N$ shows that the correspondence $\eta \leftrightarrow 1/N$ between the two problems holds asymptotically, for $\eta \to 0$ and $N \to \infty$, but that at finite $N$ it is affected by subleading corrections. These make the effective decay with $N$ slightly steeper than predicted by Eq.~\eqref{eq:FSS} with $a = 1$, and decrease only slowly with $N$. They have the same origin as the slow drift with $N$ of the constant $A$ relating the two master curves in Fig.~\ref{fig:overlay}, and as the excess of the fitted $\kappa_N$ over $\kappa_\eta$ in Fig.~\ref{fig:kappa}.

Equation~\eqref{eq:FSS} also provides a practical extrapolation procedure. Fitting the exact-diagonalization data alone with $N \ge 2^{11}$, with $I_2^\infty(W)$ left free at each $W$ and a common $\Phi_0$ and $a$, gives $I_2^\infty(W)$ in agreement with the population-dynamics values to within $2$ to $5 \times 10^{-4}$ for all $W$ between $20.5$ and $34$. The critical value $I_2^{\infty,{\rm crit}}$ is less constrained by these data: the population-dynamics values follow $I_2^{\infty,{\rm crit}} + c_{I_2}\,(W - W_c)^{1/2}$ closely over $20.5 \le W \le 34$, with $I_2^{\infty,{\rm crit}} \approx 0.30$, but the deviations from the pure square root, of order $5 \times 10^{-3}$, exceed the statistical errors by more than an order of magnitude, and the value of $I_2^{\infty,{\rm crit}}$ extracted from this range depends on how they are modeled. A precise determination of the critical value, and a test of the critical branch $x \to 0$ of Eq.~\eqref{eq:FSS}, require data closer to $W_c$.

In summary, the finite-size scaling of the IPR in the localized phase is controlled by the variable $(2\beta - 1)\ln N$, i.e.\ by a length $\xi_\kappa = 1/(2\beta - 1) \propto (W - W_c)^{-1/2}$, and crosses over to a $1/\ln N$ correction at the critical point; extrapolations to the thermodynamic limit should use Eq.~\eqref{eq:FSS} with $\kappa = 2\beta - 1$, which is known independently.\footnote{The statistical errors used in the fits treat the eigenstates of the same sample as independent and are probably somewhat underestimated, which does not affect these conclusions.}

\subsection{Consequences for previous finite-size analyses} \label{sec:previous}

Since Eq.~\eqref{eq:FSS} gives the size dependence of the IPR for all $N$ and $W$ in the localized phase and at the critical point, it allows one to reexamine analytically the finite-size quantities on which previous studies of the Anderson transition on random regular graphs were based. Besides the choice of the scaling variable, with $\nu = 1/2$ and $\zeta = 1$, discussed after Eq.~\eqref{eq:FSSlinear}, we consider here two of them: the renormalization-group (RG) flow of a running fractal dimension~\cite{vanoni2023renormalization}, and the fractal dimensions of finite graphs and their derivative with respect to disorder~\cite{pino2020scaling}. We work with $I_2$, for which the scaling form has been tested, but the whole picture below holds for any moment $I_q$ with $q > \beta$. The correction to the bulk of the distribution of the amplitudes factorizes into an $N$-dependent amplitude times a fixed function (Secs.~\ref{sec:zeromode} and~\ref{sec:Tu}), whose moment $\int g_\eta(\hat m)\, \hat m^{1-q}\, \de\hat m$ converges at large $\hat m$ for $q > \beta$. For these $q$, $I_q - I_q^\infty$ therefore has the same scaling function, the same scale $1/\kappa$ and the same crossover to the critical $1/\ln N$ behavior as $I_2 - I_2^\infty$, and differs from it only by a $q$-dependent amplitude [for integer $q$ this also follows from Eq.~\eqref{eq:spectralq}, and is checked numerically for $q = 3$ and $4$ in Sec.~\ref{sec:corrtests}]. At leading order the running dimensions $D_q = -{\rm d}\ln I_q/[(q-1)\,{\rm d}\ln N]$, including $D_1$, obtained from the participation entropy ($q \to 1$), and $D_\infty$, are thus all proportional to each other, with $N$-independent ratios, and they have the same flow function $\mathcal{B}$ defined below: the flow of $D_q$ is that of $D_2$ derived below, up to a $q$-dependent value of the constant $g$, and in particular the line of fixed points, reached at $\mathcal{B} = -(2\beta - 1)$, is the same for all $q > \beta$.

\begin{figure}[t]
\centering
\includegraphics[width=0.46\textwidth]{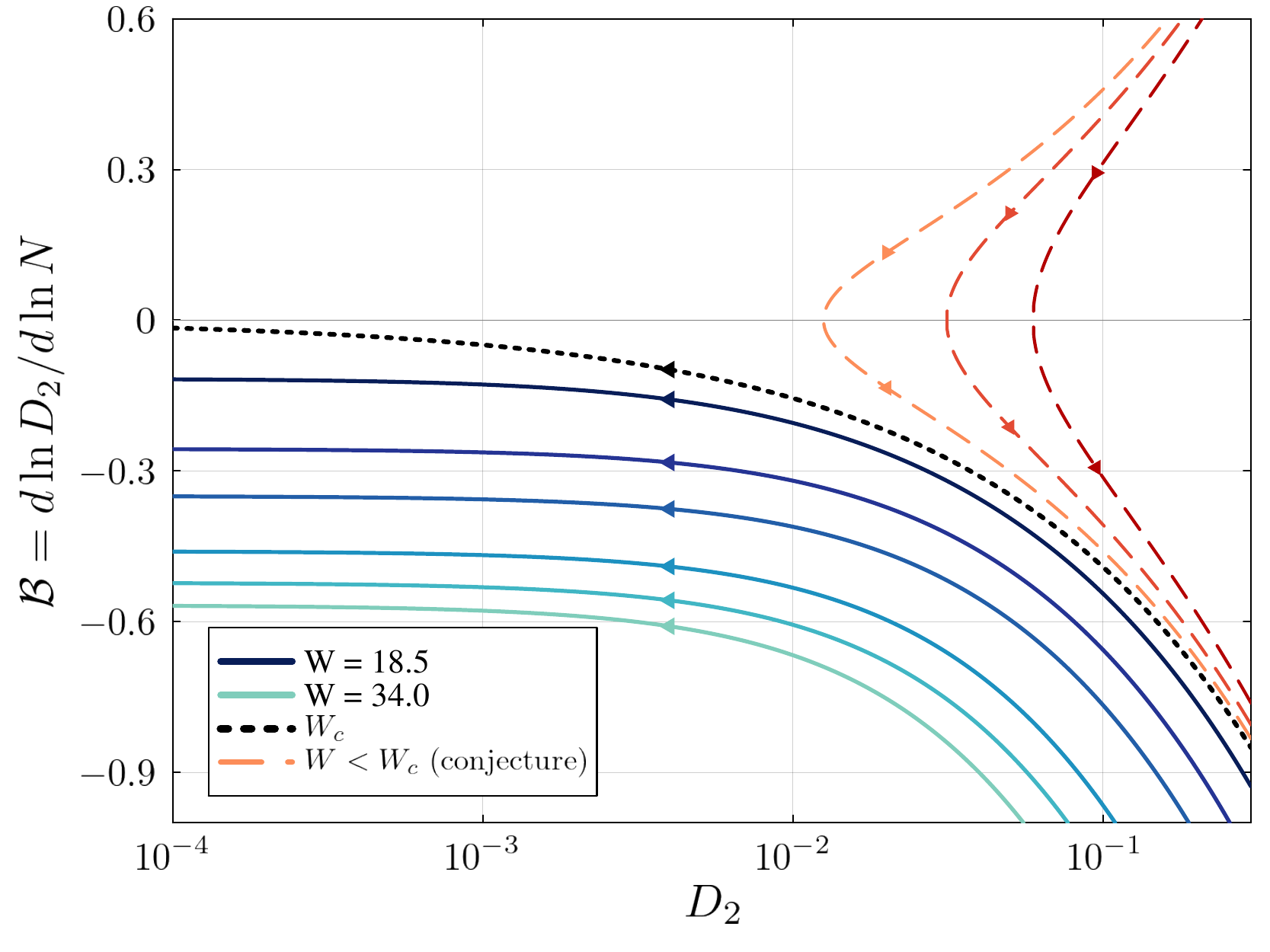}
\caption{Flow of the running fractal dimension $D_2 = -{\rm d}\ln I_2/{\rm d}\ln N$ in the $(D_2, \mathcal{B})$ plane, with $\mathcal{B} = {\rm d}\ln D_2/{\rm d}\ln N$, from Eq.~\eqref{eq:flowinv}. No data are shown; in the localized phase the parameters $\kappa = 2\beta - 1$, $I_2^\infty$ and $\Phi_0$ are those obtained from the population dynamics at each $W$ (Sec.~\ref{sec:FSS}). Solid lines: localized phase, $W = 18.5$, $20$, $22$, $26$, $30$ and $34$ (top to bottom); each trajectory ends on the line of fixed points $D_2 = 0$ at $\mathcal{B} = -(2\beta - 1)$. Dotted line: critical point, $\mathcal{B} = -(2gD_2)^{1/2}$. Dashed lines: conjectured continuation to the extended side, $\kappa^2 = -\tilde\kappa^2$, for $W = 17.5$, $16.5$ and $15$ (left to right), with $\tilde\kappa^2 = 4c_\beta^2 (W_c - W)$ and $g$ fixed at its value at $W_c$: they cross $\mathcal{B} = 0$ at the minimum $D_A$ of $D_2(N)$, which is not a fixed point, and then flow towards the ergodic fixed point $D_2 = 1$, $\mathcal{B} = 0$, outside the range of the figure and of validity of Eq.~\eqref{eq:FSS}. Triangles indicate the direction of increasing $N$.}
\label{fig:flow}
\end{figure}

\paragraph{RG flow of the running fractal dimension.}
Following Ref.~\cite{vanoni2023renormalization}, we define a running fractal dimension and its flow function,
\begin{equation} \label{eq:D2run}
    D_2(N) = -\frac{{\rm d} \ln I_2}{{\rm d} \ln N} \, , \qquad \mathcal{B} = \frac{{\rm d} \ln D_2}{{\rm d} \ln N} \, .
\end{equation}
Reference~\cite{vanoni2023renormalization} uses $D_1$, i.e.\ the participation entropy instead of $\ln I_2$, which only changes the value of the constant $g$ below. Inserting Eq.~\eqref{eq:FSS} with $a = 1$ and $L = \ln N$, we obtain, at leading order in $(I_2 - I_2^\infty)/I_2^\infty$,
\begin{equation} \label{eq:D2flow}
    D_2 = \frac{\Phi_0}{I_2^\infty}\, \frac{\kappa^2}{4 \sinh^2(\kappa L/2)} \, , \qquad \mathcal{B} = -\kappa \coth\frac{\kappa L}{2} \, .
\end{equation}
Eliminating $L$ gives
\begin{equation} \label{eq:flowinv}
    \mathcal{B}^2 - 2 g\, D_2 = \kappa^2 \, , \qquad \frac{{\rm d}\mathcal{B}}{{\rm d} \ln N} = g\, D_2 \, , \qquad g = \frac{2 I_2^\infty}{\Phi_0} \, .
\end{equation}
These are the flow equations proposed phenomenologically in Ref.~\cite{vanoni2023renormalization} [their Eqs.~(15) and~(16), $\frac{1}{2}\mathcal{B}^2 = c\,(D - D_A)$, with a constant $c$ taken there as independent of $W$]. Here they follow from the theory, in the localized phase and at the critical point, with constants fixed by $\kappa$, $I_2^\infty$ and the amplitude $\Phi_0$ of the scaling function:
\begin{itemize}
    \item $g$, which increases from $g \approx 1.2$ at $W_c$ to $g \approx 6$ at $W = 34$ (with $\Phi_0$ from the fits of Sec.~\ref{sec:FSS});
    \item $D_A = -\kappa^2/(2g) = -\Phi_0 \kappa^2/(4 I_2^\infty)$, which is negative and vanishes linearly in $W - W_c$, since $\kappa^2 \propto W - W_c$, i.e.\ as the continuation to the localized side of the linear dependence found on the extended side in the fits of Ref.~\cite{vanoni2023renormalization} [their Eq.~(13)].
\end{itemize}
The resulting flow is shown in Fig.~\ref{fig:flow}. In the localized phase each trajectory ends on the line of fixed points $D_2 = 0$, where $\mathcal{B} = -\kappa$, i.e.\ $D_2 \propto N^{-(2\beta - 1)}$.

Reference~\cite{vanoni2023renormalization} anticipated this structure by analogy with the low-temperature phase of the Berezinskii--Kosterlitz--Thouless (BKT) transition~\cite{berezinskii1971destruction,berezinskii1972destruction,kosterlitz1973ordering,kosterlitz1974critical}, to which the Anderson transition on random regular graphs had also been compared in Refs.~\cite{garcia2022critical,garcia2026connecting}, and the analogy is in fact exact. With $x = -\mathcal{B}$, $y = (2 g D_2)^{1/2}$ and $\ell = \frac{1}{2}\ln N$, Eq.~\eqref{eq:flowinv} becomes ${\rm d}x/{\rm d}\ell = -y^2$, ${\rm d}y/{\rm d}\ell = -xy$, which are the RG equations of the BKT transition~\cite{kosterlitz1974critical,jose1977renormalization}, with the invariant $x^2 - y^2 = \kappa^2$:
\begin{itemize}
    \item the localized phase is the analogue of the low-temperature phase, with the fixed line $y = 0$, $x = \kappa$;
    \item the critical point is the separatrix $x = y$, on which $D_2 \simeq \Phi_0/(I_2^{\infty,{\rm crit}} \ln^2 N)$ and $\mathcal{B} = -2/\ln N = -(2 g D_2)^{1/2}$ (dotted line in Fig.~\ref{fig:flow}). This is the marginal flow, with the $1/\ln^2 N$ behavior of the running fractal dimension at $W_c$, found in Ref.~\cite{vanoni2023renormalization}.
\end{itemize}
What the theory adds to Ref.~\cite{vanoni2023renormalization} is the identification of their parameter $D_A$, which fixes the value of $\mathcal{B}$ on the line of fixed points, $-(-2gD_A)^{1/2} = -(2\beta - 1)$. This is the only point on which our results differ from those of Ref.~\cite{vanoni2023renormalization}, where this value was related to a localization length, $\mathcal{B} = -1/(\xi \ln k)$, with $\xi$ defined from the expected number of resonances. This is an estimate of the same type as the one based on the average decay discussed in Sec.~\ref{sec:resonances}. With $\xi = \xi_{\rm loc}$ it gives $0.16$ instead of $0.35$ at $W = 22$, and close to $W_c$ it vanishes linearly in $W - W_c$ instead of as $(W - W_c)^{1/2}$. This linear behavior is also difficult to reconcile with the numerical results of Ref.~\cite{vanoni2023renormalization} themselves: if their flow equations hold on both sides of the transition, with the linear dependence of $D_A$ on $W - W_c$ found in their fits on the extended side continued to the localized one, the value on the fixed line, $-(-2gD_A)^{1/2}$, has a square-root behavior, i.e.\ the exponent $2\beta - 1$. The same conclusion has been reached independently in Ref.~\cite{scardicchio2026conformality}.

The identification of $D_A$ also clarifies the structure of the flow. In the localized phase the flow cannot be reduced to a function $\mathcal{B}(D_2)$ of a single variable: each trajectory is labeled by $D_A$, i.e.\ by $\kappa$, which is conserved along it. The two-parameter flow~\eqref{eq:flowinv} is thus the differential form of Eq.~\eqref{eq:FSS}. Although $N$ enters Eq.~\eqref{eq:FSS} only through the variable $\kappa\ln N$, besides the critical factor $1/\ln N$, $\mathcal{B}$ is not a function of $D_2$ alone, since $\mathcal{B} \to -\kappa(W)$ as $D_2 \to 0$: it is the line of fixed points, i.e.\ the continuously varying exponent $\kappa$, that makes the flow two-parameter. In the variables $(x, y)$ the flow is autonomous, while in the $(D_2, \mathcal{B})$ plane of Fig.~\ref{fig:flow} the constant $g$ enters as a $W$-dependent parameter, which does not flow. The trajectories~\eqref{eq:D2flow} have exactly the form of those of the low-temperature phase of the BKT transition and, as there, a collapse in a single variable obtained once the critical behavior has been divided out does not make the flow one-parameter.
Since $\mathcal{B}$ involves the second logarithmic derivative of $I_2$, it is a much less precise numerical diagnostic than the collapse of Fig.~\ref{fig:FSS}; our finite-$N$ and finite-$\eta$ data are consistent with Eq.~\eqref{eq:flowinv} within their accuracy (not shown).

\paragraph{A conjecture for the extended side.}
For $W < W_c$ the two solutions $\beta$ and $1 - \beta$ of $\lambda(\beta) = 1$ [Eq.~\eqref{eq:fbeta}], which merge at $W_c$, become complex, $\beta = \frac{1}{2} \pm \frac{\rm i}{2}\, \tilde\kappa$, so that $\kappa^2 = (2\beta - 1)^2 = -\tilde\kappa^2 < 0$. Close to $W_c$, $\kappa^2 \simeq 4 c_\beta^2\,(W - W_c)$ continues analytically through the transition. At leading order, $D_2$ and $\mathcal{B}$ in Eq.~\eqref{eq:D2flow} depend on $\kappa$ only through $\kappa^2$: the term $-\Phi_0 \kappa/2$ of Eq.~\eqref{eq:FSS}, odd in $\kappa$, does not depend on $L$ and drops out of $D_2$.

It is then natural to conjecture that Eq.~\eqref{eq:flowinv} continues to hold close to $W_c$ on the extended side, with $\kappa^2 = -\tilde\kappa^2$ and $g$ taken at $W_c$ (since $I_2^\infty = 0$ for $W < W_c$), i.e.\ $D_2 = \tilde\kappa^2/[2g\sin^2(\tilde\kappa L/2)]$ and $\mathcal{B} = -\tilde\kappa \cot(\tilde\kappa L/2)$ (dashed lines in Fig.~\ref{fig:flow}):
\begin{itemize}
    \item as $N$ grows, $D_2$ first decreases as at the critical point;
    \item it reaches a minimum $D_A = \tilde\kappa^2/(2g) > 0$, where $\mathcal{B} = 0$, at $\ln N_A = \pi/\tilde\kappa \propto (W_c - W)^{-1/2}$;
    \item it then increases, towards the ergodic value $D_2 = 1$.
\end{itemize}
This is the behavior found numerically in Ref.~\cite{vanoni2023renormalization} on the extended side [their Eqs.~(13), (19) and~(20)], and the length $\ln N_A$ has the exponent $\nu = 1/2$ of the correlation volume on the delocalized side~\cite{tikhonov2019critical}. The point $D_2 = D_A$, $\mathcal{B} = 0$ is not a fixed point, as already noted in Ref.~\cite{vanoni2023renormalization}, since ${\rm d}\mathcal{B}/{\rm d}\ln N = g D_A > 0$ there: the flow crosses it and proceeds towards the ergodic fixed point $D_2 = 1$, $\mathcal{B} = 0$. The minimum of $D_2$ around $N_A$ is thus a finite-size plateau, which can mimic multifractality at accessible sizes (for instance $D_A \approx 0.03$ at $\ln N_A \approx 11$ for $W = 16.5$), and not an intermediate phase: in the thermodynamic limit $D_2 = 1$ for all $W < W_c$, as in the exact solution~\cite{tikhonov2019critical}. The divergence of $D_2$ at $\tilde\kappa L = 2\pi$ signals that the expansion breaks down when $D_2 = O(1)$; the subsequent flow towards ergodicity is not described by Eq.~\eqref{eq:FSS}. The merging of the two roots $\beta$ and $1 - \beta$ at $W_c$ and their continuation into the complex plane is the scenario known in field theory as ``conformality lost''~\cite{kaplan2009conformality}; a derivation of the flow on both sides of the transition along these lines, from the linearized cavity equations, will be presented in Ref.~\cite{scardicchio2026conformality}.

\paragraph{Fractal dimensions at fixed $N$ and their derivative with respect to disorder.}
Several works characterize finite graphs by the fractal dimensions $\tilde D_q(N) = \ln I_q/[(1 - q)\ln N]$, and Ref.~\cite{pino2020scaling} locates the transition from their derivative ${\rm d}\tilde D_q/{\rm d}W$.
\begin{itemize}
    \item \emph{$\tilde D_2$ itself.} In the localized phase Eq.~\eqref{eq:FSS} gives
\begin{equation} \label{eq:D2tilde}
    \tilde D_2(N) = -\frac{\ln I_2^\infty}{\ln N} - \frac{\Phi(\kappa \ln N)}{I_2^\infty \ln^2 N} + \dots \, .
\end{equation}
    Unlike the running $D_2$ of Eq.~\eqref{eq:D2run}, $\tilde D_2$ contains the term $-\ln I_2^\infty/\ln N$, and therefore vanishes only as $1/\ln N$. For example, at $W = 20.5$, $\tilde D_2 \approx 0.08$ for $N = 3 \times 10^4$ and still $\approx 0.06$ for $N = 10^6$. Finite values of $\tilde D_q$ at accessible sizes on the localized side are thus a logarithmic finite-size effect, and do not signal multifractality of the localized states.
    \item \emph{Its derivative.} The same holds for ${\rm d}\tilde D_2/{\rm d}W \simeq -(\ln N)^{-1}\, {\rm d}\ln I_2^\infty/{\rm d}W$. Through the square-root singularity of $I_2^\infty$, for $\kappa\ln N \gg 1$ it behaves as $(W - W_c)^{-1/2}/\ln N$, which diverges at $W_c$ but with a prefactor that vanishes in the thermodynamic limit. At finite $N$ the singularity is rounded off over the critical window $\kappa \ln N \lesssim 1$. At the edge of this window, $|{\rm d}\tilde D_2/{\rm d}W| \approx c_{I_2} c_\beta / I_2^{\infty,{\rm crit}}$, a few $10^{-2}$,, independently of $N$.
\end{itemize}
A divergence of ${\rm d}\tilde D_q/{\rm d}W$ is therefore not expected at any accessible size, even on the localized side of the transition, where the singularity is known exactly. Its absence, observed in Ref.~\cite{pino2020scaling}, cannot be taken as evidence against a direct transition from localized to ergodic states, or in favor of a multifractal metal.

\section{Conclusions and perspectives} \label{sec:conclusions}

We have derived analytically the anomalous finite-regulator corrections to the IPR in the localized phase of the Anderson model on the Bethe lattice, $I_2(\eta) - I_2^\infty \propto \eta^{2\beta - 1}$, by analyzing a toy NL$\sigma$M on the hyperbolic superplane that shares the non-compact graded symmetry and, as shown by Gruzberg and Mirlin~\cite{Gruzberg_1996}, the criticality of the true Anderson transition on the BL. The mechanism that emerges is simple and, we believe, general. In the $\eta \to 0^+$ ($H \to 0$) limit the cavity equation linearized around its fixed point has two power-law solutions at the tail, with exponents $\beta$ and $1-\beta$ symmetric about $1/2$; the unperturbed solution is controlled by the larger one. A small regulator removes a weight $\propto H^{\beta}$ (respectively $\eta^\beta$) from the tail of the distribution, in a boundary layer at the scale set by the regulator. This perturbation cannot propagate into the bulk along the dominant mode, which is fixed by the matching with the unperturbed solution; it propagates along the subdominant one, $e^{(1-\beta)t}$ (respectively $\hat m^{-(2-\beta)}$), and is amplified by $H^{-(1-\beta)}$ between the boundary layer and the bulk. The bulk correction is therefore a zero mode of the linearized recursion, with a universal shape and an amplitude $\propto H^{2\beta - 1}$, and so is the correction to any observable dominated by the bulk, the IPR in particular. At the critical point the two roots merge, the zero mode acquires a logarithmic prefactor, and the correction becomes $\propto 1/|\ln H|$; the crossover between the two behaviors occurs at $|\ln H| \sim (\beta - 1/2)^{-1}$. In the toy model every step of this chain is explicit, all constants are computed from the cavity equation, and the predictions are verified against the numerical solution of the cavity equation with no adjustable parameter.

For the Anderson model we have shown that the same chain of arguments goes through under three explicit assumptions: the existence of a zero mode of the linearized $\eta = 0$ recursion, its decay with the subdominant root $1 - \beta$ of the Abou-Chacra--Anderson--Thouless condition, and the dominance of a single neighbor in the tails. The second assumption is the one with the richest consequences, and the numerical results of Sec.~\ref{sec:numerics} support it in two independent ways: the finite-$\eta$ correction to the distribution of the LDoS at the resonant sites collapses onto a single master function when rescaled by $\eta^{2\beta-1}$, and the finite-size correction to the distribution of the eigenstates' amplitudes decays at small $u$ as $u^{-1-\beta}$, the image of $\hat m^{-(2-\beta)}$ under $u = 1/\hat m$; the two scaling functions, compared directly at the same disorder strength, turn out to be the same function in the appropriate variables, up to a multiplicative constant close to one, which drifts slowly with $N$ because of subleading finite-size corrections (Sec.~\ref{sec:FSS}). As a by-product, the structure of the recursion beyond the boundary layer, where it acts as an inversion rather than as an amplification, predicts the exponent $\beta - 3$ for the far tail of the LDoS distribution at the resonant sites at finite $\eta$, in agreement with the numerics, and the full shape of this tail in terms of the $\eta = 0$ solution.

The anomalous corrections also have a direct real-space interpretation. By the spectral identities~\eqref{eq:spectral} and~\eqref{eq:spectralq} and the sum rule~\eqref{eq:sumrule}, the finite-$\eta$ corrections of all the moments $I_q$, and the finite-$N$ corrections of the IPR, are entirely due to pairs of eigenstates close in energy: a finite $\eta$ does not resolve the pairs with splittings smaller than $\eta$, and a finite graph lacks the rarest ones. These relations are exact: the exponent $\kappa = 2\beta - 1$ translates into the exponent $\mu = 2(1-\beta)$ of their correlation function, $N^2\mathcal{C}(\omega) \propto \omega^{-\mu}$, and of its generalizations to higher powers of the amplitudes, and the amplitude of the correction into that of the correlation function, so that the derivation of $\kappa$ is also a derivation of $\mu$. Exact diagonalization and population dynamics directly at $\eta = 0^+$ confirm this prediction, including the amplitude, with no adjustable parameter, and in the toy model continuing the cavity solution to complex $H$ gives the same law, with its amplitude, analytically (Sec.~\ref{sec:complexH}). What neither calculation reveals is the physical origin of the corrections, which a phenomenological, Mott-type counting of resonances makes transparent, although it fixes only their exponent, not their amplitude. The tail of the LDoS distribution, which fixes $\beta$, is produced by single eigenstates localized far from a given site, whose exponential tail reaches it with a small weight $u$, so that $\beta$ is the exponent of the small-amplitude tail $T_\infty(u) \propto u^{\beta-2}$ of the amplitude distribution; a state localized at the site hybridizes with such a distant state when their energy separation is smaller than their coupling $\sim \sqrt{u}$, and counting these resonances with the exact amplitude statistics reproduces $\mu = 2(1-\beta)$ --- the estimate of Ref.~\cite{tikhonov2021eigenstate}, with the average decay of the wavefunctions replaced by their full statistics. Exact diagonalization confirms this picture eigenstate by eigenstate: the resonant pairs share two localization centers at a distance that grows as $\ln(1/\omega)$, and their density follows $\omega^{-2(1-\beta)}$; on a finite graph the doubly-peaked eigenstates whose centers would lie farther apart than the typical inter-site distance $\sim \ln N/\ln k$ are missing, replaced by compact states with a larger IPR, and this population shift accounts for most of the finite-size correction. The rare events in the tail of the LDoS distribution and the rare Mott-like resonances between distant localized eigenstates are thus two faces of the same phenomenon. Finally, the number of resonant partners with splitting smaller than $\eta$, $\propto \eta^{2\beta-1}$, vanishes as $\eta \to 0$ only for $\beta > 1/2$: in this language the localization condition $\beta > 1/2$ is the stability of the localized states against the proliferation of resonances~\cite{anderson1958absence,abou1974self}.

Finally, the same mechanism fixes the finite-size scaling of the IPR in the localized phase and at the critical point, Eq.~\eqref{eq:FSS}: the scaling variable is $(2\beta - 1)\ln N$, i.e.\ the scaling length is $1/\kappa \propto (W - W_c)^{-1/2}$, the critical value is approached as $1/\ln N$, and the same form describes the finite-$\eta$ and the finite-$N$ data. Its differential form is the two-parameter renormalization-group flow of the running fractal dimension proposed phenomenologically in Ref.~\cite{vanoni2023renormalization}, which therefore turns out to be exact, and exactly of the Kosterlitz type. The exponent $\kappa = 2\beta - 1$ is conserved along each trajectory, as the invariant of the Kosterlitz equations, so that the flow can be integrated exactly; its constants are fixed by $\kappa$, $I_2^\infty$ and the amplitude $\Phi_0$ of the scaling function, and each trajectory ends on the line of fixed points $D_2 = 0$, reached at $\mathcal{B} = -(2\beta - 1)$ (Sec.~\ref{sec:previous}). In particular, the parameter $D_A$ of the phenomenological flow is identified as $-\kappa^2/(2g)$, and it is this value that fixes the exponent on the line of fixed points. The same form shows that finite fractal dimensions of the localized states at accessible sizes, and the absence of a divergence of their derivative with respect to disorder at $W_c$, are logarithmic finite-size effects, which do not signal an intermediate multifractal phase.

Several directions are left open. On the analytical side, the natural next step is to repeat the scaling analysis for the supersymmetric NL$\sigma$M of the Anderson transition~\cite{efetov1985anderson,zirnbauer1986localization,mirlin1991localization} and, ultimately, for the Anderson model itself, where the self-consistent equations act on a two-variable distribution and the existence of the zero mode, which we have assumed, should be established; now that the mechanism is understood, this seems within reach, and would also give access to the amplitudes $c_\eta$ and $c_N$ of the power laws, which we find empirically to depend only weakly on $W$ deep in the localized phase, and which vanish $\propto 2\beta - 1$ close to $W_c$, where the power-law regime $\ln N \gg 1/\kappa$ moves to ever larger sizes (Sec.~\ref{sec:FSS}). The extension to the higher moments $I_q$ is straightforward as long as the moment $\int g_\eta(\hat m) \hat m^{1-q} \de \hat m$ converges at large $\hat m$, which with the tail $\hat m^{-(2-\beta)}$ requires $q > \beta$: all integer moments then share the exponent $2\beta - 1$ [as found from the eigenstate correlations, Eq.~\eqref{eq:spectralq}, and checked numerically for $q = 3$ and $4$ in Sec.~\ref{sec:corrtests}], while for $q < \beta$ the correction is dominated by the boundary layer and acquires a different, $q$-dependent exponent. The continuation of the scaling form to the extended side, which we have conjectured from the analytic continuation $\kappa^2 \to -\tilde\kappa^2$ of the flow and which reproduces the non-monotonic behavior of the running fractal dimension observed numerically there~\cite{vanoni2023renormalization}, would extend it to both sides of the transition; a derivation, within the ``conformality lost'' scenario~\cite{kaplan2009conformality}, will be presented in Ref.~\cite{scardicchio2026conformality}. The critical branch of the scaling form, the $1/\ln N$ law, counterpart of Eq.~\eqref{eq:logcorr} for the Anderson model, and the precise value of the IPR at $W_c$ deserve a dedicated numerical study of the Anderson model closer to $W_c$, since they bear directly on the long-standing difficulty of extrapolating IPR data to the thermodynamic limit near the transition~\cite{tikhonov2019statistics,garcia2022critical,10.21468/SciPostPhys.15.2.045,garcia2017scaling,chen2023quantum,chen2024describing,vanoni2023renormalization,garcia2026connecting,de2014anderson,altshuler2016nonergodic,tikhonov2016anderson}. Finally, the correspondence between finite-$\eta$ and finite-$N$ corrections, which we have exploited here at the level of the bulk zero mode, is discussed from the real-space point of view --- the loops of a finite graph exciting the same zero mode as a regulator of order $1/N$ --- in the companion Letter~\cite{letter}; understanding how the amplitude $c_N$ is fixed in the finite-$N$ boundary layer, where the two problems differ, would close the circle. In real space, how the cutoff of the resonances on a finite graph (Sec.~\ref{sec:finiteN}) fits with the long loops that generate the finite-$N$ corrections in the one-loop scheme~\cite{letter}, why it sets in at splittings of the order of the mean level spacing $1/(N\avg{\rho})$, and how it produces precisely $N^{-\kappa}$, remain to be understood analytically, as does a controlled version of the resonance counting itself.

\begin{acknowledgments}
We thank W.~Chen, V.~E.~Kravtsov, G.~Lemari\'e, A.~D.~Mirlin, A.~Scardicchio and C.~Vanoni for invaluable discussions. We acknowledge financial support from the ANR research grant ManyBodyNet ANR-24-CE30-5851.
\end{acknowledgments}

\appendix

\section{Numerical solution of the cavity equations of the toy model} \label{app:numerics}

In this appendix we describe the algorithms used to compute the functions $P_H(t)$, $P_0(t)$, $\ft(t)$ and $\fm(\Delta t)$ of Secs.~\ref{sec:scaling} and~\ref{sec:critical}. The codes (two Mathematica notebooks, for $J<J_c$ and for $J=J_c$, and a faster C++ implementation of the critical-point algorithm (Algorithm~5), used for the runs at the smallest fields, $\ln H = -50$ and $-54$) are available in the public repository of Ref.~\cite{coderepo}. All results are for $k=2$; the only place where this choice enters explicitly is the expansion of $P^k$ in Algorithm~5, whose generalization to any $k$ is straightforward. The two constants that characterize the linearized problem are obtained by root finding: $\beta$ from Eq.~\eqref{eq:betaJ} and $J_c$ from $k\,K_0(J_c) = K_{-1/2}(J_c)$, which give $\beta = 0.7530159$ for $J=0.01$ and $J_c = 0.0265402$.

All equations below are written in terms of the kernel $K_H(t,t')$ of Eq.~\eqref{eq:KH}; at $H = 0$ it reduces to $L_0(t-t')$, Eq.~\eqref{eq:L0}.

\subsection{Common ingredients}

\begin{enumerate}
\item \textbf{Grid.} $t_i = t_{\rm min} + i\, \delta t$, $i = 0, \ldots, M$, on the interval $[t_{\rm min}, t_{\rm max}]$, with $t_{\rm max} = 10$ and $\delta t = |t_{\rm min}|/N_t$ chosen close to a target value ($\delta t \simeq 0.08$ in most runs), so that both $t_{\rm min}$ and $t=0$ are grid points.

\item \textbf{Exponent $y$.} The definition $y = J e^{t+t'}(s s' - 1)/H^2$, with $s \equiv s(t)$, $s' \equiv s(t')$, is the difference of two terms of order $J e^{t+t'}/H^2$, which diverges as $H \to 0$. It is evaluated in the equivalent form
\begin{equation}
    y(t,t') = - J \, \frac{e^{t-t'} + e^{t'-t} - H^2 e^{-t-t'}}{1 + s(t)\, s(t')} \, ,
\end{equation}
which is free of cancellations and makes the limit $y \to -J\cosh(t-t')$ manifest.

\item \textbf{Bessel factor.} $e^{F(x)} = \sqrt{2\pi x}\, e^{-x}\, \tilde I_1(x)$ is computed as $\exp[\ln \tilde I_1(x) - x + \tfrac12 \ln(2\pi x)]$ for $\ln x < 15$, and from the asymptotic expansion $e^{F(x)} \simeq 1 - 3/(8x)$ for $\ln x \ge 15$; this avoids the overflow of $\tilde I_1(x)$ at $x = O(H^{-2})$.

\item \textbf{Endpoint $t' = t_{\rm min}$.} There $s(t') = 0$ and $x = 0$: the integrand is of the form $0/0$ (the factor $s'^{-3/2}$ diverges while $\tilde I_1(x) \simeq x/2$ vanishes). Its finite limit is inserted analytically,
\begin{equation}
    K_H(t,t_{\rm min}) = \tfrac12 \, e^{J} s^2(t) \left( \frac{J e^t}{H} \right)^2 e^{-J e^t/H} \, .
\end{equation}

\item \textbf{Quadrature.} Integrals over $t'$ are computed with an end-corrected  trapezoidal rule, $\int \de t' g(t') \simeq \delta t \sum_j w_j\, g(t_j)$, with weights $w_j = \tfrac{95}{288}, \tfrac{317}{240}, \tfrac{23}{30}, \tfrac{793}{720}, \tfrac{157}{160}$ on the first five nodes $j=0,\ldots,4$ and $w_j = 1$ elsewhere. No correction is needed at $t_{\rm max}$, where the integrand is suppressed by $e^{-e^{t'}}$. With three corrected weights ($\tfrac38, \tfrac76, \tfrac{23}{24}$) the error of the solution scales as $\delta t^4$, as we checked with $\delta t = 0.08$, $0.04$, $0.02$ at $t_{\rm min} = -6$; the five-weight rule is expected to be accurate to $O(\delta t^6)$. The exact identity $1 = e^{J(1-e^t/H)} + \int_{t_{\rm min}}^\infty \de t' \, K_H(t,t')$ [the statement that $P \equiv 1$ solves Eq.~\eqref{eq:PtH} when the boundary value and the factor $e^{-e^{t'}}$ are set to one, cf.\ Eq.~\eqref{eq:zeroth}] provides a stringent test of the endpoint treatment and of the weights: it is satisfied to $10^{-8}$ on the grid $\delta t = 0.08$ at $t_{\rm min} = -20$.

\item \textbf{Fixed-point iteration.} Picard iteration $P^{(n+1)} = \mathcal{T}_H[P^{(n)}]$, where $\mathcal{T}_H$ is the discretized right-hand side of Eq.~\eqref{eq:PtH}, stopped when $\max_i |P^{(n+1)}(t_i) - P^{(n)}(t_i)| < \epsilon$.
\end{enumerate}

\subsection{Algorithm 1: $P_H(t)$ for $J < J_c$}

\begin{enumerate}
\item Initialize $P^{(0)}(t_i) = e^{H/(k-1)} e^{-e^{t_i}}$, or a converged solution at a nearby $H$.
\item Fix the boundary value $P(t_0) = P(t_{\rm min}) = e^{H/(k-1)}$.
\item For $i = 1, \ldots, M$ update
\begin{equation*}
\begin{aligned}
    P^{(n+1)}(t_i) = \, & e^{J(1-e^{t_i}/H)} e^{\frac{H}{k-1}} \\
    & + \delta t \sum_{j=0}^{M} w_j \, K_H(t_i,t_j)\, e^{-e^{t_j}} \big[P^{(n)}(t_j)\big]^k ,
\end{aligned}
\end{equation*}
using steps 2--4 of the common ingredients for the kernel.
\item Iterate until $\epsilon = 10^{-10}$; compute $I_2(H) = \int_{t_{\rm min}}^\infty \de t \, e^{-e^t} e^t P_H^{k+1}(t)$ on the same grid [this is Eq.~\eqref{eq:I2toy} in the variable $t$, up to the regular factor $e^{-2H/(k-1)} = 1 - O(H)$, which is dropped].
\end{enumerate}
The runs of Figs.~\ref{fig:1mpt}, \ref{fig:res1mPtdt} and~\ref{fig:DI2vsH} use $J=0.01$, $\delta t \simeq 0.08$ and $t_{\rm min} = -6, \ldots, -16$.

The same algorithm, run in complex arithmetic, gives the solution at complex $H = |H|\,e^{-{\rm i}\chi}$ of Sec.~\ref{sec:complexH} and Appendix~\ref{app:complexH}. We keep the real variable $t = \ln(|H|\cosh\theta)$ and the real kernel $K_{|H|}(t,t')$, so that $H$ enters only through the boundary value and the source term, both $\propto e^{H/(k-1)}$, and through the factor $\exp(-e^{t'-{\rm i}\chi})$ that replaces $e^{-e^{t'}}$ and oscillates at large $t'$; the grid spacing is then reduced as $1/(e^{t'}\sin\chi)$ at large $t'$ (about ten points per oscillation, with the end-corrected weights applied in the mapped variable), the Picard iteration is replaced by Newton's method, and the factor $e^{-2H/(k-1)}$ is kept in $I_2(H)$, as required for $I_2(H) \to 1$ at large $H$. At $\chi = 0$ this reproduces the real-$H$ solution to twelve digits, refining the grid changes $I_2$ by less than $10^{-8}$, and we have used it up to $\chi = 0.496\pi$.

\subsection{Algorithm 2: $P_0(t)$ and the constant $c$}

At $H=0$ the source term vanishes, $s \equiv 1$, $e^F \equiv 1$ and $K_H(t,t') \to L_0(t-t')$, Eq.~\eqref{eq:L0}. Equation~\eqref{eq:Pt0} is then posed on the whole real axis, and the left boundary condition must be supplied through the asymptotic form~\eqref{eq:P0asympt}, whose prefactor $c$ is unknown.
\begin{enumerate}
\item Grid on $[t_-, t_{\rm max}] = [-60, 10]$ with $\delta t = 0.16$; initialize $P^{(0)}(t) = e^{-e^t}$.
\item Choose a trial value of $c$ and impose $P_0(t) = 1 - c\, e^{\beta t}$ on the strip $t_- \le t \le t_- + 10$ at every iteration.
\item Outside the strip update $P^{(n+1)}(t_i) = \delta t \sum_j L_0(t_i - t_j)\, e^{-e^{t_j}} [P^{(n)}(t_j)]^k$ (the integrand is negligible at both ends, so that the plain rectangle rule is accurate).
\item Iterate until $\epsilon = 10^{-7}$.
\item Check the junction at $t_- + 10$: the logarithmic derivative of $1-P_0(t)$ must remain equal to $\beta$ across it. Adjust $c$ and repeat until this holds; for a wrong $c$ the iteration still converges, but to a solution in which the local exponent of $1-P_0(t)$ deviates from $\beta$ within a kernel range of the junction.
\end{enumerate}
This gives $c = 0.8551$ for $J = 0.01$. The strip must lie deep enough in the tail, because $1 - P_0$ also contains a subleading term $\propto e^{t}$, whose size relative to $c\,e^{\beta t}$, $e^{(1-\beta)t}$, decays slowly: with the junction at $t = -30$ one would find $c \approx 0.853$. At $J = J_c$ the same procedure, with $\beta = 1/2$, gives $c = 0.67573$. Since $P_0$ is computed on a coarser grid than $P_H$, the difference $\Delta P(t,H) = P_H(t) - P_0(t)$ of Figs.~\ref{fig:1mpt} and~\ref{fig:DpDP0crit} is formed with $P_0$ interpolated onto the grid of Algorithm 1.

\subsection{Algorithm 3: the bulk zero mode $\ft(t)$}

\begin{enumerate}
\item Same grid as in Algorithm 2, with $P_0$ from Algorithm 2.
\item Impose $\ft(t) = e^{(1-\beta)t}$ on the strip $t_- \le t \le t_- + 10$; this selects the subdominant root and fixes the normalization~\eqref{eq:ftypasympt}.
\item Outside the strip iterate $\ft^{(n+1)}(t_i) = k\, \delta t \sum_j L_0(t_i-t_j)\, e^{-e^{t_j}} P_0^{k-1}(t_j)\, \ft^{(n)}(t_j)$, Eq.~\eqref{eq:lindp}, until $\epsilon = 10^{-7}$.
\end{enumerate}
For $J=0.01$ this gives $\max_t \ft(t) = 0.5710$ and $(k+1)\int \de t\, e^{-e^t} e^t P_0^k \ft = 1.08772$, the constant entering Eq.~\eqref{eq:I2scal}. At $J = J_c$ the functions $\ft$ and $\fm$ (Algorithm 4) are computed in exactly the same way, with $\beta = 1/2$ and with the critical asymptotic forms of Sec.~\ref{sec:critical}, $\ft(t) \simeq -t\, e^{t/2}$ and $\fm(\Delta t) \simeq \Delta t\, e^{\Delta t/2}$, imposed on the strips.

\subsection{Algorithm 4: the boundary-layer function $\fm(\Delta t)$ and $c_3$}

In the variables $\Delta t = t - t_{\rm min}$ the kernel is exactly $H$-independent (Sec.~\ref{sec:boundary}). Equation~\eqref{eq:linfmin} is therefore solved with the machinery of Algorithm 1 at $H=1$, $t_{\rm min}=0$, so that $t \equiv \Delta t$, dropping the factor $e^{-e^{t'}}$ (which tends to one in the boundary layer).
\begin{enumerate}
\item Grid on $[0, \Delta t_{\rm max}] = [0, 40]$ with step $0.16$; three corrected weights ($\tfrac38, \tfrac76, \tfrac{23}{24}$).
\item Choose a trial value of $c_3$ and impose $\fm(\Delta t) = e^{\beta \Delta t} - c_3\, e^{(1-\beta)\Delta t}$, Eq.~\eqref{eq:fminsub}, for $\Delta t > \Delta_\ast = 30$ at every iteration.
\item For $0 < \Delta t \le \Delta_\ast$ iterate $\fm^{(n+1)}(\Delta t_i) = k\, \delta t \sum_j w_j\, K_{H=1}(\Delta t_i, \Delta t_j)\, \fm^{(n)}(\Delta t_j)$, with $K_{H=1}$ the kernel~\eqref{eq:KH} at $H=1$, until $\epsilon = 10^{-5}$. The boundary value is updated as $\fm^{(n+1)}(0) = k e^{-1} \fm^{(n)}(0)$, a contraction that drives it to $\fm(0) = 0$.
\item Check $[e^{\beta \Delta t} - \fm(\Delta t)]\, e^{-(1-\beta)\Delta t}$: it must be flat up to $\Delta_\ast$. Adjust $c_3$ and repeat; for a wrong $c_3$ this function bends close to $\Delta_\ast$.
\end{enumerate}
This gives $c_3 = 0.969181$ for $J=0.01$. Note that in the boundary layer $\tilde x = J e^{\Delta t + \Delta t'} \tilde s \tilde s' = O(1)$ for $\Delta t, \Delta t' = O(1)$, so that $e^{F(\tilde x)}$ differs from one by $O(1)$: the full Bessel function is required here, the asymptotic branch of step 3 of the common ingredients taking over only at large $\Delta t + \Delta t'$. In a run of Algorithm 1 at small $H$ the same happens for $t, t' = O(t_{\rm min})$, while $x = O(H^{-2})$ in the bulk; the switch at $\ln x = 15$ handles both regimes within the same run.

\subsection{Algorithm 5: $P_H(t)$ at $J = J_c$} \label{app:critical}

At the critical point the asymptotic regime of Eq.~\eqref{eq:I2scalcrit} is reached only for $-\ln H \gtrsim 50$ (Fig.~\ref{fig:fmincrit}). In the region $t_{\rm min} < t \le 0$ the deficit $1 - P_H(t) \sim e^{t/2}$ is then as small as $e^{-27} \approx 10^{-12}$: Algorithm 1, which iterates on $P$, computes $1-P$ as the difference of two numbers close to one and retains only a few significant digits there, while the quantities of interest are differences $P_H - P_0$ of such deficits, which must be resolved to better than that. Algorithm 1 can be used directly at $J = J_c$ for $|t_{\rm min}| \lesssim 30$, where $1 - P_H \gtrsim 10^{-7}$; for smaller $H$ we iterate, for $t \le 0$, on the rescaled deficit
\begin{equation}
    \delta\psi(t) \equiv [1 - P_H(t)]\, e^{-t/2} = O(1) \, .
\end{equation}
Its equation follows from the exact normalization of the kernel, $1 = e^{J(1-e^t/H)} + \int_{t_{\rm min}}^\infty \de t' \, K_H(t,t')$, which expresses the fact that $P \equiv 1$ solves Eq.~\eqref{eq:PtH} when the factor $e^{-e^{t'}}$ and the boundary value are set to one [cf.\ Eq.~\eqref{eq:zeroth}]. Subtracting Eq.~\eqref{eq:PtH} from it,
\begin{equation} \label{eq:deficiteq}
\begin{aligned}
    1 - P_H(t) = \, & e^{J(1-e^t/H)} \big[1 - e^{H/(k-1)}\big] \\
    & + \int_{t_{\rm min}}^\infty \de t' \, K_H(t,t') \big[ 1 - e^{-e^{t'}} P_H^k(t') \big] \, ,
\end{aligned}
\end{equation}
and for $k=2$ the bracket is written without cancellations as
\begin{equation*}
\begin{aligned}
    1 - e^{-e^{t'}} P_H^2 = \, & 2\, e^{-e^{t'}/2} \sinh\!\big(e^{t'}/2\big) \\
    & + e^{-e^{t'}} \big[ 2\, e^{t'/2} \delta\psi(t') - e^{t'} \delta\psi^2(t') \big] \, ,
\end{aligned}
\end{equation*}
with $1 - e^{H/(k-1)} = -2\, e^{H/(2k-2)} \sinh[H/(2k-2)]$.
\begin{enumerate}
\item Grid as in Algorithm 1 with $J = J_c$; $\delta t = 0.08$ for $|t_{\rm min}| \le 30$ and $\delta t = 0.02$ for larger $|t_{\rm min}|$.
\item Initialize from a converged solution at a nearby $t_{\rm min}$ (interpolated on the new grid), or from the $H=0$ solution of Algorithm 2 at $J_c$ as $\delta\psi^{(0)}(t) = \delta\psi_0(t)\,(1 - t/t_{\rm min})$ for $t<0$, which already vanishes at $t_{\rm min}$ as required by the matching~\eqref{eq:matching_critical}.
\item For $t_i \le 0$ update $\delta\psi$ with Eq.~\eqref{eq:deficiteq} multiplied by $e^{-t_i/2}$, and set $P_H(t_i) = 1 - e^{t_i/2}\delta\psi(t_i)$; the boundary value is $\delta\psi(t_{\rm min}) = -2 H^{-1/2} e^{H/(2k-2)}\sinh[H/(2k-2)]$.
\item For $t_i > 0$ update $P_H$ with Algorithm 1, step 3, and set $\delta\psi(t_i) = [1 - P_H(t_i)]e^{-t_i/2}$.
\item Iterate until $\epsilon = 10^{-10}$.
\end{enumerate}

\begin{figure*}[t]
\panel{a}{\includegraphics[width=0.48\textwidth]{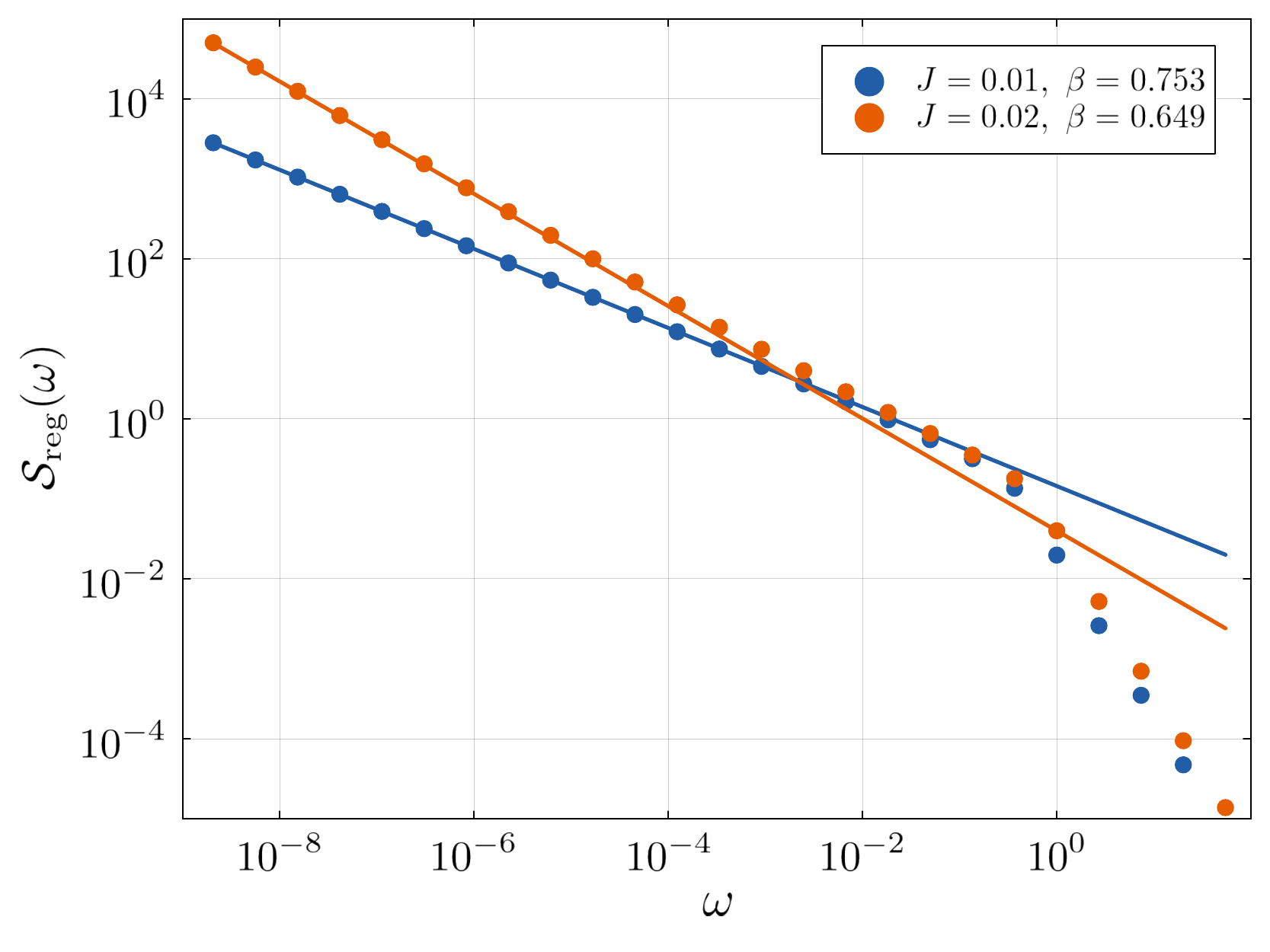}}\hfill
\panel{b}{\includegraphics[width=0.48\textwidth]{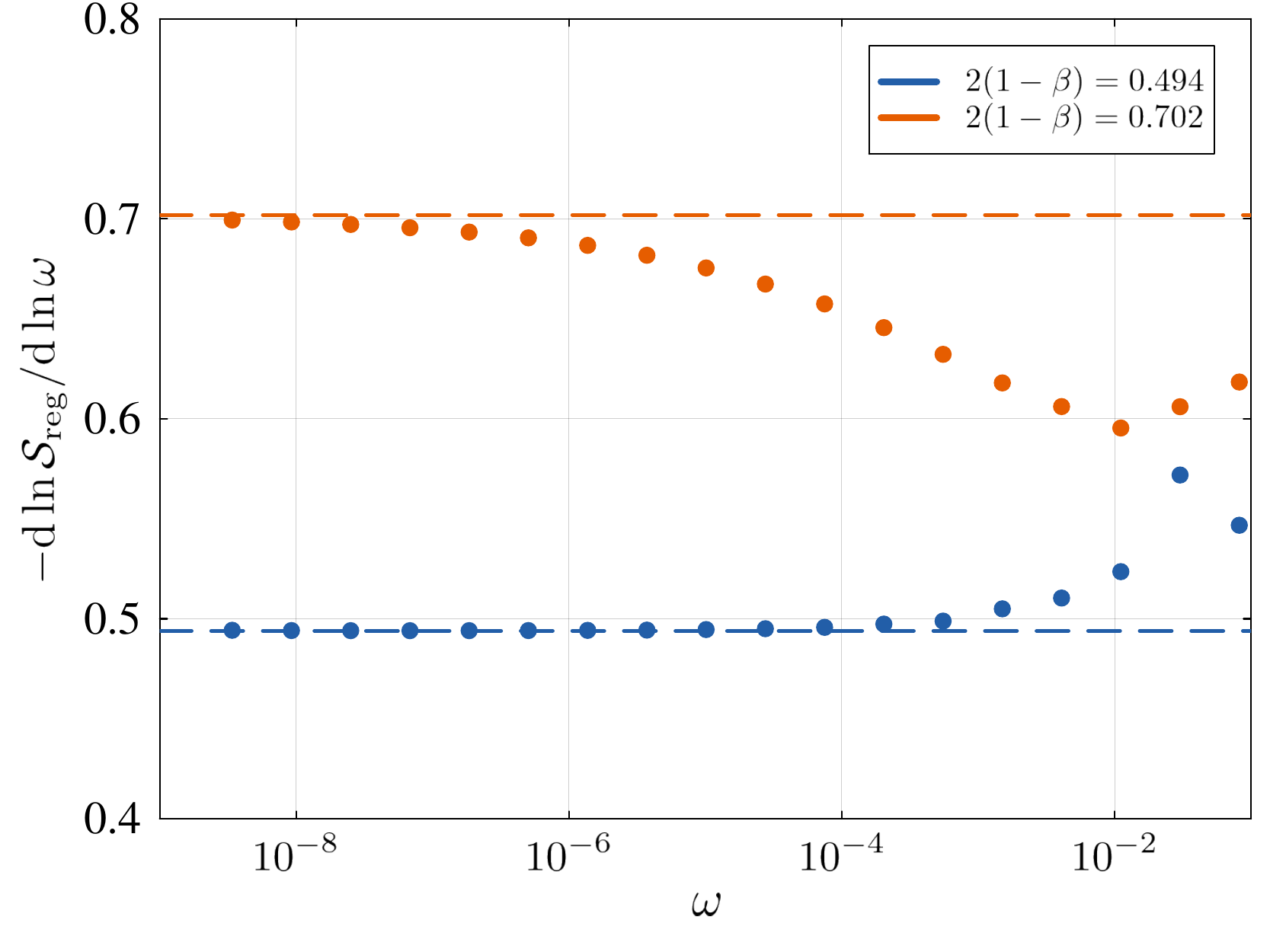}}
\caption{Toy model at complex field $H = \eta - {\rm i}\omega/2$, for $k = 2$ and $J = 0.01$ ($\beta = 0.753$, blue) and $J = 0.02$ ($\beta = 0.649$, orange). (a) Regular part $\mathcal{S}_{\rm reg}(\omega)$ of the spectral weight of $I_2(H)/H$, Eq.~\eqref{eq:spectraltoy}, the toy-model counterpart of $\avg{\rho}N^2\mathcal{C}(\omega)$ [Fig.~\ref{fig:N2C}(a)], from the complex version of Algorithm~1 extrapolated to $\eta \to 0^+$ (symbols). Lines: parameter-free prediction~\eqref{eq:Stoy}. (b) Local exponent $-{\rm d}\ln\mathcal{S}_{\rm reg}/{\rm d}\ln\omega$ between consecutive points of (a); dashed lines: $\mu = 2(1-\beta)$.}
\label{fig:Stoy}
\end{figure*}

\subsection{Delicate points}

The quantities that enter the scaling analysis are small differences between solutions at different $H$ [$\Delta P(t,H)$ ranges from a few times $10^{-3}$ down to $10^{-6}$ in Fig.~\ref{fig:1mpt}(b)], which must be resolved over many decades in $t$, and at the critical point over $|t_{\rm min}|$ up to about $54$. The algorithms above are designed around four sources of error. (i) \emph{Cancellations}: $y(t,t')$, $\tilde I_1(x)$, $1-e^{-e^{t}}$, $1-e^{H/(k-1)}$ and, at criticality, $1-P_H(t)$ must never be computed as differences of large or nearly equal numbers; with a naive implementation the errors are larger than $\Delta P$ itself. (ii) \emph{The endpoint $t_{\rm min}$}: the integrand of Eq.~\eqref{eq:PtH} is finite and smooth at $t' = t_{\rm min}$ (in the original variable $\theta'$ it is an even analytic function), but the factorized form~\eqref{eq:KH} of the kernel is an indeterminate $0/0$ there, since $s(t') \to 0$ while $x \to 0$; the limit must be inserted analytically, and the endpoint carries an $O(1)$ contribution that must be integrated with a high-order rule. Earlier versions of the code, which omitted the endpoint or treated it with the plain trapezoidal rule, had errors of order $\delta t$ and $\delta t^2$ respectively, visible as spurious bumps in $\Delta P(t,H)$. (iii) \emph{Boundary conditions at infinity}: Eqs.~\eqref{eq:Pt0}, \eqref{eq:lindp} and~\eqref{eq:linfmin} admit families of solutions, and the physical one is selected by imposing the asymptotic form on a strip wider than the range of the kernel [$L_0(t)$ is cut off at $|t| \simeq \ln(2/J)$ by the factor $e^{-J\cosh t}$] and by tuning the free constant ($c$ or $c_3$) until the solution joins the imposed form smoothly. (iv) \emph{Convergence at $J_c$}: the Picard iteration converges slowly there and requires many iterations, so that initializing each run from the solution at the previous $t_{\rm min}$ is essential.

\section{Toy model at complex field: numerical test} \label{app:complexH}

We have solved the cavity equation at complex $H = |H|\,e^{-{\rm i}\chi}$ with the complex version of Algorithm~1 (Appendix~\ref{app:numerics}), for $k = 2$ and $J = 0.01$ and $0.02$ ($\beta = 0.7530$ and $0.6490$, $c_3 = 0.9692$ and $0.9783$, $c_H = 0.9014$ and $0.341$). Along rays $0 \le \chi \le 0.495\pi$ the ratio $[I_2(H) - I_2(0)]/(c_H H^{2\beta-1})$, extrapolated to $|H| \to 0$, equals one within $10^{-4}$ for every $\chi$: $c_H$, hence $c_3$, does not depend on $\chi$, and the small-$H$ law of Eq.~\eqref{eq:I2scal} holds in the whole right half plane. Close to the imaginary axis, up to $\chi = 0.496\pi$, we have computed $I_2(H)/H$ at $H = \eta - {\rm i}\omega/2$ for $e^{-20} \le \omega \le e^{4}$ and extrapolated to $\eta \to 0^+$. The resulting $\mathcal{S}_{\rm reg}(\omega)$ agrees very well with the parameter-free prediction~\eqref{eq:Stoy} (Fig.~\ref{fig:Stoy}). It is also positive at all $\omega$, satisfies the sum rule $\int\de\omega\, \mathcal{S}_{\rm reg} = 1 - I_2(0)$ to better than $0.5\%$ and, inserted back into Eq.~\eqref{eq:spectraltoy} at real $H = \eta$, reproduces $I_2(\eta) - I_2(0)$ from the direct solution to within $0.2\%$.

\bibliography{references}

\end{document}